\pdfoutput=1
\documentclass[12pt]{article}
\usepackage[utf8]{inputenc}
\usepackage[margin=1in]{geometry} 
\usepackage{amsmath,amsthm,amssymb, graphicx, multicol, array, sgame, setspace, threeparttable, tabularx, rotating, booktabs, multirow, dcolumn}
\usepackage[square,sort&compress, authoryear, nonamebreak, round]{natbib}
\usepackage[table]{xcolor}
\usepackage{titlesec}

\usepackage{makecell}
\usepackage{bbm}
\usepackage{float}
\usepackage{mathtools}
\setcitestyle{round}
\usepackage{tikz}
\usepackage{pgfplots}
\usetikzlibrary{plotmarks}
\usepackage[ruled, linesnumbered]{algorithm2e}

\usepackage{ulem}
\definecolor{sangre}{rgb}{0.6,0.18,0.19}
\definecolor{dullmagenta}{rgb}{0.4,0,0.4}
\definecolor{darkblue}{rgb}{0,0,0.6}

\ifx\pdfoutput\undefined
\usepackage[backref=none,hypertex,colorlinks,urlcolor = darkblue,citecolor = sangre,linkcolor=blue]{hyperref}
\else
\usepackage[backref=none,hypertexnames=false,colorlinks,urlcolor = darkblue,citecolor =  sangre,linkcolor=blue]{hyperref}
\fi

\usepackage[capitalize]{cleveref}

\newcommand{\yyr}[1]{{\color{black}  #1 }}

\usepackage{color-edits}

\usepackage{algorithmicx}
\algnewcommand\COMMENT[1]{\hfill\(\triangleright\) #1}

\usepackage{pdflscape}

\usepackage{booktabs}
\usepackage{tabularx}

\usepackage{subcaption}
\usepackage{color,soul}
\usepackage{longtable}

\usetikzlibrary{decorations.pathreplacing} % Load library for path replacing

\usepackage{enumitem}
\newlist{steps}{enumerate}{1}
\setlist[steps, 1]{label = Step \arabic*:}

\hypersetup{
    colorlinks=true,
    linkcolor=blue,
    filecolor=magenta,      
    urlcolor=cyan,
    pdftitle={Overleaf Example},
    pdfpagemode=FullScreen,
    }
\title{AI Adoption in S\&P 500 Firms\thanks{
The authors gratefully acknowledge the support of the Alfred P. Sloan Foundation Grant No. G-2025-25158. The views expressed in this paper are solely those of the authors. }
}

\author{
Yang Yu\thanks{MIT FutureTech. Email: \texttt{yyu7@mit.edu}}
\and
Martin Fleming\thanks{MIT FutureTech. Email: \texttt{marti264@mit.edu}}
\and
Lucy Hampton\thanks{MIT FutureTech. Email: \texttt{lhampton@mit.edu}}
\and Christophe Combemale\thanks{Carnegie Mellon University. Email: \texttt{ccombema@andrew.cmu.edu}}
\and Neil Thompson\thanks{MIT FutureTech. Email: \texttt{neil\_t@mit.edu}}
}

\date{\today}

\begin{document}
\maketitle
% \begin{center}
% \Large\href{add google drive link}{[Click here for the latest version]}
% \end{center}
\begin{abstract}

The adoption of artificial intelligence (AI) by large enterprises is an important potential source of aggregate productivity improvement and labor market impact. We study AI adoption of S\&P 500 firms over the period 2016 to 2025, estimating adoption at the enterprise level. While generative AI tools are useful for personal and professional applications, our focus is on the deep integration of AI in the business processes of large enterprises which are bellwethers for firm adoption more broadly. We develop a novel measure to assess deep AI adoption (and distinguish it from AI hype) that is based on SEC 10-K filings, where laws and regulations ``prohibit companies from making materially false or misleading statements." In 2025, 11\% of S\&P 500 enterprises had AI deeply integrated into their business processes, and a further 10\% were using AI in the production of goods and delivery of services. AI adoption has more than quadrupled from 5\% in 2022 with slowly accelerating adoption among non-technology firms but very aggressive adoption in the technology sector which accounts for two-thirds of deeply integrated enterprise adoption. Firm profitability shows a “J-curve” as firms move from no adoption to deep adoption, but we observe no differences in capex or productivity. Among technology firms, but not others, AI adoption is higher for firms with more employees and higher values of Tobin's q.

\end{abstract}

\section{Introduction}

The advent of well-known artificial intelligence (AI) models has generated attention and headlines around the globe. Offerings such as Open AI's ChatGPT and Anthropic's Claude are fascinating investors, business leaders, and consumers. At the same time, these models have  renewed fears that workers can be replaced in a wide range of occupations. Despite remarkable progress in the training and deployment of large language models (LLMs), enterprise adoption remains limited. In late April 2026, the U.S. Census Bureau reported that 19.8\% of U.S enterprises used AI ``in any of its business functions'' in the last two weeks. In addition, during the next six months, 23.0\% of U.S. businesses ``think this business will be using'' AI ``in any of its business functions.''\footnote{See \url{https://www.census.gov/hfp/btos/data_downloads} for detailed data. Beginning with the first November 2025 survey, the Census Bureau altered the wording of both questions in the bi-monthly Business Trends and Outlook Survey from a question that asked for the use of AI in ``the production of goods and services'' to the use of AI for ``any business function.'' The second October 2025 survey had reported that 10\% of respondents used AI in ``the production of goods and services'' in the last two weeks and 14\% expected to use AI in ``the production of goods and services'' in the next six months. From the discontinued historic data, AI ``use in the production of goods and services'' gained 2.3 percentage points between October 2023 and October 2024 and 4.0 percentage points between October 2024 and October 2025.}

In this paper, we develop a novel rubric to measure AI adoption among S\&P 500 firms. We find that in 2025 21\% of S\&P 500 enterprises had AI deeply integrated (11\%) into their business processes or were using AI in the production of goods and delivery of services (10\%). AI adoption has more than quadrupled from 5\% in 2022 with slow but accelerating adoption among non-technology firms but very aggressive adoption in the technology sector. Regressing adoption on financial outcome measures recovers ``J-curve'' effects among firms achieving deep integration in business processes on profitability but with little impact on capex and productivity. Among technology sector firms, high values of Tobin's q and employment levels are associated with enterprises with more advanced AI adoption. 

Both financial markets and central banks have been eager for AI adoption benefits, while workers have been fearful of labor market impacts.\footnote{Financial markets, of course, are seeking to identify possible enterprise valuation improvement while central banks are watching for aggregate productivity gains to ease inflationary pressure.} With adoption progressing, three considerations govern the pace of AI adoption:

First, scaling laws apply to the AI models required for enterprise-class adoption \citep{bryan2026economic}. To execute tasks typically performed by workers, AI models must deliver results with an acceptably high level of accuracy. To achieve the necessary accuracy, models need to scale to considerable size and data volumes \citep{rosenfeld2019constructive}. Consequently, adoption requires enterprises to either support the high fixed costs for model training and data (if they build their own systems) or wait until general-purpose systems achieve sufficient accuracy. As \cite{svanberg2024beyond} showed for computer vision deployments, most potential AI deployments are unattractive if firms must themselves build the models. As \cite{mertens2026crashing} have shown, even cutting-edge foundation models cannot achieve average worker performance for most tasks, never mind the full set of tasks needed to automate processes. Consequently, independently of whether firms are building their own models or adopting foundation models, the challenging economics of scaling laws will delay adoption for many applications.
%Identifying tasks exposed to AI cannot predict the pace of adoption since it attempts to measure an overall potential for AI, not the technical feasibility and economic attractiveness of building such systems \citep{eloundou2024gpts}.

%\cite{svanberg2024beyond} present a new type of AI task automation model that is end-to-end, estimating the level of technical performance needed to do a task, the characteristics of an AI system capable of performance, and the economic choice of whether to build and deploy such a system. The result is an estimate of which tasks are technically feasible and economically attractive to automate---and which are not.

%cite{svanberg2024beyond} focus on computer vision, where cost modeling is more developed and data are plentiful. They find that at 2023's costs U.S. businesses would choose not to automate most vision tasks that have ``AI Exposure,'' and that only 23\% of worker wages paid for vision tasks would be attractive to automate. The slower roll-out of AI can be accelerated if costs fall rapidly or if it is deployed as an AI-agent or via AI-as-a-service platforms that have greater scale than individual firms. However, the AI-agent and AI-as-a-service platform markets remain in a nascent stage, limiting AI adoption \citep{goldmansachs2025agentic,shao2025future}.

Second, while adopting new technologies can improve profits \citep{aghion1992growth,aghion2001competition,aghion2005competition}, the adoption of AI requires organizational transformation and skill reconfiguration, which can be challenging and introduce risks. These reconfigurations are crucial, as the economics literature has established that management practices are a central source of productivity differences across firms \citep{bloom2010management} and that the returns to digital technologies are larger when firms pair information technology with complementary organizational practices and skills \citep{bresnahan2002it}. Adopting AI will require firms to adapt their practices, potentially causing an initial period of mismatch between their technology and management practices until they can be aligned. This can result in J-curve responses, with AI adoption initially producing productivity losses prior to producing gains \citep{mcelheran2025rise}. 

%are necessary for AI adoption as is the willingness to take on the risk of change. Transformation can grow out of the need for innovation or competitive pressure. In some instances, there is a desire to move to a new technology for profit improvement \citep{aghion1992growth,aghion2001competition,aghion2005competition}. 

%In addition, the economics literature has established that management practices are a central source of productivity differences across firms \citep{bloom2010management}. The literature has also found that the returns to digital technologies are larger when firms pair information technology with complementary organizational practices and skills. \cite{bresnahan2002it}, using firm-level data, found that IT-enabled organizational change, rather than just technology adoption, constitutes key skill-biased technical change, boosting productivity. Firms adopting the combination of IT and new organizational forms (``co-invention'') significantly increase their demand for skilled workers, who are better at navigating new, autonomous roles. The combination of technology and structural reorganization also leads to higher efficiency, with the highest productivity gains appearing in firms that implement both simultaneously.

Getting adoption right can also be challenging even if firms want to. \cite{makridis2025organizational} examines how managers lead adoption of genAI and links leadership to employee outcomes. He finds that ``the realized impact of genAI depends not only on tool availability, but also on the managerial infrastructure that reallocates tasks, legitimizes experimentation, and mitigates perceived downside risks.'' The empirical insight from \cite{makridis2025organizational} is that the perceived clarity of an organization's AI strategy is the most meaningful correlate of adoption. Workers who report a clear understanding of their organization's plan for integrating AI are more likely to be frequent adoptees, even after controlling for demographics and time effects.

Third, with a transformative technology, new risks are arise. \cite{acemoglu2023regulating} show that gradual enterprise adoption is optimal when facing such risks, because it enables greater learning. If all sectors immediately adopt the new technology and disaster strikes, many would not be able to avoid negative consequences. Gradual adoption instead allows enterprises, and society more generally, to gain updated knowledge as to the nature of the risks. As time passes without a mishap, the belief that there will be setbacks declines. As optimism increases, the technology is adopted across a larger number of sectors. \cite{acemoglu2024regulating} show that under reasonable conditions the adoption path is slow and convex, accelerating only after there is increased certainty that a mishap will not occur. They also show that adoption should be slower when the new technology has a higher growth rate and damages from setbacks are potentially large. This is exactly the case with AI, where \cite{thompson2024the} have shown that there are many important risks from AI, and \cite{delphi2026the} have shown that experts believe that the damages from AI risks could be considerable.

While internal enterprise risk is present, network risk is also a concern slowing AI adoption as business leaders react cautiously in the early stages of industry AI adoption. Production networks, service providers networks, and supply chains face risks as interdependencies require accurate and quality outcomes from AI models to accommodate network partners success.

The network perspective offers insight in how shocks occurring along the network can propagate across an economy, creating systemic risk with AI adoption within interconnected industries. \cite{carvalho2014micro} describes an economy that is an intricate web of specialized service providers and goods producers, each relying on input from suppliers to deliver their own service or produce their own good which, in turn, is routed to other downstream enterprises. 

Understanding these drivers, we estimate AI adoption among S\&P 500 firms. Large publicly held firms offer the advantage of providing publicly available annual financial results accompanied by detailed commentary.\footnote{To that extent AI adoption is proceeding, it is more likely found among large firms. For enterprises in the largest employment size category---250 employees or more---in late March 2026 the Census Bureau reported 35.3\% of firms using AI ``in any business function'', the highest adoption rate among employee size classes.} With available structured and unstructured data, we develop a novel rubric to classify a firm's level of AI adoption on a scale from no mention (1=lowest) to deep integration (5=highest). We extract AI adoption-related text from SEC EDGAR 10-K filings and create a two-step procedure to quantify AI adoption for each firm in each year. We provide both the AI-related paragraphs from the 10-K filings and the rubric to GPT-5-mini and prompt it to perform the classification. Our primary dataset for firms' basic firmographic information and financial performance comes from Compustat. For each of the S\&P 500 firms, we match basic firmographic information, financial performance, and AI adoption data over the period 2016 to 2025.\footnote{Late 2022 was a point of discontinuity when ChatGPT, based on the GPT-3.5 architecture, was released by OpenAI on November 30, 2022.}

Our rubric is designed to measure enterprise deployment of genAI and deep learning models in the performance of tasks in the execution of business processes. The use of genAI models as a stand-alone work tool may increase worker productivity, but by itself might not be adequate in all cases to perform a wide range of tasks, for instance searching for new molecules in the pharmaceutical industry and managing supply chains in the retail industry. For this reason, firms that score a 2 in the rubric are only exploring the possibly of adoption and those scoring a 3 are pilot-testing AI in products or processes but are not yet emphasizing financial results. Scores of 4 and 5 mean use of AI in production (4) and deeply embedded in business processes (5). See Appendix \ref{apx:sp500scores} for the complete list of S\&P 500 firms and their AI adoption scores from 2016 to 2025.  

With our novel rubric, we show that in 2025, 11\% of all S\&P 500 firms scored a 5 in the AI adoption rubric and 21\% of firms scored a 4 or 5. No mention of AI adoption is found for 18\% of firms. Among those firms scoring a 4 or 5 in 2025, 67\% were technology sector firms for a 62\% technology sector adoption rate.\footnote{The technology sector in this paper includes semiconductor, semiconductor equipment, software, technology hardware and equipment, and telecommunications firms. There are 76 such firms included.} Only 5\% of technology sector firms provide no mention of AI adoption in 2025.\footnote{While the Census Bureau's Business Trends and Outlook Survey does not provide sufficient detail to measure AI adoption for the technology sector, in late March 2026, 45\% of firms in the information services sector reported using AI ``in any business function.'' The information services sector includes software vendors and reports the highest adoption rate among all two-digit NAICS sectors.}

Regressing adoption on financial outcome shows differential effects across measures and sectors. Across all sectors, those firms using AI in production and having deeply integrated AI in business processes report higher net profits margins with a J-curve effect. Conversely, there is no significant association between AI adoption and productivity, as measured by revenue per employee. There is also no significant correlation between AI adoption and capex. We show that the well-known substantial technology sector capex expenditures are limited to fewer than a half dozen firms over two years in a data set of more than 500 firms over ten years. Finally, not surprisingly, there a statistically significant relationship between technology sector AI adoption and Tobin’s q.

The paper proceeds as follows. Section two describes the data, how the data set was constructed, and AI adoption scores.  Section three describes the econometric methods. Section 4 presents the results of the relationships among the various financial measures and AI adoption. Section five concludes. Throughout we are careful to observe that we are not making statements about causality, only descriptive relations among AI adoption and financial measures.

\section{Data}

We focus on the universe of S\&P 500 firms spanning from 2016 to 2025. The data set comprises 510 unique firms, reflecting additions to and deletions from the index over time. For example, Airbnb was added to the index effective on September 18, 2023 as part of a quarterly rebalancing, replacing Newell Brands in the index. The complete list of S\&P 500 firms in our sample is provided in the Appendix \ref{apx:sp500scores}. 
Our primary dataset on firms' basic information and financial performance comes from \emph{Compustat}. 
This database tracks publicly traded firms and provides standardized firm-level data, including (1) firm characteristics (e.g., location, employment, and industry classifications based on the North America Industrial Classification System(NAICS) and Global Industrial Classification System (GICS)); (2) financial statement variables (e.g., revenue, net income, capital expenditures, and total assets); and (3) stock prices and market valuations.

We complement Compustat with several other data. First, to construct a firm-year–level measure of AI adoption, we extract the textual content of firms' 10-K filings from \emph{SEC EDGAR} and create a two-step procedure to quantify AI adoption for each firm in each year. More precisely, in the first step, we employ a predefined list of AI-related keywords to extract paragraphs containing AI-related content from 10-K filings. This filtering step substantially reduces the data size and accelerates subsequent computations. In the second step, we develop a novel rubric to classify a firm’s level of AI adoption on a scale from 1 (lowest) to 5 (highest). We then provide both the AI-related paragraphs and the rubric to GPT-5-mini and prompt it to perform the classification. We then manually validate and refine the rubric and prompt to improve the algorithm’s performance. Details are provided in Section~\ref{sec:aiadoption}. As a result, we obtain 4,463 AI adoption scores covering 510 firms from 2016 through 2025. A small number of the companies were not publicly traded for the entire ten-year period.

Our data set consists of 510 firms, corresponding to 4,561 firm-year observations. We classify firms into three groups—technology, financial, and other—using the 4-digit GICS industry group codes. Specifically, we define firms in the \emph{Technology Hardware \& Equipment}, \emph{Semiconductors \& Semiconductor Equipment}, and \emph{Software \& Services} groups as the technology sector. Firms in \emph{Banks}, \emph{Insurance}, and \emph{Diversified Financials} are classified as the financial sector. All remaining firms, including \emph{Pharmaceuticals, Biotechnology \& Life Sciences,
Health Care Equipment \& Services,
Consumer Services,
Capital Goods,
Utilities,
Retailing,
Energy,
Real Estate,
Commercial \& Professional Services,
Food, Beverage \& Tobacco, 
Materials, 
Telecommunication Services,
Transportation,
Household \& Personal Products,
Automobiles \& Components,
Consumer Durables \& Apparel, 
Consumer Staples Distribution \& Retail}, are categorized as the other sector.

One caveat is that a firm’s business activities may span multiple sectors. Moreover, GICS codes in Compustat are analyst-curated—based primarily on a firm’s main revenue source—rather than self-reported, and may not always reflect the most  relevant classification for our purposes. To address this concern, we manually review the sector assignments and make adjustments where appropriate. For example, we reclassify Amazon and Tesla as technology firms, although their GICS industry groups are \emph{Retailing} and \emph{Automobiles \& Components}, respectively.

\cref{tab:summarystat} reports the summary statistics of our sample. 
\emph{Net Profit Margin}, \emph{Capex-to-Revenue}, and \emph{Tobin's Q} each capture distinct dimensions of a firm's financial performance. \emph{Net Profit Margin} reflects current profitability—measuring net income per dollar of sales—and thus captures operating efficiency, cost structure, and pricing power. \emph{Capex-to-Revenue} measures investment intensity relative to firm size, indicating the extent to which firms reinvest to support growth. Traditional capital-intensive industries, such as manufacturing and utilities, typically exhibit high \emph{Capex-to-Revenue} ratios due to substantial investment in physical infrastructure. In contrast, modern AI-intensive and digital firms may also display higher ratios, reflecting increased spending on computing power and data center infrastructure. \emph{Tobin's Q} reflects market valuation relative to the book value of assets and serves as a forward-looking proxy for growth opportunities and expected future profitability.

Notably, \emph{Headcount} and \emph{Tobin's Q} exhibit pronounced right skewness, indicating the presence of a small number of “superstar” firms. To mitigate the influence of outliers, we use the logarithmic transformation of these variables in the regression analysis below. We also provide the more detailed summary statistics by AI scores in the Appendix \ref{apx:detailstats}.

\begin{table}[htbp]
\caption{\bf Summary Statistics}
    \centering
    \begin{tabular}{lllllllll}
\toprule
 & count & mean & std & min & 25\% & 50\% & 75\% & max \\
\midrule
Year & 4561 & 2020.56 & 2.88 & 2016 & 2018 & 2021 & 2023 & 2025 \\
Sector & 4561 & 0.16 & 0.36 & 0 & 0 & 0 & 0 & 1 \\
Headcount & 4561 & 57.03 & 139.73 & 0.01 & 9.00 & 20.00 & 55.06 & 2300.00 \\
Net profit margin & 4561 & 0.12 & 0.25 & -8.54 & 0.06 & 0.11 & 0.18 & 4.39 \\
Capex-to-Revenue & 4561 & 0.08 & 0.13 & 0.00 & 0.02 & 0.04 & 0.07 & 2.74 \\
Tobin's Q & 4561 & 2.27 & 2.72 & 0.00 & 0.73 & 1.43 & 2.76 & 47.75 \\
AI score & 4470 & 1.77 & 1.15 & 1 & 1 & 1 & 3 & 5 \\
\bottomrule
\end{tabular}

    \captionsetup{font=footnotesize}
    \vspace{0.1in}
    \caption*{\textit{Note: }\emph{Headcount} is in thousand, \emph{Net profit margin} is net income-to-revenue, \emph{Tobin's Q} is market value-to-book value.}
    \label{tab:summarystat}
\end{table}
\subsection{AI Adoption Metric}\label{sec:aiadoption}

As described above, we employ a two-step procedure to measure firm-year–level AI adoption. In the first step, we extract from each 10-K all paragraphs that contain at least one term from an AI-related keyword list. This is because 10-K filings are typically very long—often approaching one million tokens—while information on AI adoption is sparsely distributed throughout the text. Feeding such lengthy inputs directly into a language model may lead to attention dilution and increase hallucination-like behavior. We therefore restrict the rubric-based classification to extracted, AI-relevant text. 

We compile a list of AI-relevant keywords that is provided in Appendix~\ref{apx:aikeywords}. This list is primarily drawn from the AI-related skill taxonomy documented in \cite{acemoglu2022artificial}. We supplement it with selected terms from the NeurIPS paper submission keywords and the top 15 AI methods identified in \cite{gao2024quantifying}.\footnote{NeurIPS keywords: \url{https://neurips.cc/Conferences/2017/PaperInformation/Keywords}}
We manually validate the coverage of this list and find that most AI-relevant paragraphs contain at least one of four core keywords—Artificial Intelligence, Machine Learning, AI, and ML—while the remaining terms incrementally improve the list's coverage.

In the second step, we start by developing a rubric for classifying AI adoption, which is presented in \cref{tab:rubric_aiadoption}. This rubric is ordinal, ranging from 1---not mentioning current AI adoption---to 5---AI is a core component of the firm's strategy and financial performance. We feed the rubric and AI-relevant paragraphs to GPT-5-mini and ask it to perform the classification using the following prompt: \emph{You are an expert analyst. 
    Using the rubric below, assign a single integer score (1–5) that best represents the company's level of AI adoption. Respond with only the score (1, 2, 3, 4, or 5).
}

\begin{table}[htbp]
    \centering
    \caption{\bf AI Adoption Rubric}
    \label{tab:rubric_aiadoption}
    \begin{tabularx}{\textwidth}{c X}
        \toprule
        Score & Description \\
        \midrule
        \bf 1 & No mention of AI, or references to AI only in a competitor or industry context, or acknowledgment of AI as a potential future necessity or risk, with no evidence of current adoption. \\[0.5em]
        
        \bf 2 & The firm is exploring the possibility of AI adoption or building adoption capacity, including early-stage implementation in limited areas (e.g., investments in data and analytics to support artificial intelligence, or early efforts to develop AI-related capabilities to improve productivity and efficiency), without clear evidence that AI is embedded in current products or operations. \\[0.5em]
        
        \bf 3 & AI is integrated into specific products or operational processes, but financial impacts or strategic dependence are not yet emphasized. \\[0.5em]
        
        \bf 4 & AI is deployed at a production level, with explicit expectations of financial outcomes such as cost savings or revenue contributions. \\[0.5em]
        
        \bf 5 & AI is a core component of the firm’s strategy and financial performance, deeply embedded across business functions and operations. \\
        \bottomrule
    \end{tabularx}
\end{table}

\cref{tab:example_aiadoptionscores} presents one example of AI-relevant paragraphs for each score level. For instance, Meta’s 10-K emphasizes that AI is a core strategic component across its ecosystem of products and services and that it has improved the firm’s efficiency and productivity; accordingly, it receives a score of 5. In contrast, the Everest example merely acknowledges changes in the regulatory environment driven by AI, with no evidence of firm-level adoption, and is therefore assigned a score of 1. We manually double-checked the scoring outcomes and find them to be largely consistent with common expectations. For example, since 2023, all firms in the “Magnificent Seven” receive a score of 5, with the exception of Apple Inc., which is widely viewed as lagging behind its peers in AI adoption.\footnote{\url{https://www.artificialintelligence-news.com/news/why-apple-is-playing-it-slow-with-ai/}} See Appendix~\ref{apx:sp500scores} for the complete list of AI adoption scores for S\&P 500 firms from 2016 to 2025. 

\begin{table}[htbp]
    \centering
    \caption{\bf Examples of AI Adoption Scores}
    \label{tab:example_aiadoptionscores}
    \begin{tabularx}{\textwidth}{c X c}
        \toprule
        Company & AI Paragraphs & Score\\
        \midrule
        \bf Meta & Across our work, we are innovating in artificial intelligence (AI) technologies to build new experiences that help make our platform more social, useful, and immersive. Our AI investments support initiatives across our products and services, helping power the systems that rank content in our apps, our discovery engine that recommends relevant content, the tools advertisers use to reach customers, the development of new generative AI experiences, and the tools that make our product development more efficient and productive. & 5 \\[0.5em]
        
        \bf Fedex & Leveraging the capabilities of FedEx Dataworks, developments in data and technology, including artificial intelligence and machine learning, are facilitating the execution of our DRIVE transformation by creating new opportunities to improve our operational efficiency ... Federal Express is also testing autonomous, driverless technologies in the handling of large, non-conveyable packages, as well as artificial intelligence-enabled robotic product sortation systems to sort small packages. & 4\\[0.5em]
        
        \bf Fiserv & We currently use AI in a variety of ways, including to enable higher quality customer service experiences, platform analytics, and fraud mitigation across a number of solutions. & 3 \\[0.5em]
        
        \bf Best Buy & We expanded our learning campaigns, reaching thousands of employees for new products to include a closer partnership with our vendors, more exposure for our employees and in-depth training in artificial intelligence to help customers with the products they use. & 2 \\[0.5em]
        
        \bf Everest & The cybersecurity regulatory environment is evolving, in particular with respect to emerging technologies, such as artificial intelligence, and it is likely that the costs of complying with new or developing regulatory requirements will increase. & 1 \\
        \bottomrule
    \end{tabularx}
    \captionsetup{font=footnotesize}
    \caption*{\textit{Note: }All these scores are corresponding to 2024 scores.}
\end{table}

\subsubsection{Validation}

We do two validation exercises. First, we compare our metric to AI adoption estimates from the US Census Bureau's Business Trends and Outlook Survey (BTOS). 
The BTOS sample consists of approximately 1.2 million businesses drawn from the Census Business Register, including both privately held and publicly traded companies.
We validate our AI adoption measure using responses to Question 7 of the survey: ``\textit{In the last two weeks, did this business use Artificial Intelligence (AI) in producing goods or services? (Examples of AI: machine learning, natural language processing, virtual agents, voice recognition, etc.)}." 
Since BTOS data are only available at the industry level, we aggregate our 10-K AI scores by taking a simple average within each industry. We then compare these to mean BTOS responses over 2023–2024, the period for which data are available, noting that direct comparability is limited given our sample is restricted to S\&P 500 firms. 

\begin{figure}[H]
    \centering
    \caption{\bf Industry-Level AI Adoption Validations}
    \begin{subfigure}[t]{0.49\linewidth}
        \centering
        \includegraphics[width=\linewidth]{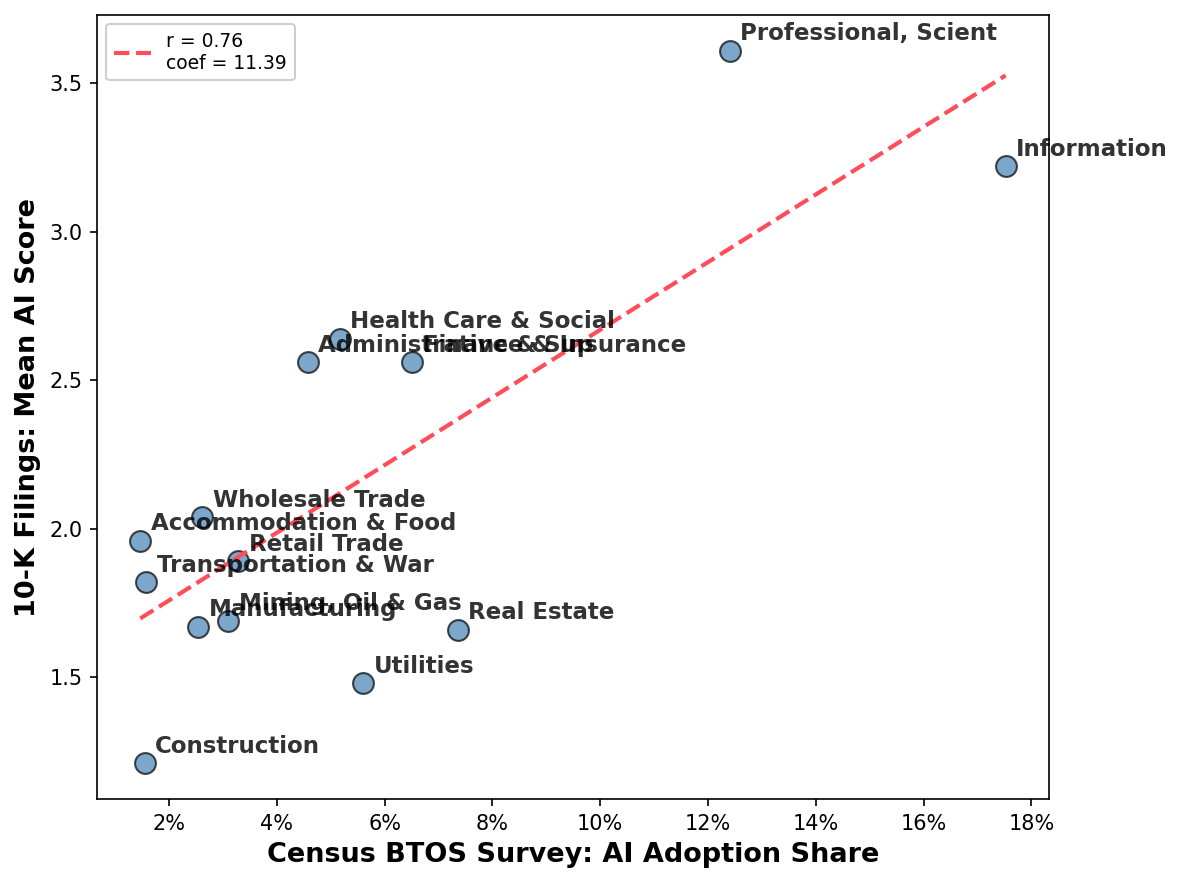}
        \caption{10-K vs Business Trends and Outlook Survey}
        \label{fig:ai_validation_btos}
    \end{subfigure}
    \hfill
    \begin{subfigure}[t]{0.49\linewidth}
        \centering
        \includegraphics[width=\linewidth]{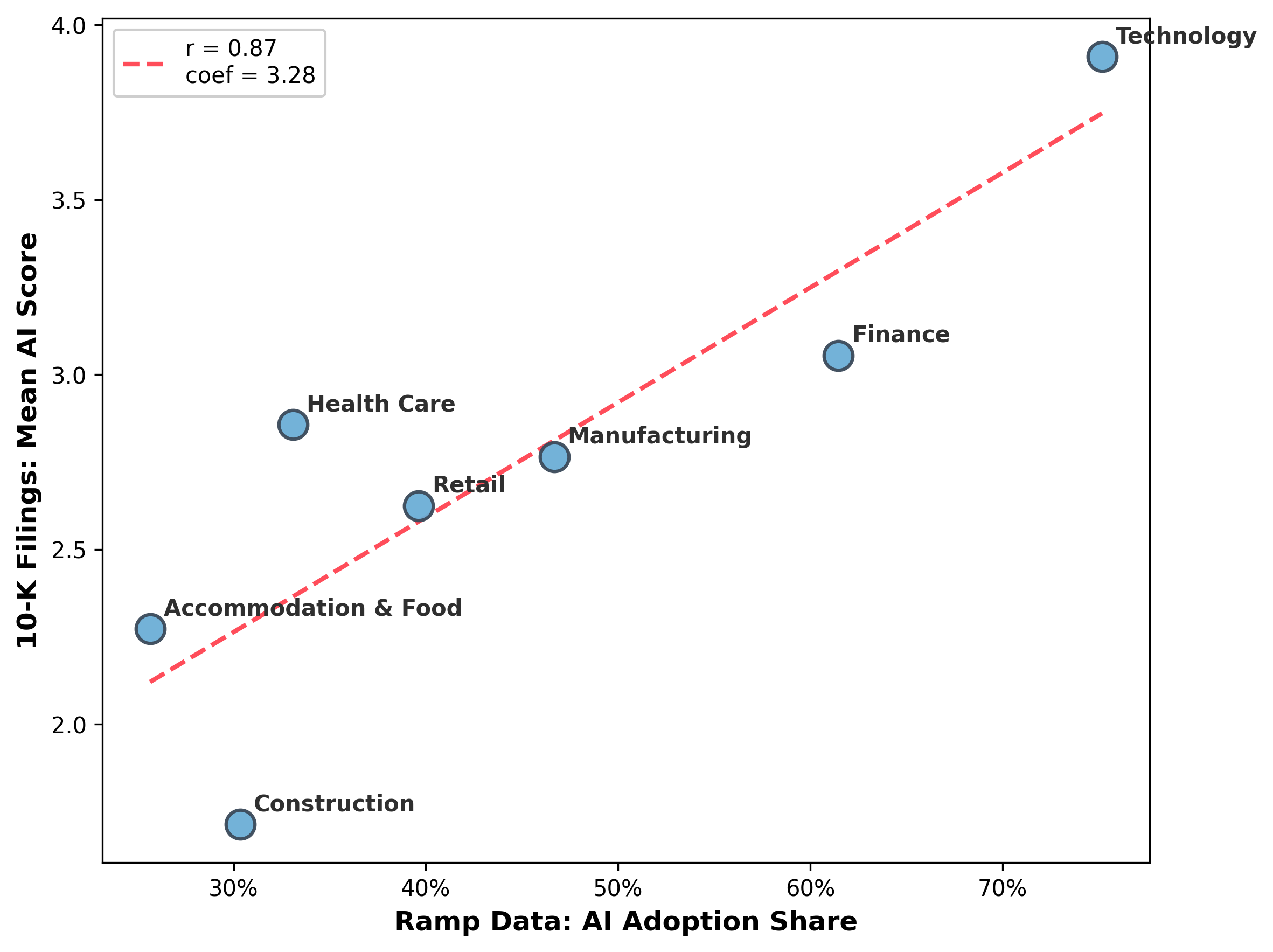}
        \caption{10-K vs Ramp}
        \label{fig:ai_validation_ramp}
    \end{subfigure}
\end{figure}

We present our mean-score comparison in \cref{fig:ai_validation_btos}.
As an example, among BTOS-sampled businesses in Information sector, approximately 18\% report using AI in the production of goods or services. In contrast, according to our metric, the mean AI score of S\&P 500 firms in the same sector over the same period is around 3.3. Our AI adoption metric is positively correlated with BTOS responses across industries, with a correlation coefficient of 0.76, suggesting that our 10-K–based measure captures similar underlying patterns of AI adoption as the BTOS survey, despite differences in sample composition and measurement approaches.
% We defer the threshold-based validation plots (AI score $>$ 1, 2, and 3) in Appendix \ref{sec:metric_validation}.
% Across all choices of AI score cutoffs, the two measures are strongly correlated, with the strongest correlation observed at a threshold of 4 or higher ($\rho = 0.82$).

An alternative data source for validation is Ramp, a US-based fintech company that provides corporate card and pay bill platforms to tens of thousands of businesses in the US. Starting from 2023, they periodically publish AI adoption share---the percentage of businesses in a given sector that had a positive transaction amount for an AI product or service in a given time period---based on the transaction data from over 50,000 US business clients.\footnote{\url{https://docs.ramp.com/developer-api/v1/ramp-data/ai-index}} 
Again, we compare the aggregated (to sector level) mean AI scores from our 10-K–based metric to the Ramp AI adoption share in the same sector and year. \cref{fig:ai_validation_ramp} shows a strong correlation of 0.87 between the two measures in 2025 (with correlations of 0.72 and 0.82 in 2023 and 2024, respectively), further reinforcing the validity of our AI adoption metric. This comparison has the same caveat as the BTOS comparison, in that Ramp's sample skews towards smaller companies and has limited coverage of the S\&P 500, so the comparison is not apples-to-apples.

\subsection{Empirical Observations}\label{sec:eobs}

\subsubsection{Enhanced AI Adoption}

\begin{figure}[H]
    \caption{\bf AI Adoption at Firm-Year Level}
    \centering
    \begin{subfigure}{1.11\textwidth}
        \centering
        % \caption{All}
        \includegraphics[width=\linewidth]{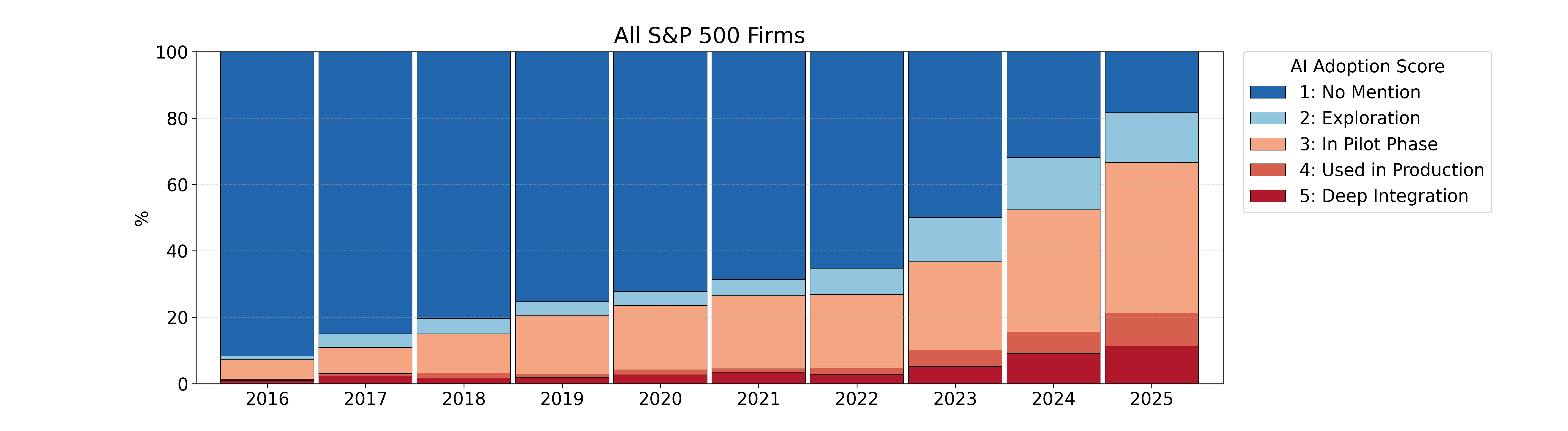}
        \label{fig:all}
    \end{subfigure}

    \vspace{-2em}

    \begin{subfigure}{1.11\textwidth}
        \centering
        % \caption{Tech}
        \includegraphics[width=\linewidth]{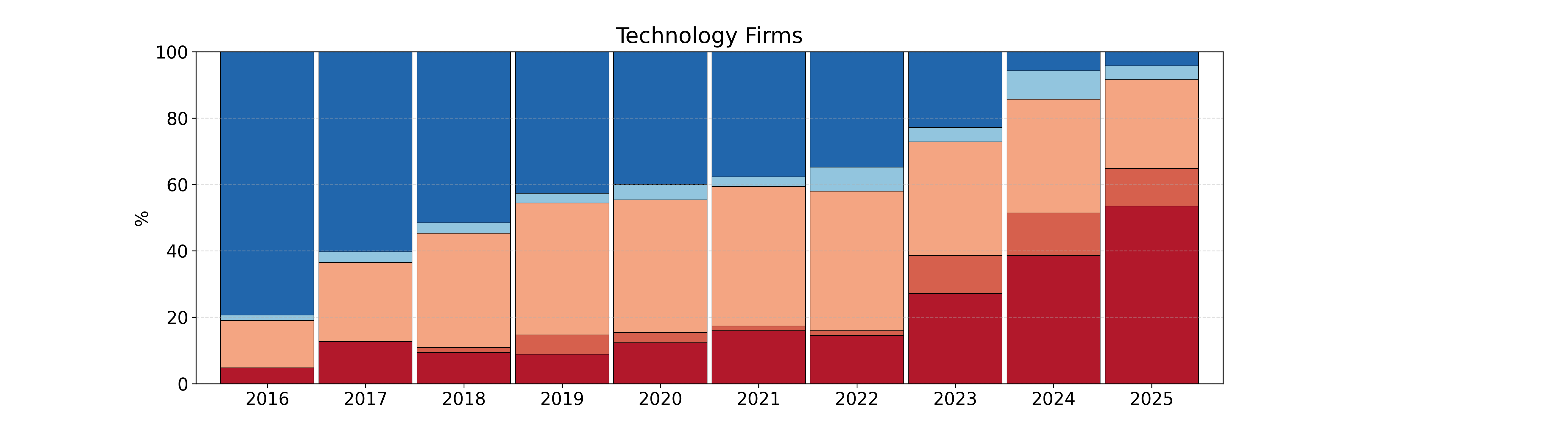}
        \label{fig:tech}
    \end{subfigure}

    \vspace{-2em}

    \begin{subfigure}{1.11\textwidth}
        \centering
        % \caption{Financial}
        \includegraphics[width=\linewidth]{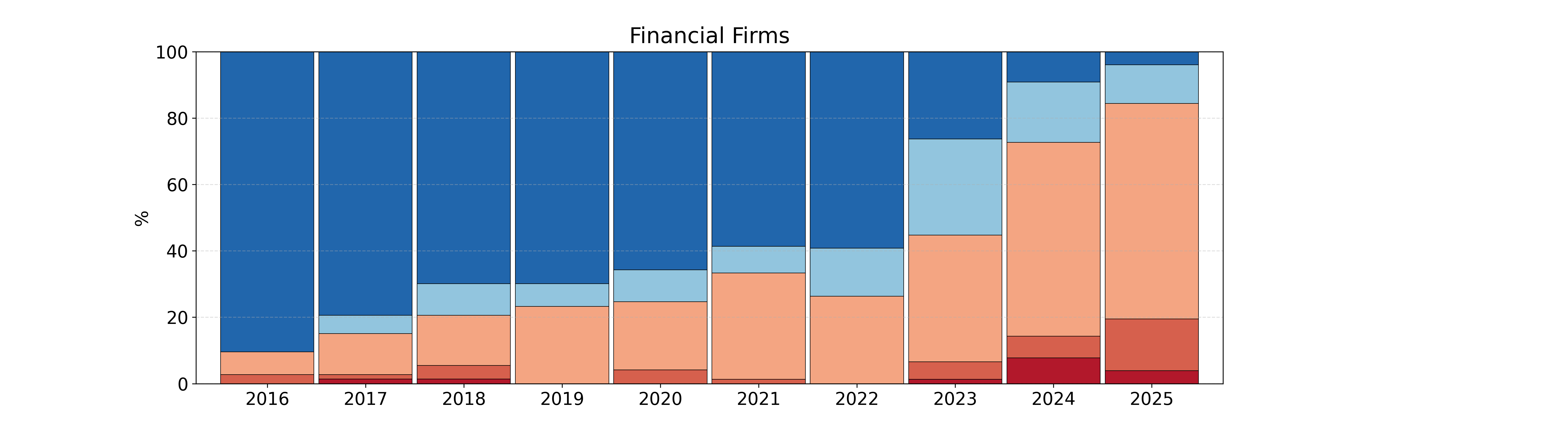}
        \label{fig:financial}
    \end{subfigure}

    \vspace{-2em}

    \begin{subfigure}{1.11\textwidth}
        \centering
        % \caption{Other}
        \includegraphics[width=\linewidth]{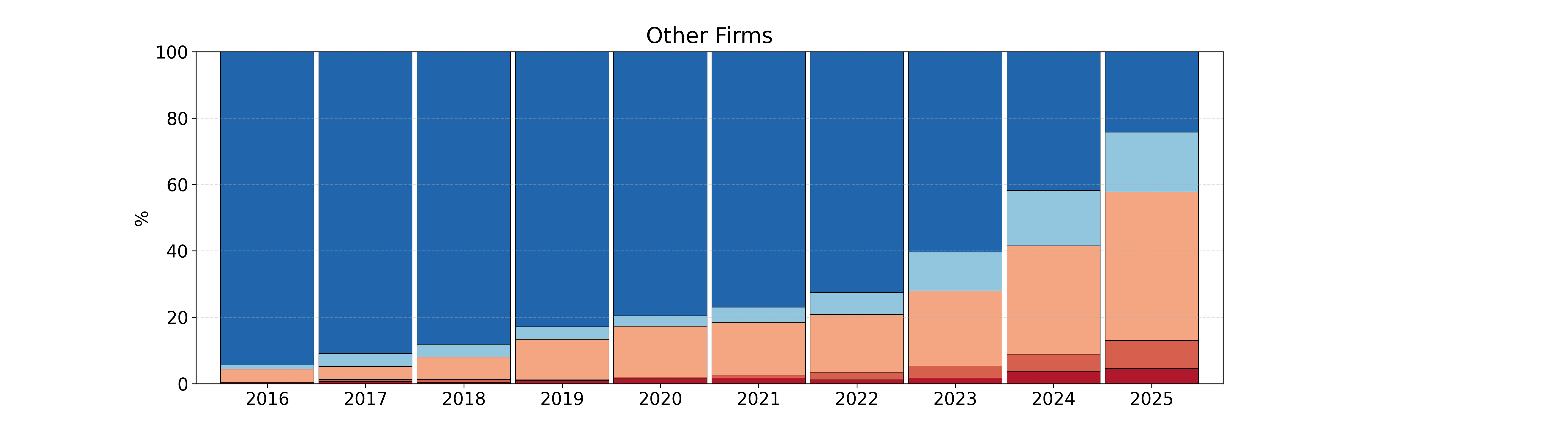}
        \label{fig:other}
    \end{subfigure}
    \captionsetup{font=scriptsize}
    \caption*{\textit{Note: }Each bar shows the proportion of firms falling into each AI Adoption level from 1 to 5. The detailed values are reported in \cref{tab:score_distribution_sector_2021_2025} in Appendix \ref{sec:aiscore_value_bysector}.}
        \label{fig:aiadoption}
\end{figure}

\cref{fig:aiadoption} illustrates the evolution of AI adoption over the past decade for S\&P 500 companies as a whole as well as at sector level. Overall, AI adoption rises sharply beginning in 2023, following the release of ChatGPT at the end of 2022, and continues to increase rapidly thereafter. This pattern is evident both at the aggregate level and across sectors. Consistent with expectations, the technology sector leads in AI adoption, while sectors such as retailing and transportation exhibit relatively lower speed of adopting AI.

Viewed by industrial sector in 2025, the S\&P 500 firms fall into three tiers (\cref{tab:ggroup_score_distribution_2025}). The clear leading-edge sectors are software \& services, semiconductors and technology hardware and equipment, with over 50\% of the firms scoring 4 or 5 on the rubric, suggesting that AI models are deeply embedded across business functions and operations, or have been deployed at a production level and with explicit expectations of financial outcomes such as cost savings or revenue contributions. There are no other sectors in which the combined scores of 4 and 5 exceed 50\%. The next closet sector is telecommunication services with combined 4 and 5 score of 50\%.

In a second tier of 13 sectors, 40\% to 70\% of firms scored a 3 in 2025. Among these firms, AI is in advanced pilot deployment with integration into specific products or operational processes, but with financial impacts or strategic dependence are not yet emphasized. Banks, diversified financial services firms, and insurance firms are among the leading sectors in this tier. Commercial and professional services and consumer services are also among those sectors with a majority of firms with advanced pilots or earlier deployment. 

Among a third tier of sectors are consumer staples; household and personal products; and food, beverage, and tobacco products. In these sectors 25 \% of firms have no mention of AI-related activity in 2025. With the exception of the food, beverage,and tobacco sector, all other sectors in this tier have less than 10 \% of firms with a score of 4 or 5.

\begin{table}[H]
    \caption{\bf AI Adoption Score Distribution by Industry Group in 2025}
    \centering
    \makebox[\textwidth][l]{\hspace*{-2.1cm}\begin{tabular}{lrrrrr}
\toprule
Sector & No Mention & Exploration & Pilot & \shortstack{Used in\\Production} & \shortstack{Deep\\Integration} \\
\midrule
Software and Services & 0.00 & 0.00 & 23.33 & 6.67 & 70.00 \\
Technology Hardware and Equipment & 4.76 & 4.76 & 23.81 & 23.81 & 42.86 \\
Semiconductors and Semiconductor Equipment & 10.00 & 10.00 & 30.00 & 10.00 & 40.00 \\
Automobiles and Components & 25.00 & 0.00 & 50.00 & 0.00 & 25.00 \\
Commercial and Professional Services & 7.14 & 14.29 & 42.86 & 21.43 & 14.29 \\
Pharmaceuticals, Biotechnology and Life Sciences & 8.00 & 16.00 & 64.00 & 4.00 & 8.00 \\
Transportation & 15.38 & 15.38 & 46.15 & 15.38 & 7.69 \\
Diversified Financial & 2.44 & 7.32 & 60.98 & 21.95 & 7.32 \\
Retailing & 23.53 & 11.76 & 52.94 & 5.88 & 5.88 \\
Healthcare Equipment and Services & 5.71 & 11.43 & 51.43 & 25.71 & 5.71 \\
Energy & 54.55 & 13.64 & 22.73 & 4.55 & 4.55 \\
Real Estate & 10.34 & 27.59 & 51.72 & 6.90 & 3.45 \\
Capital Goods & 24.49 & 18.37 & 51.02 & 4.08 & 2.04 \\
Telecommunication Services & 0.00 & 25.00 & 25.00 & 50.00 & 0.00 \\
Consumer Services & 26.32 & 10.53 & 47.37 & 15.79 & 0.00 \\
Insurance & 4.35 & 26.09 & 60.87 & 8.70 & 0.00 \\
Banks & 7.69 & 0.00 & 84.62 & 7.69 & 0.00 \\
Household and Personal Products & 31.82 & 27.27 & 36.36 & 4.55 & 0.00 \\
Utilities & 48.39 & 12.90 & 35.48 & 3.23 & 0.00 \\
Consumer Durable and Apparel & 25.00 & 16.67 & 58.33 & 0.00 & 0.00 \\
Food, Beverage and Tobacco & 28.57 & 14.29 & 57.14 & 0.00 & 0.00 \\
Materials & 37.04 & 25.93 & 37.04 & 0.00 & 0.00 \\
Consumer Staples Distribution and Retail & 28.57 & 42.86 & 28.57 & 0.00 & 0.00 \\
\bottomrule
\end{tabular}
}
    \captionsetup{font=footnotesize}
    \caption*{\textit{Note: }Each row is a GICS industry group. Values are the share (in \%) of firms in that group with AI adoption score 1 through 5 in year 2025.}
    \label{tab:ggroup_score_distribution_2025}
\end{table}

\subsubsection{Net Profit Margin}

\begin{figure}[H]
    \centering
    \caption{\bf AI Adoption and Net Profit Margin}
    \includegraphics[width=0.95\linewidth]{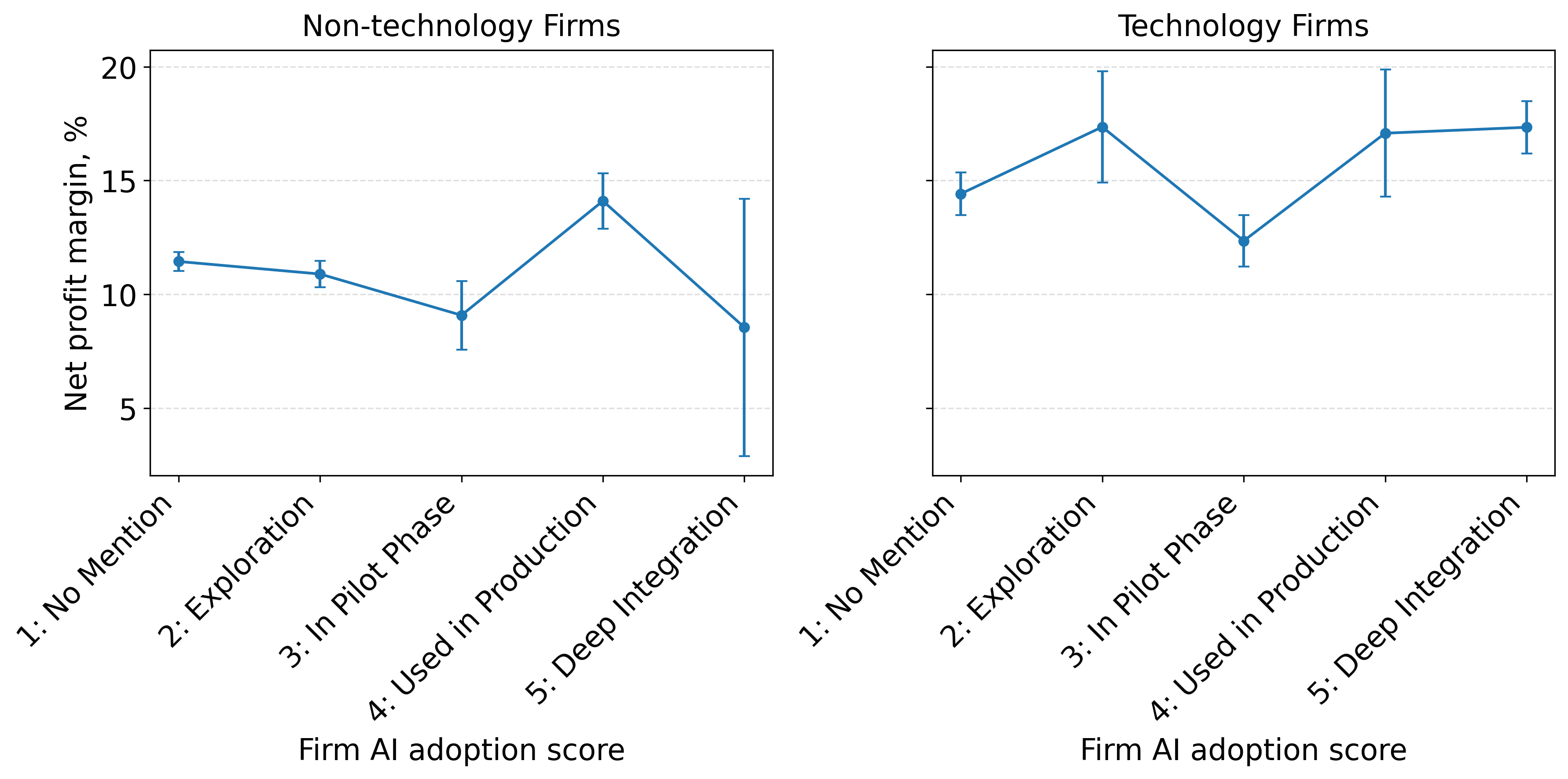}
    \label{fig:aiscore_npm}
    \captionsetup{font=footnotesize}
    \caption*{\textit{Note: }This figure reports the mean net profit margin of firms at each level of AI adoption from 2016 to 2025. Error bars represent one standard error of the mean estimates. The error bar for AI score 5 in the non-technology sector is notably wide due to the small number of observations in that category.}
\end{figure}

\cref{fig:aiscore_npm} reports the mean net profit margin of firms at each level of AI adoption from 2016 to
2025. In the technology sector, firms at AI score 1 (no or very limited adoption), report an average net profit margin of 14.4\%. Among those firms with an AI score of 3, the product-level deployment phase, the average profit margin is 12.4\%. However, among those firms with deep adoption and score a 5 the profit margin is 17.4\%. On average, the net profit margin is 3 percentage points higher among those firms who have deeply integrated AI models into their business processes than those who have taken little or no action.

\cref{fig:aiscore_npm} also shows among non-technology sector firms in the early stages of adoption (AI score = 3)  have low profit margins than those in more advanced stages of deployment (AI score = 4). With only 19 non-technology firms in the most advanced stage of deployment and a very wide confidence interval, comparison is not meaningful.

\subsubsection{Capex-to-Revenue Ratio}

\begin{figure}[H]
    \centering
    \caption{\bf AI Adoption and Capex-to-Revenue}
    \includegraphics[width=0.95\linewidth]{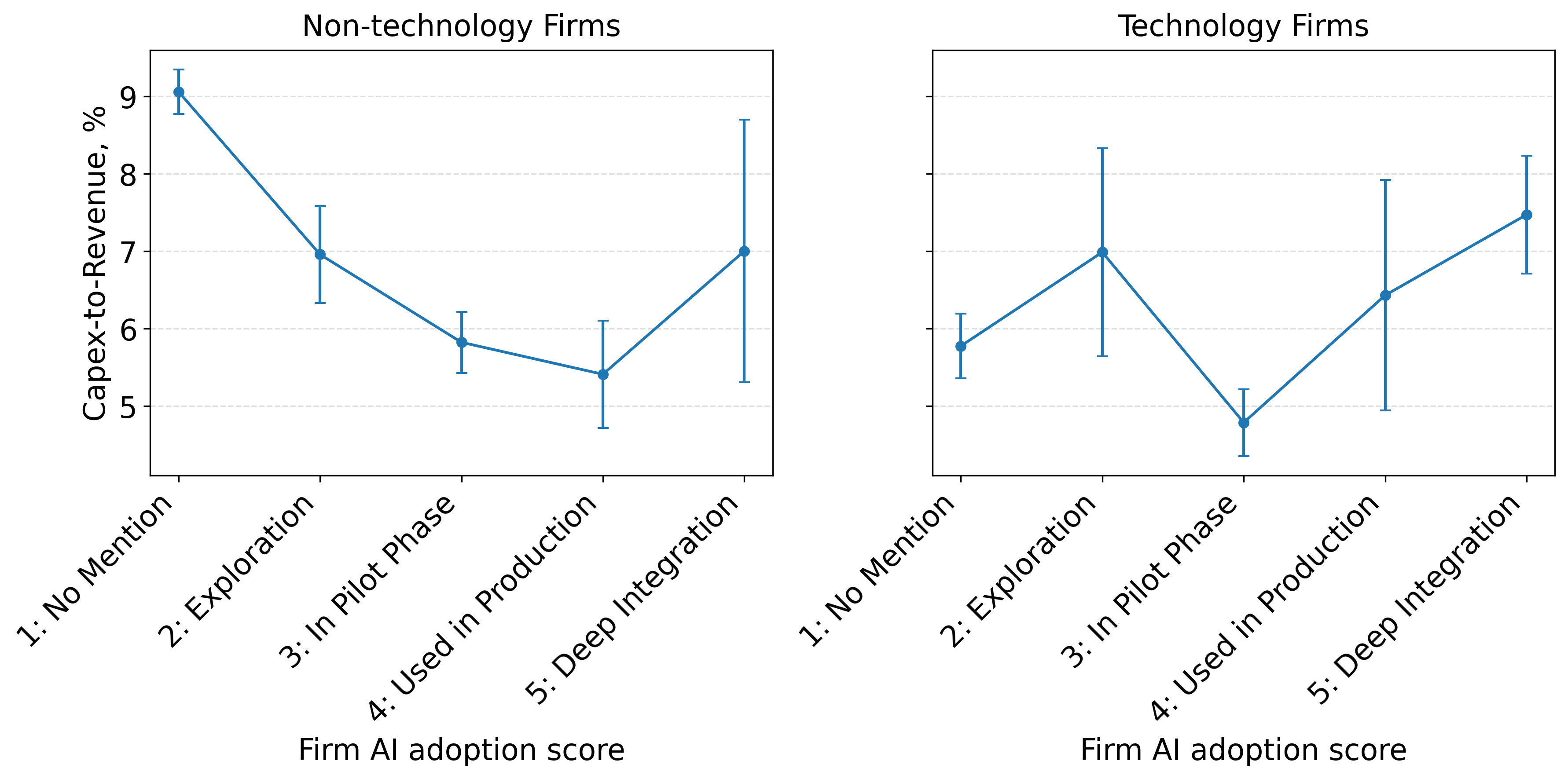}
    \label{fig:aiscore_capx_revt}
    \captionsetup{font=footnotesize}
    \caption*{\textit{Note: }This figure reports the mean capex-to-revenue ratio of firms at each level of AI adoption from 2016 to 2025. Error bars represent one standard error of the mean estimates. The error bar for AI score 5 in the non-technology sector is notably wide due to the small number of observations in that category.}
\end{figure}

\cref{fig:aiscore_capx_revt} reports the mean capex-to-revenue ratio of firms at each level of AI adoption from 2016 to
2025. In neither the technology nor the non-technology sector do we observe a clear monotonic relationship between AI adoption and capex intensity. A potential explanation is that most firms, rather than building in-house AI infrastructure, act primarily as consumers of AI services through APIs or cloud platforms, making their AI adoption largely disconnected from capital expenditures.

% \yang{add description of the table below.}
% \begin{table}
%     \centering
%     \caption{\bf Capex-to-Revenue in Non-Technology Sector}
%     \input{../tables/su_capx_revt_nontech_by_aiscore.tex}
% \end{table}

% \begin{table}
%     \centering
%     \caption{\bf Capex-to-Revenue in Technology Sector}
%     \input{../tables/su_capx_revt_tech_by_aiscore.tex}
% \end{table}

% [Before we discuss Figure 5, can we add a figure like figure 4 for capex-to-revneue?]

\begin{figure}[H]
    \centering
    \caption{\bf Capex-to-Revenue Ratio of Magnificent Seven}
    \includegraphics[width=0.85\linewidth]{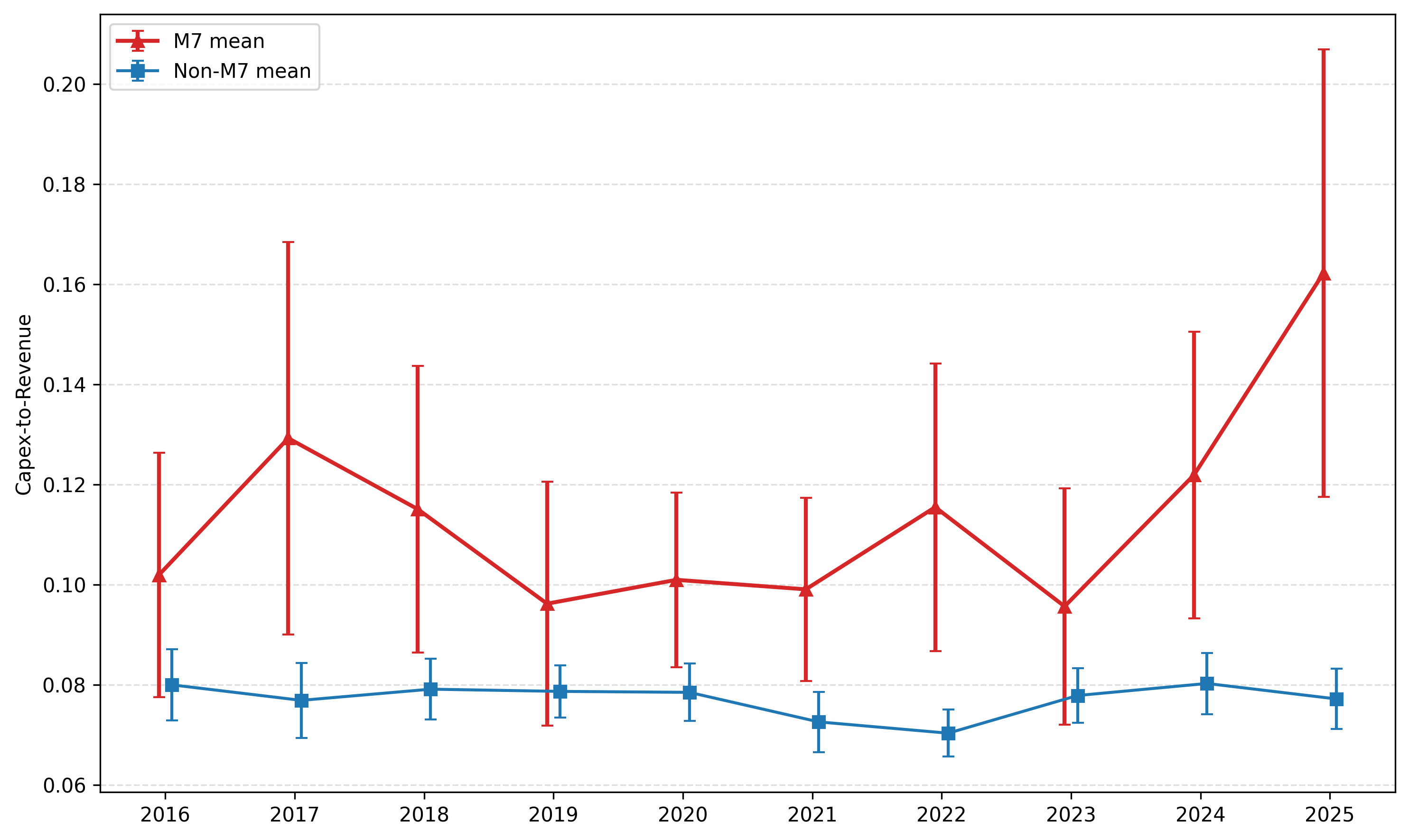}
    \label{fig:aiscore_capxrev}
    \captionsetup{font=footnotesize}
    \caption*{\textit{Note: }This figure reports the mean capex-to-revenue ratio of firms inside and outside the Magnificent Seven from 2016 to 2025, the error bars represent one standard error. 
    % The black dashed line represents the median for firms outside the Magnificent Seven. 2017 was one of the most investment-heavy years in Tesla’s history due to Model 3 ramp-up and massive manufacturing expansion. Meta's CAPEX surged in 2018, 2022 and 2025 due to data center expansion, Metaverse and the construction of AI-optimized data centers, and building AI infrastructure during the AI supercomputing arms race. 
    }
\end{figure}

Since AI-related capital expenditures are likely concentrated among a small number of leading technology firms, we next focus on the Magnificent Seven to examine their capex patterns over the past decade. \cref{fig:aiscore_capxrev} delivers two key messages. First, compared with the firms outside the Magnificent Seven, these leading technology firms have substantially higher capex-to-revenue ratios, reflecting their greater scale of AI-related investments. Second, there is a regime shift beginning in 2023: leading technology firms such as Meta, Google, Amazon, and Microsoft substantially increased AI-related capex, signaling the opening of an AI infrastructure race. For these firms, capex is primarily directed to hyperscale data-center expansion, AI-optimized servers and networking equipment, high-end GPU and accelerator capacity, and cloud infrastructure needed for large-model training and inference at scale. 

Forward-looking guidance suggests these firms expect another meaningful capex step-up in 2026, especially Google, Meta, and Amazon. Google is expecting capex of between \$175--\$185 billion USD in 2026, approximately double that of 2025, and a large portion of it will go to AI infrastructure.\footnote{\url{https://www.datacenterdynamics.com/en/news/google-estimates-2026-capex-of-up-to-185bn/}} Meta is estimating 2026 full-year capex of between \$115--\$135 billion USD, up from \$72.22 billion in 2025, and this spending growth is driven by increased investment to support Meta Superintelligence Labs efforts and core business.\footnote{\url{https://finance.yahoo.com/news/heres-why-metas-135-billion-192122813.html}} Amazon projected a surge of more than 50\% in capital expenditures in 2025 and signaled around \$200 billion to further boost AI efforts in 2026, joining peers in a broad spending cycle to build out AI infrastructure.\footnote{\url{https://finance.yahoo.com/news/amazon-200-billion-ai-spending-153341517.html}} 

At the same time, the pattern is not uniform within the Magnificent Seven. Apple appears to be a notable exception, with a relatively cautious AI investment strategy and no comparably large step-up in AI-oriented capex intensity. This heterogeneity is also reflected in equity market pricing: in early June 2025, when market sentiment toward AI was broadly optimistic, the AI highlights in Apple's WWDC product announcements were viewed as relatively underwhelming, leading to a notable decline in Apple's share price; by contrast, during episodes of AI-bubble concerns at the end of 2025 and start of 2026, several AI-linked mega-cap stocks saw sharper corrections while Apple's share price remained comparatively resilient.\footnote{\url{https://www.bitget.com/en-CA/wiki/does-apple-stock-go-up-after-wwdc}}

% [Do we have a reference for these paragraphs? I will try to find one.]

\subsubsection{Tobin's Q}

Tobin's Q measures the relationship between a company's market value and the replacement cost of it's assets. A higher Tobin's Q can therefore reflect two different forces: stronger expected operating efficiency and growth potential, or market overvaluation relative to fundamentals.

\begin{figure}[H]
    \centering
    \caption{\bf AI Adoption and Tobin's Q}
    \includegraphics[width=0.95\linewidth]{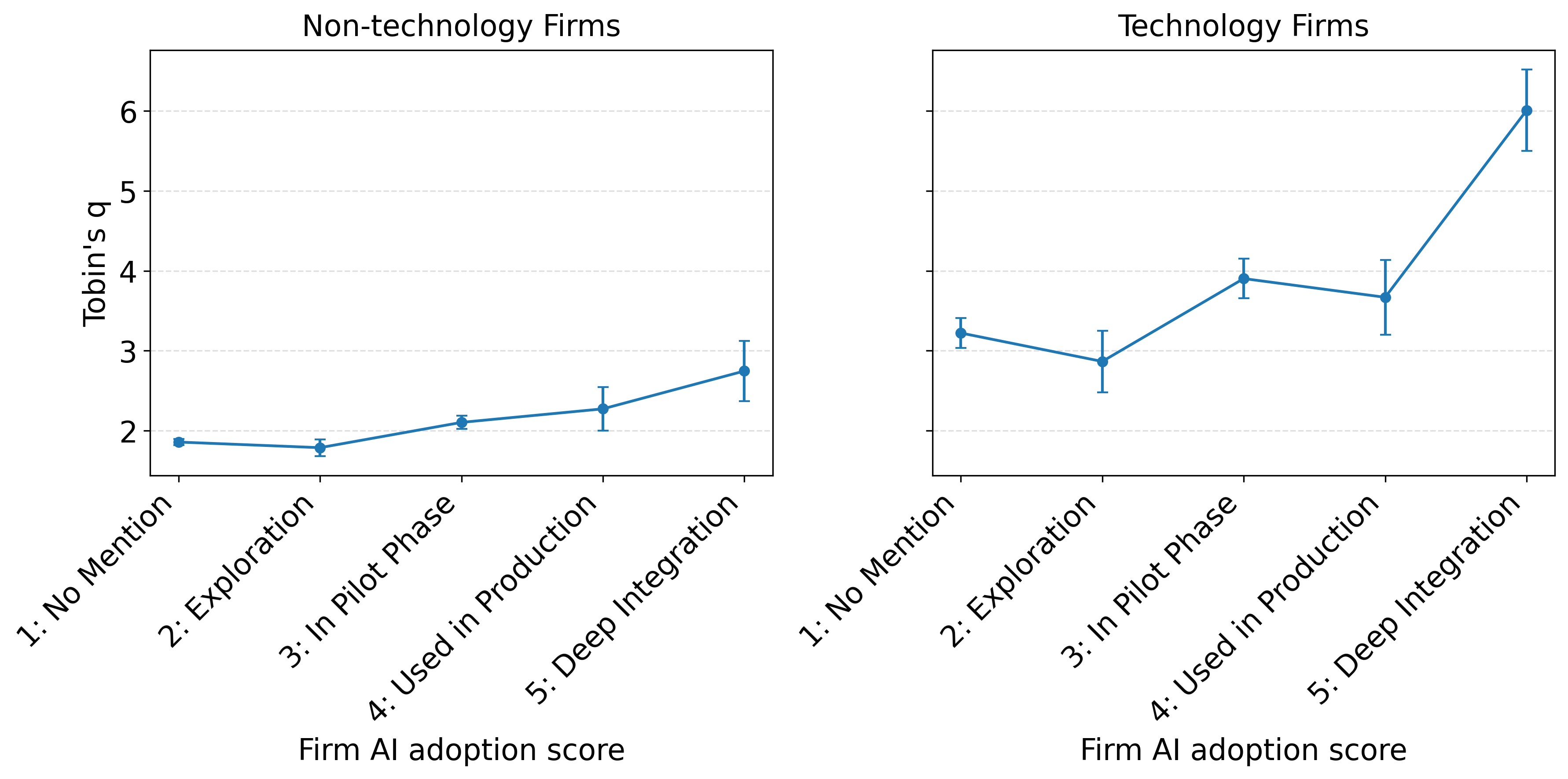}
    \label{fig:aiscore_tobinsq}
    \captionsetup{font=footnotesize}
    \caption*{\textit{Note: }This figure reports the mean Tobin's Q of firms at each level of AI adoption from 2016 to 2025. Error bars represent one standard error of the mean estimates.}
\end{figure}

\cref{fig:aiscore_tobinsq} reports the mean Tobin's Q of firms at each level of AI adoption from 2016 to 2025. We observe two salient patterns. First, technology firms on average have a higher Tobin's Q than non-technology firms across all AI adoption levels. Second, Tobin's Q tends to be higher for firms with deeper integration of AI capabilities for both technology and non-technology firms. This pattern is consistent with the view that investors expect deeper AI adoption to improve firm efficiency and future value creation by, for example, automating routine tasks with lower costs or augmenting human capabilities delivering productivity gains and decision-making improvement. However, it is important to note that the observed positive relationship between AI adoption and Tobin's Q does not necessarily imply a causal effect of AI on firm valuation. It is possible that firms with higher growth potential or stronger fundamentals are more likely to adopt AI. Further analysis would be needed to disentangle these effects and establish causality.

% \begin{table}[H]
%     \centering
%     \caption{\bf Summary statistics of Tobin's Q for non-tech sector}
% \input{../tables/su_tobinsq_nontech_by_aiscore.tex}
% \end{table}

% \begin{table}[H]
%     \centering
%     \caption{\bf Summary statistics of Tobin's Q for tech sector}
% \input{../tables/su_tobinsq_tech_by_aiscore.tex}
% \end{table}

\subsubsection{Employment}\label{subsec:obs_hc}

\begin{figure}[H]
    \centering
    \caption{\bf AI Adoption and Headcount}
    \includegraphics[width=0.95\linewidth]{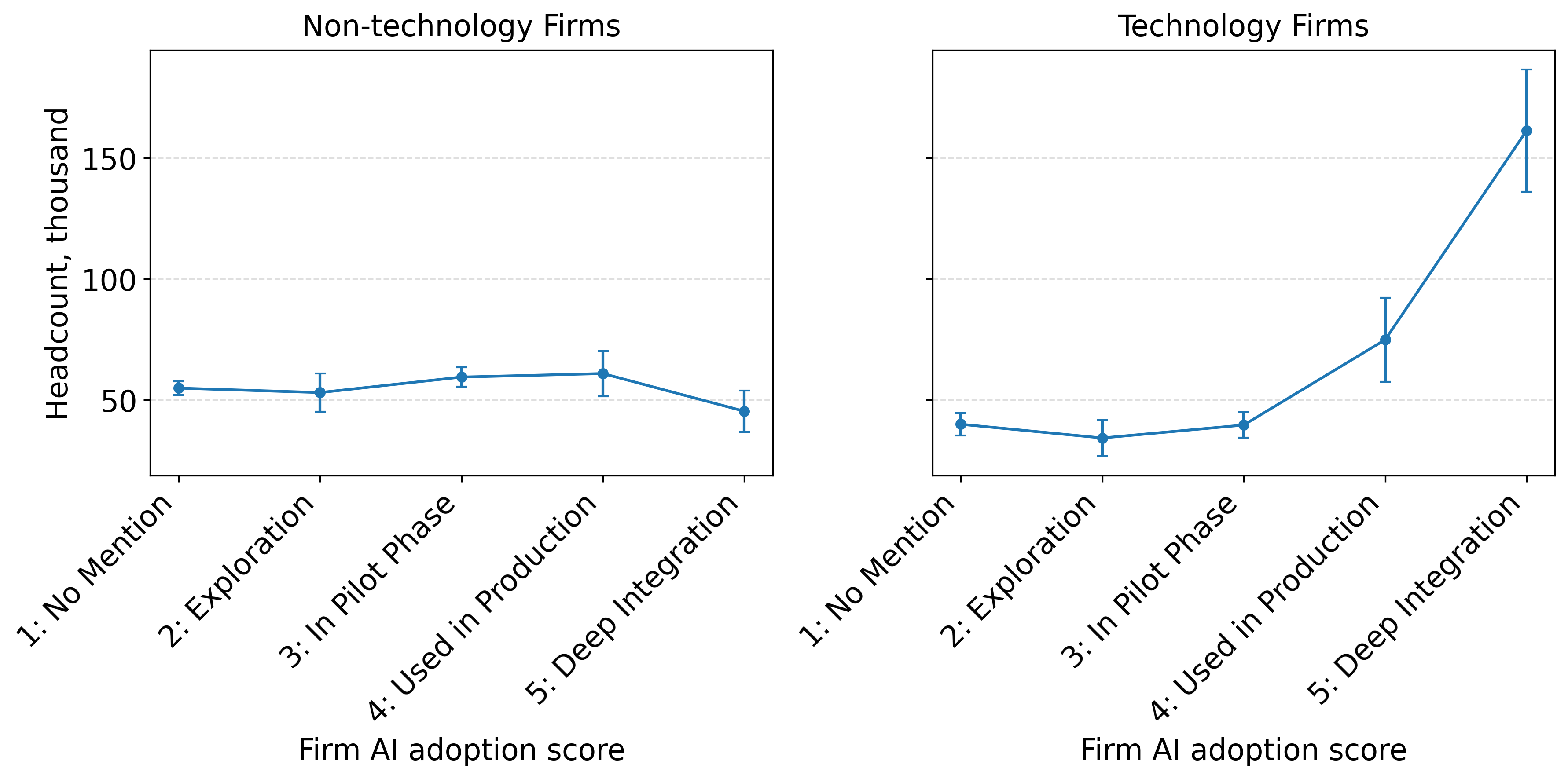}
    \label{fig:aiscore_emp}
    \captionsetup{font=footnotesize}
    \caption*{\textit{Note: }This figure reports the headcount of firms at each level of AI adoption from 2016 to 2025. Error bars represent one standard error of the mean estimates. The unit of headcount is in thousands.}
\end{figure}

\cref{fig:aiscore_emp} reports the headcount of firms at each level of AI adoption from 2016 to 2025. It shows a clear cross-sector difference in the relationship between AI adoption and firm headcount. In the non-technology sector, the association between AI score and headcount is weak: average headcount remains relatively stable across AI levels. In contrast, the technology sector exhibits a much stronger positive pattern, especially among those firms with more advanced when AI adoption (scores 4 or 5), where average headcount is significantly higher. This suggests that, in technology sector, deeper AI adoption is more concentrated among larger firms.

One interpretation is that large technology firms are better positioned to adopt AI at scale because they already have more complete data architecture and data pipelines, stronger compute and cloud capacity, mature machine learning operations (MLOps) capabilities, and organizational experience integrating new technologies across products. These advantages make it easier to realize scale economies from AI. Competitive dynamics and peer pressure in the AI race, together with workforce expansion during the COVID period, may also reinforce this size-adoption gradient in technology sector. By contrast, many non-technology firms lack AI DNA and are primarily consumers of AI services rather than infrastructure builders, so AI adoption is less tightly linked to their own headcount size \citep{bryan2026economic}.

A prominent concern is that AI adoption may lead to massive layoffs. In the first quarter of 2026 alone, job cuts include approximately 30,000 at Oracle, 16,000 at Amazon, 11,000 at Dell, 4,000 at Block, and over 2,000 at Meta, almost all cited AI-related restructuring as the primary reason.\footnote{\url{https://www.trueup.io/layoffs}} However, our aggregate firm-level headcount patterns do not yet show large-scale contraction. One possible reason is compositional churn: firms may cut tasks and roles more exposed to AI automation while simultaneously hiring for new AI-related roles \citep{eloundou2024gpts}. A more definitive assessment therefore requires job-posting and job-transition dynamics. It is also possible that the observed headcount patterns reflect a lagged effect, and that more significant layoffs may materialize in the future as AI adoption expands and its impact on labor market becomes more pronounced. Continued monitoring of employment trends will be important to understand the full labor market implications of AI adoption.

\subsubsection{Productivity}

To approximate firm productivity, we use revenue per employee as a tractable proxy. 
\yyr{Revenue per employee is a widely used indicator of workforce productivity in benchmarking and management analyses \citep{syverson2011determines}. More structural measures such as total factor productivity require substantially richer data and stronger modeling assumptions and are beyond the scope of this analysis.}

\begin{figure}[H]
    \centering
    \caption{\bf AI Adoption and Revenue per Employee}
    \includegraphics[width=0.95\linewidth]{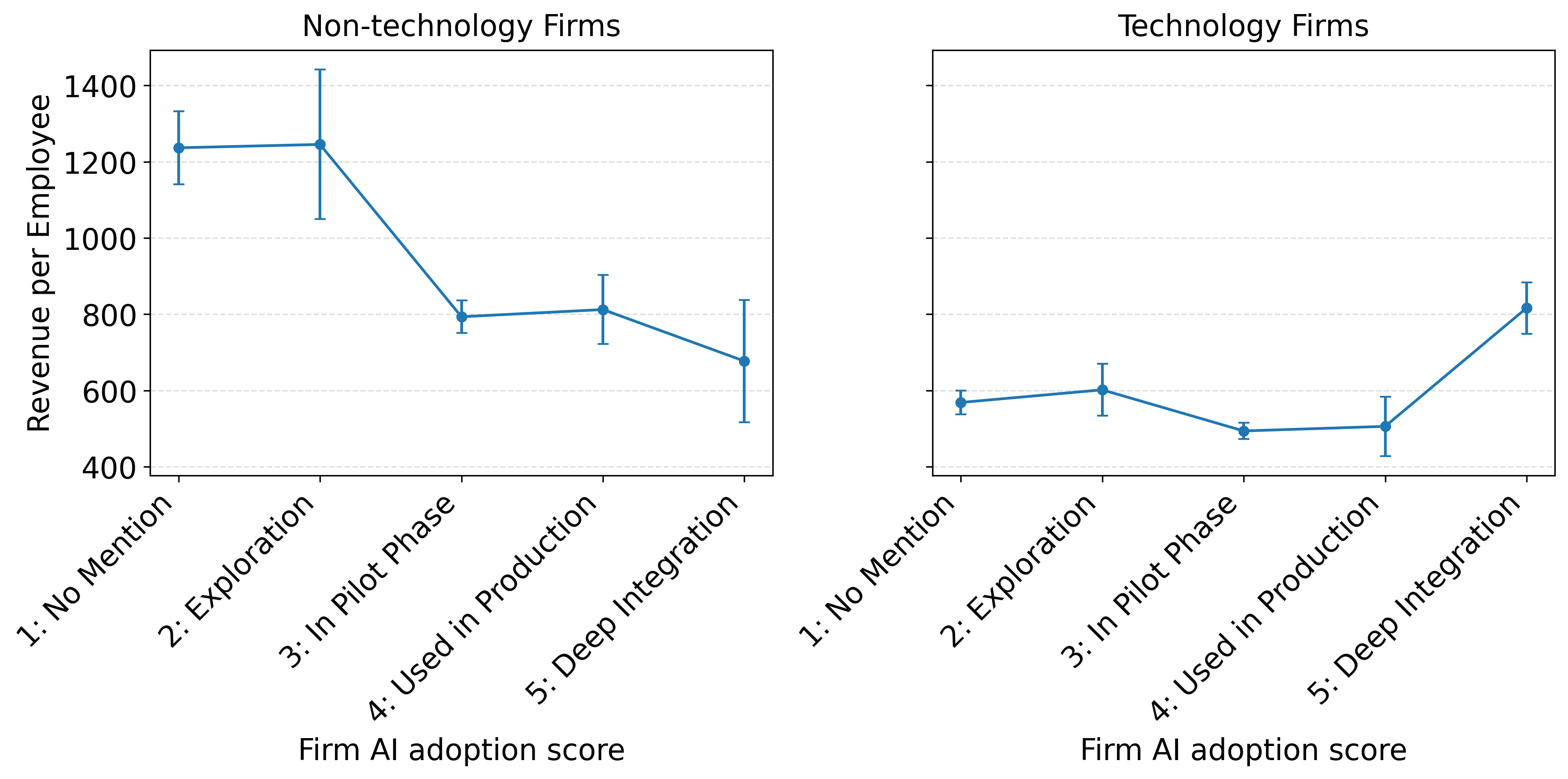}
    \label{fig:aiscore_prod}
    \captionsetup{font=footnotesize}
    \caption*{\textit{Note: }This figure reports the revenue per employee of firms at each level of AI adoption from 2016 to 2025. Error bars represent one standard error of the mean estimates. The unit of revenue per employee is in thousand dollars.}
\end{figure}

\cref{fig:aiscore_prod} shows that firms with deeper AI adoption do not uniformly exhibit higher revenue per employee. Across sectors and AI-score bins, high-adoption firms are not always associated with clearly higher revenue per employee. This provides suggestive (but not causal) evidence that the productivity gains from AI adoption may not yet have fully materialized in realized firm outcomes over the sample window. \yyr{Another explanation is reverse causality---outside the technology sector, firms with less productive employees could see AI as more attractive.}

% \subsubsection{More Headcounts Lower Tobin's Q}
% % \yang{move to appendix}

% \begin{figure}[H]
%     \centering
%     \caption{\bf Headcounts and Tobin's Q}
%     \includegraphics[width=1\linewidth]{../figures/emp_tobinsq_by_sector.png}
%     \label{fig:emp_tobinsq}
%     \captionsetup{font=footnotesize}
%     \caption*{\textit{Note: }This figure is a scatterplot of log(headcount) and log(Tobin's Q). The dashed line represents the fitted linear trend.}
% \end{figure}

\section{Econometric Methods}\label{sec:method}

We examine how different levels of AI adoption (1–5) relate to firm outcomes, and whether this relationship differs for technology vs. non-technology firms.
The baseline specification is as follows:

\begin{equation}\label{eq:baseline}
    % \hspace{-0.3in}
    Y_{it} = \alpha + \sum_{k=2}^{5} \beta_k \cdot \mathbf{1}\{AI_{it} = k\} + 
    \sum_{k=2}^{5} \phi_k \cdot \mathbf{1}\{AI_{it} = k\}\times Tech_{i}
     + \gamma_i + \delta_t + \epsilon_{it}
\end{equation}
where $Y_{it}$ is the firm-level outcome variable for firm $i$ at year $t$, including net profit margin, capex-to-revenue ratios, Tobin's Q, headcount, and revenue per employee. 
Net profit margin, capex-to-revenue ratios, and Tobin's Q are normalized (by firm scale) financial metrics to ensure comparability across firms.
$AI_{it}$ is the AI adoption score introduced above, we interact it with $Tech_i$, a binary variable indicating whether the firm is in the technology sector or not, to capture potential heterogeneous effects of AI adoption across technology and non-technology sectors.
We include firm fixed effects $\gamma_i$ to control for time-invariant firm characteristics (e.g., industry, location, founding history), which absorbs the lowest level of AI adoption (i.e., $k=1$) as the reference group so that $\beta_1$ and $\phi_1$ cannot be identified,  
and year fixed effects $\delta_t$ to control for year shocks such as macro trends and AI hype cycles that impact all firms at the same extent. $\epsilon_{it}$ captures firm-year varying unobserved heterogeneity such as management quality or firm culture.

With the inclusion of firm and year fixed effects, $\beta_k$ and $\phi_k$ are identified by comparing a firm to itself over time and to other firms in the same year. To interpret, among non-technology firms, moving from AI adoption level 1 to $k$ is associated with a change of $\beta_k$ in outcome $Y$; among technology firms, the same change in AI adoption is associated with a change of $\beta_k + \phi_k$ in the outcome variable.

It's worth noting that the current specification is likely suffer from endogeneity, as unobserved time-varying factors that affect both AI adoption and firm performance (e.g., innovation, organizational transformation, management quality, firm culture) are not controlled for. 
In addition, the correlation could also be driven by reverse causality --- high-performing firms have better access to financial capital and therefore can afford AI adoption.
Therefore, the coefficients are merely correlations and should not be interpreted as causal effects.

To reduce the concern of reverse causality, we also estimate the following specification with lagged AI adoption:

\begin{equation}\label{eq:laggedAI}
    % \hspace{-0.3in}
    Y_{it} = \alpha + \sum_{k=2}^{5} \beta_k \cdot \mathbf{1}\{AI_{i,t-1} = k\} + 
    \sum_{k=2}^{5} \phi_k \cdot \mathbf{1}\{AI_{i,t-1} = k\}\times Tech_{i}
     + \gamma_i + \delta_t + \epsilon_{it}
\end{equation}
where $AI_{i,t-1}$ is the AI adoption score at year $t-1$. To make it clear, although \cref{eq:laggedAI} helps with addressing simultaneity, it does not fix the above-mentioned endogeneity as well as at least the following two issues:
\begin{enumerate}
    \item \textbf{Forward-looking behaviors}: Firms may adopt AI because they expect upcoming demand shocks and thus higher performance in the future, which restores the reverse causality of expected future outcomes on current AI adoption decisions.
    \item \textbf{Serial correlation}: If errors are persistent, i.e., $Cov(\epsilon_{i,t}, \epsilon_{i,t-1}) \neq 0$, and past AI adoption is response to past shocks, then past AI adoption may be correlated with current shock, biasing the estimates.
\end{enumerate}
That said, we still view \cref{eq:laggedAI} as a meaningful robustness check as it rules out the most direct form of reverse causality.

Next, we replace year fixed effects with sector-year fixed effects to control for any shock common to all firms in the same sector in the same year, such as sector-level demand shock and regulation amendments. The specification is as follows:

\begin{equation}\label{eq:sector-year-fe}
    % \hspace{-0.3in}
    Y_{it} = \alpha + \sum_{k=2}^{5} \beta_k \cdot \mathbf{1}\{AI_{it} = k\} + 
    \sum_{k=2}^{5} \phi_k \cdot \mathbf{1}\{AI_{it} = k\}\times Tech_{i}
     + \gamma_i + \lambda_{s(i),t} + \epsilon_{it}
\end{equation}
where $\lambda_{s(i),t}$ is a fixed effect for firm's sector $s(i)$ in year $t$, which captures sector-specific shocks and trends.
Now the AI adoption coefficients are identified by comparing a firm to (1) itself over time; and (2) other firms in the same sector-year
rather than relying on comparisons across sectors.
This is useful if AI adoption is correlated with sector-specific trends. For example, software firms may adopt AI faster and may also achieve better performance for unrelated reasons.
Without sector-year fixed effects, AI adoption coefficients may pick up those sector-specific time trends.

% \yyr{I tried the following specification:
% \begin{equation}
%     % \hspace{-0.3in}
%     \Delta Y_{it} = \alpha + \sum_{k=-3}^{4} \beta_k \cdot \mathbf{1}\{\Delta AI_{it} = k\} + 
%     \sum_{k=-3}^{4} \phi_k \cdot \mathbf{1}\{\Delta AI_{it} = k\}\times Tech_{i}
%      + \Delta\lambda_{s(i),t} + \Delta\epsilon_{it}
% \end{equation}
% It doesn't solve the co-movement of $\epsilon_{it}$ and $AI_{it}$, and the results don't show clear patterns.
% }

A caveat is that although the inclusion of sector-year fixed effects helps with sector-level time-varying confounding, it does not solve the problem of firm-level time-varying omitted variables. Hence, the estimates still cannot be interpreted as causal even though \cref{eq:sector-year-fe} provides a strong robustness improvement.

\section{Results}\label{sec:results}
Our main results are presented in \cref{tab:reg_npm_10yrs} to \cref{tab:reg_prod_10yrs}, corresponding to the specifications in \cref{eq:baseline} and \cref{eq:sector-year-fe}. In each regression table presented below, our preferred columns are (4) and (6) that estimate for technology and non-technology firms separately and include year and firm fixed effects.
% The specification with lagged AI adoption show similar patterns and are deferred to Appendix \ref{sec:robustness} for the ease of presentation.

\paragraph{Net Profit Margin. }As is shown in column (4) of \cref{tab:reg_npm_10yrs}, non-technology firms in the early stages of AI adoption (AI score = 2 or 3) have 2 to 3 percentage points lower profit margins than those with no or minimal AI adoption (AI score = 1), while those in the most advanced stage of deployment (AI score = 5) have 12.6 percentage points higher profit margins. These effects are statistically significant and suggest a J-curve relationship between AI adoption and net profit margin \citep{brynjolfsson2021productivity,mcelheran2025rise}. 

In contrast, technology firms exhibit a more modest relationship between AI adoption and profitability and we don't see a clear J-curve pattern. Firms with AI adoption scores of 2–4 report net profit margins that are approximately 3–4 percentage points higher than those of firms with minimal AI adoption, although only the estimate for adoption level 3 is statistically significant. Firms at the highest level of AI integration (AI score = 5) achieve significantly higher net profit margins, by approximately 5 percentage points. However, this effect is considerably less transformative than that observed among non-technology firms. 

After controlling for sector-year fixed effects (column (6) in \cref{tab:reg_npm_10yrs}), we don't find significant differences in net profit margins across AI adoption levels 1 to 4 (although the effect directions persist as column (4)), while the positive relationship between deep AI integration and net profit margin remains significant in both technology and non-technology sectors, with 15\% higher net profit margin for non-technology firms and
(only) 2.6\% higher net profit margin for technology firms.

The intuition behind the J-curve phenomenon is that firms in the early stages of adoption experience margin compression. While their average net profit margins are positive, as shown in \cref{fig:aiscore_npm}, AI deployment requires substantial upfront investment, compared to firms with no deployment. Costs for digital infrastructure, data pipelines, model integration, and organizational reconfiguration are incurred before efficiency gains are fully realized, and short-run disruption to existing workflows can further weigh on profitability. As AI adoption matures, however, firms with AI in use for production and in deep integration begin to benefit from accumulated learning, process automation, and better resource allocation, which gradually improve operating efficiency. Net profit margins therefore stabilize and then rise in the mature stage, consistent with delayed but increasing returns to AI investment.

One possibility for the gap in profit gain from deep integration is that there are greater opportunities for cost reduction and process automation in traditionally less-digitized environments. This is because non-tech firms generally start from a lower technological baseline and often adopt AI as a cost-reducing tool rather than a core business activity, they may capture larger efficiency gains while avoiding many of the infrastructure and R\&D costs borne by technology firms. In contrast, competitive pressures and substantial AI-related investment expenditures may attenuate the effect of AI adoption on margins within the technology sector. Another possibility is that non-techology firms that expect to benefit most from AI adoption choose to adopt AI more aggressively.
% These patterns persist after controlling for sector-year fixed effects (columns (5) and (6) in \cref{tab:reg_npm_10yrs}), suggesting that the relationship is not driven by sector-specific time trends.

\begin{table}[p]
    \centering
    \caption{\bf AI Adoption and Net Profit Margin from 2016 to 2025}
    % \hspace*{-2.5cm}
\scalebox{0.65}{
{
\def\sym#1{\ifmmode^{#1}\else\(^{#1}\)\fi}
\begin{tabular}{l*{6}{D{.}{.}{-1}}}
\toprule
\multicolumn{7}{c}{Dependent Variable: Net Profit Margin}\\
            &\multicolumn{1}{c}{(1)}         &\multicolumn{1}{c}{(2)}         &\multicolumn{1}{c}{(3)}         &\multicolumn{1}{c}{(4)}         &\multicolumn{1}{c}{(5)}         &\multicolumn{1}{c}{(6)}         \\
\midrule
AI(2): Exploration&      -0.002         &      -0.005         &      -0.011         &      -0.017         &      -0.013         &      -0.016         \\
            &     (0.014)         &     (0.015)         &     (0.015)         &     (0.016)         &     (0.015)         &     (0.016)         \\
\addlinespace[1.1em]
AI(3): In Pilot Phase&      -0.019\sym{**} &      -0.024\sym{**} &      -0.017         &      -0.027\sym{**} &      -0.011         &      -0.017         \\
            &     (0.009)         &     (0.010)         &     (0.012)         &     (0.013)         &     (0.013)         &     (0.014)         \\
\addlinespace[1.1em]
AI(4): Used in Production&       0.032         &       0.027         &       0.028         &       0.024         &       0.026         &       0.032         \\
            &     (0.022)         &     (0.026)         &     (0.024)         &     (0.028)         &     (0.025)         &     (0.029)         \\
\addlinespace[1.1em]
AI(5): Deep Integration&       0.038\sym{**} &      -0.029         &       0.057\sym{**} &       0.126\sym{***}&       0.062\sym{**} &       0.150\sym{***}\\
            &     (0.018)         &     (0.038)         &     (0.027)         &     (0.042)         &     (0.029)         &     (0.045)         \\
\addlinespace[1.1em]
AI(2) $\times$ Tech&                     &       0.035         &                     &       0.045         &                     &       0.025         \\
            &                     &     (0.049)         &                     &     (0.047)         &                     &     (0.050)         \\
\addlinespace[1.1em]
AI(3) $\times$ Tech&                     &       0.003         &                     &       0.054\sym{*}  &                     &       0.030         \\
            &                     &     (0.025)         &                     &     (0.028)         &                     &     (0.033)         \\
\addlinespace[1.1em]
AI(4) $\times$ Tech&                     &       0.000         &                     &       0.019         &                     &      -0.012         \\
            &                     &     (0.051)         &                     &     (0.050)         &                     &     (0.057)         \\
\addlinespace[1.1em]
AI(5) $\times$ Tech&                     &       0.058         &                     &      -0.076         &                     &      -0.124\sym{*}  \\
            &                     &     (0.046)         &                     &     (0.053)         &                     &     (0.064)         \\
\addlinespace[1.1em]
Tech        &                     &       0.030\sym{*}  &                     &                     &                     &                     \\
            &                     &     (0.016)         &                     &                     &                     &                     \\
\addlinespace[1.1em]
constant    &       0.117\sym{***}&       0.114\sym{***}&       0.117\sym{***}&       0.116\sym{***}&       0.117\sym{***}&       0.117\sym{***}\\
            &     (0.005)         &     (0.005)         &     (0.005)         &     (0.005)         &     (0.005)         &     (0.006)         \\
\midrule
Observations&\multicolumn{1}{c}{4470}         &\multicolumn{1}{c}{4470}         &\multicolumn{1}{c}{4465}         &\multicolumn{1}{c}{4465}         &\multicolumn{1}{c}{4301}         &\multicolumn{1}{c}{4301}         \\
R\textsuperscript{2}&\multicolumn{1}{c}{0.003}         &\multicolumn{1}{c}{0.006}         &\multicolumn{1}{c}{0.338}         &\multicolumn{1}{c}{0.339}         &\multicolumn{1}{c}{0.416}         &\multicolumn{1}{c}{0.417}         \\
Firm FE     &\multicolumn{1}{c}{No}         &\multicolumn{1}{c}{No}         &\multicolumn{1}{c}{Yes}         &\multicolumn{1}{c}{Yes}         &\multicolumn{1}{c}{Yes}         &\multicolumn{1}{c}{Yes}         \\
Year FE     &\multicolumn{1}{c}{No}         &\multicolumn{1}{c}{No}         &\multicolumn{1}{c}{Yes}         &\multicolumn{1}{c}{Yes}         &\multicolumn{1}{c}{No}         &\multicolumn{1}{c}{No}         \\
Sector $\times$ Year FE&\multicolumn{1}{c}{No}         &\multicolumn{1}{c}{No}         &\multicolumn{1}{c}{No}         &\multicolumn{1}{c}{No}         &\multicolumn{1}{c}{Yes}         &\multicolumn{1}{c}{Yes}         \\
\bottomrule
\multicolumn{7}{l}{\footnotesize Standard errors in parentheses}\\
\multicolumn{7}{l}{\footnotesize \sym{*} \(p<0.1\), \sym{**} \(p<0.05\), \sym{***} \(p<0.01\)}\\
\end{tabular}
}

}
\vspace{0.1in}
\caption*{\footnotesize \textit{Note:} 
AI(2) is a dummy that equals 1 if the firm's AI score at a given year is 2 and otherwise 0. Tech is a dummy that equals to 1 if the firm's sector is Tech. 
}
    \label{tab:reg_npm_10yrs}
\end{table}

\paragraph{Capex-to-Revenue Ratio.} As is shown in columns (3) to (6) of \cref{tab:reg_capex_10yrs}, AI adoption, regardless of the stage, is (mostly) not significantly correlated with capex-to-revenue ratio. The lack of correlation is in fact consistent with most firms outsourcing AI infrastructure rather than building it in-house. Except for a small set of leading technology firms that internalize AI production—by purchasing GPUs, building data centers, securing long-term compute contracts, or even operating like semiconductor companies or cloud infrastructure providers—most firms treat AI-related spending as operating expenses rather than capital expenditures. This includes salaries for engineers and data scientists, fees for AI-as-a-service and API usage, and cloud computing costs (e.g., AWS, Azure, GCP), all of which enter SG\&A or COGS rather than capex. In columns (1) and (2) of \cref{tab:reg_capex_10yrs}, when firm and year fixed effects are excluded, we observe significantly lower capex-to-revenue ratio among firms with AI scores 2 to 4, likely reflecting sectoral self-selection into different levels of AI adoption.

\begin{table}[p]
    \centering
    \caption{\bf AI Adoption and Capex-to-Revenue Ratio from 2016 to 2025}
    % \hspace*{-2.5cm}
\scalebox{0.65}{
{
\def\sym#1{\ifmmode^{#1}\else\(^{#1}\)\fi}
\begin{tabular}{l*{6}{D{.}{.}{-1}}}
\toprule
\multicolumn{7}{c}{Dependent Variable: Capex-to-Revenue}\\
            &\multicolumn{1}{c}{(1)}         &\multicolumn{1}{c}{(2)}         &\multicolumn{1}{c}{(3)}         &\multicolumn{1}{c}{(4)}         &\multicolumn{1}{c}{(5)}         &\multicolumn{1}{c}{(6)}         \\
\midrule
AI(2): Exploration&      -0.018\sym{**} &      -0.021\sym{***}&      -0.000         &       0.002         &       0.001         &       0.003         \\
            &     (0.007)         &     (0.008)         &     (0.004)         &     (0.005)         &     (0.004)         &     (0.005)         \\
\addlinespace[1.1em]
AI(3): In Pilot Phase&      -0.032\sym{***}&      -0.032\sym{***}&       0.002         &       0.003         &       0.006         &       0.006\sym{*}  \\
            &     (0.005)         &     (0.005)         &     (0.003)         &     (0.004)         &     (0.004)         &     (0.004)         \\
\addlinespace[1.1em]
AI(4): Used in Production&      -0.031\sym{***}&      -0.036\sym{***}&      -0.010         &      -0.006         &      -0.003         &       0.002         \\
            &     (0.011)         &     (0.013)         &     (0.007)         &     (0.008)         &     (0.007)         &     (0.008)         \\
\addlinespace[1.1em]
AI(5): Deep Integration&      -0.014         &      -0.021         &      -0.005         &      -0.015         &       0.002         &      -0.006         \\
            &     (0.009)         &     (0.020)         &     (0.008)         &     (0.012)         &     (0.008)         &     (0.013)         \\
\addlinespace[1.1em]
AI(2) $\times$ Tech&                     &       0.033         &                     &      -0.019         &                     &      -0.024\sym{*}  \\
            &                     &     (0.025)         &                     &     (0.013)         &                     &     (0.014)         \\
\addlinespace[1.1em]
AI(3) $\times$ Tech&                     &       0.022\sym{*}  &                     &      -0.003         &                     &      -0.006         \\
            &                     &     (0.013)         &                     &     (0.008)         &                     &     (0.009)         \\
\addlinespace[1.1em]
AI(4) $\times$ Tech&                     &       0.043         &                     &      -0.014         &                     &      -0.020         \\
            &                     &     (0.026)         &                     &     (0.014)         &                     &     (0.016)         \\
\addlinespace[1.1em]
AI(5) $\times$ Tech&                     &       0.038         &                     &       0.012         &                     &       0.006         \\
            &                     &     (0.024)         &                     &     (0.015)         &                     &     (0.018)         \\
\addlinespace[1.1em]
Tech        &                     &      -0.033\sym{***}&                     &                     &                     &                     \\
            &                     &     (0.008)         &                     &                     &                     &                     \\
\addlinespace[1.1em]
constant    &       0.088\sym{***}&       0.091\sym{***}&       0.078\sym{***}&       0.078\sym{***}&       0.077\sym{***}&       0.078\sym{***}\\
            &     (0.002)         &     (0.002)         &     (0.001)         &     (0.002)         &     (0.002)         &     (0.002)         \\
\midrule
Observations&\multicolumn{1}{c}{4470}         &\multicolumn{1}{c}{4470}         &\multicolumn{1}{c}{4465}         &\multicolumn{1}{c}{4465}         &\multicolumn{1}{c}{4301}         &\multicolumn{1}{c}{4301}         \\
R\textsuperscript{2}&\multicolumn{1}{c}{0.011}         &\multicolumn{1}{c}{0.015}         &\multicolumn{1}{c}{0.801}         &\multicolumn{1}{c}{0.801}         &\multicolumn{1}{c}{0.829}         &\multicolumn{1}{c}{0.830}         \\
Firm FE     &\multicolumn{1}{c}{No}         &\multicolumn{1}{c}{No}         &\multicolumn{1}{c}{Yes}         &\multicolumn{1}{c}{Yes}         &\multicolumn{1}{c}{Yes}         &\multicolumn{1}{c}{Yes}         \\
Year FE     &\multicolumn{1}{c}{No}         &\multicolumn{1}{c}{No}         &\multicolumn{1}{c}{Yes}         &\multicolumn{1}{c}{Yes}         &\multicolumn{1}{c}{No}         &\multicolumn{1}{c}{No}         \\
Sector $\times$ Year FE&\multicolumn{1}{c}{No}         &\multicolumn{1}{c}{No}         &\multicolumn{1}{c}{No}         &\multicolumn{1}{c}{No}         &\multicolumn{1}{c}{Yes}         &\multicolumn{1}{c}{Yes}         \\
\bottomrule
\multicolumn{7}{l}{\footnotesize Standard errors in parentheses}\\
\multicolumn{7}{l}{\footnotesize \sym{*} \(p<0.1\), \sym{**} \(p<0.05\), \sym{***} \(p<0.01\)}\\
\end{tabular}
}

}
\vspace{0.1in}
% \caption*{\footnotesize \textit{Note:} 
% AI(2) is a dummy that equals 1 if the firm's AI score at a given year is 2 and otherwise 0. Tech is a dummy that equals to 1 if the firm's sector is Tech. 
% }
    \label{tab:reg_capex_10yrs}
\end{table}

\paragraph{Tobin's Q.} 
As shown in column (2) of \cref{tab:reg_tobq_10yrs}, non-technology firms with AI adoption scores of 2 to 4 do not exhibit significantly different Tobin’s Q ratios relative to firms with minimal AI adoption. In contrast, firms with deep AI integration (AI score = 5) have Tobin’s Q ratios that are 47.8\% higher. Among technology firms, however, Tobin’s Q does not differ significantly across AI adoption levels. Nevertheless, technology firms, on average, have Tobin’s Q ratios that are 64.4\% higher than those of non-technology firms.

A contrasting picture emerges once firm and year fixed effects are included. As reported in column (4), non-technology firms with AI adoption scores of 2 to 4 continue to show no significant differences in Tobin’s Q relative to firms with minimal AI adoption, whereas firms with deep AI integration exhibit Tobin’s Q ratios that are 23\% lower, and the effect is statistically significant. By contrast, among technology firms, higher levels of AI adoption are consistently associated with higher market valuations. Relative to firms with minimal AI adoption, those with AI adoption scores of 2, 3, 4, and 5 exhibit Tobin’s Q ratios that are 13.0\%, 14.3\%, 14.6\%, and 15.4\% higher, respectively.

The large negative and significant association between Tobin's Q and deep AI integration for non-technology firms is counterintuitive as the capital market is forward-looking and should have priced in the future efficiency gains from deep AI adoption. We find that this abnormality is largely driven by two outliers, Align Technology (ALGN) and Paycom (PAYC), both of which have experienced sharp declines in market valuation since 2022 due to weakened market demand or lower customer growth, coinciding with their increased adoption of AI.
\footnote{ALGN suffered from weak demand on non-essential dental spending while PAYC experienced cannibalization from the rollout of its new product and a slowdown in new customer acquisition. See details at \url{https://finance.yahoo.com/news/reasons-sharp-decline-align-technology-054229705.html?guccounter=1} and \url{https://finance.yahoo.com/news/why-paycom-payc-stock-nosediving-165806227.html}}
Omitting these firm-year level variables that are parallel with AI adoption leads to biased estimates, especially when the number of non-technology firms falling into AI level 5 is small. In column (8), we exclude ALGN and PAYC from the sample, and see that the negative relationship between deep AI integration and Tobin's Q among non-technology firms weakens while the positive relationship between AI adoption and Tobin's Q among technology firms persists in both significance level and magnitude.

However, such positive relationship observed in technology firms is greatly reduced, often to insignificance, once sector-by-year fixed effects are included (columns (6) and (10)), suggesting that the investors are giving credit to AI adoption as a sector-level phenomenon.
% positive association between AI adoption and firm valuation in the capital market is largely driven by which sector the firm is in, not necessarily within-sector differences in AI adoption. In other words, investors may value AI-exposed sectors (so that we see sector-level AI premium), but not strongly differentiate which firms inside that sector adopt more AI. 
One example is Allbirds, the sneaker company, whose stock price surged by over 600\% after it announced a rebrand to ``NewBird AI" in an effort to position itself as an AI company.

One last note is that the much less considerable relationship between AI adoption and Tobin's Q after controlling for firm and year fixed effects (column (4) compared to column (2)) provides suggestive evidence for reverse causality: highly valued firms are more likely to adopt AI.

\begin{table}[p]
    \centering
    \caption{\bf AI Adoption and Tobin's Q from 2016 to 2025}
    \hspace*{-2.2cm}
\scalebox{0.6}{
{
\def\sym#1{\ifmmode^{#1}\else\(^{#1}\)\fi}
\begin{tabular}{l*{10}{D{.}{.}{-1}}}
\toprule
\multicolumn{11}{c}{Dependent Variable: Log Tobin's q}\\
            &\multicolumn{1}{c}{(1)}         &\multicolumn{1}{c}{(2)}         &\multicolumn{1}{c}{(3)}         &\multicolumn{1}{c}{(4)}         &\multicolumn{1}{c}{(5)}         &\multicolumn{1}{c}{(6)}         &\multicolumn{1}{c}{(7)}         &\multicolumn{1}{c}{(8)}         &\multicolumn{1}{c}{(9)}         &\multicolumn{1}{c}{(10)}         \\
\midrule
AI(2): Exploration&      -0.057         &      -0.066         &       0.005         &      -0.010         &       0.023         &       0.033         &       0.005         &      -0.010         &       0.024         &       0.035         \\
            &     (0.060)         &     (0.061)         &     (0.025)         &     (0.026)         &     (0.025)         &     (0.026)         &     (0.025)         &     (0.026)         &     (0.025)         &     (0.026)         \\
\addlinespace[1.1em]
AI(3): In Pilot Phase&       0.152\sym{***}&       0.018         &      -0.006         &      -0.035         &       0.009         &       0.018         &      -0.005         &      -0.034         &       0.012         &       0.021         \\
            &     (0.038)         &     (0.041)         &     (0.020)         &     (0.022)         &     (0.020)         &     (0.022)         &     (0.020)         &     (0.022)         &     (0.020)         &     (0.022)         \\
\addlinespace[1.1em]
AI(4): Used in Production&       0.210\sym{**} &       0.051         &      -0.020         &      -0.072         &      -0.055         &      -0.015         &      -0.010         &      -0.058         &      -0.047         &      -0.006         \\
            &     (0.093)         &     (0.105)         &     (0.040)         &     (0.046)         &     (0.040)         &     (0.046)         &     (0.040)         &     (0.046)         &     (0.040)         &     (0.046)         \\
\addlinespace[1.1em]
AI(5): Deep Integration&       0.996\sym{***}&       0.478\sym{***}&      -0.046         &      -0.232\sym{***}&      -0.114\sym{**} &      -0.133\sym{*}  &      -0.009         &      -0.133\sym{*}  &      -0.073         &      -0.027         \\
            &     (0.076)         &     (0.155)         &     (0.044)         &     (0.069)         &     (0.047)         &     (0.072)         &     (0.045)         &     (0.072)         &     (0.047)         &     (0.074)         \\
\addlinespace[1.1em]
AI(2) $\times$ Tech&                     &       0.027         &                     &       0.140\sym{*}  &                     &      -0.082         &                     &       0.139\sym{*}  &                     &      -0.084         \\
            &                     &     (0.197)         &                     &     (0.077)         &                     &     (0.080)         &                     &     (0.077)         &                     &     (0.079)         \\
\addlinespace[1.1em]
AI(3) $\times$ Tech&                     &       0.172\sym{*}  &                     &       0.178\sym{***}&                     &      -0.051         &                     &       0.178\sym{***}&                     &      -0.054         \\
            &                     &     (0.100)         &                     &     (0.046)         &                     &     (0.052)         &                     &     (0.046)         &                     &     (0.052)         \\
\addlinespace[1.1em]
AI(4) $\times$ Tech&                     &       0.148         &                     &       0.218\sym{***}&                     &      -0.161\sym{*}  &                     &       0.203\sym{**} &                     &      -0.171\sym{*}  \\
            &                     &     (0.209)         &                     &     (0.083)         &                     &     (0.091)         &                     &     (0.082)         &                     &     (0.091)         \\
\addlinespace[1.1em]
AI(5) $\times$ Tech&                     &       0.087         &                     &       0.386\sym{***}&                     &      -0.027         &                     &       0.285\sym{***}&                     &      -0.132         \\
            &                     &     (0.186)         &                     &     (0.087)         &                     &     (0.102)         &                     &     (0.089)         &                     &     (0.104)         \\
\addlinespace[1.1em]
Tech        &                     &       0.644\sym{***}&                     &                     &                     &                     &                     &                     &                     &                     \\
            &                     &     (0.066)         &                     &                     &                     &                     &                     &                     &                     &                     \\
\addlinespace[1.1em]
constant    &       0.241\sym{***}&       0.183\sym{***}&       0.324\sym{***}&       0.316\sym{***}&       0.318\sym{***}&       0.321\sym{***}&       0.317\sym{***}&       0.309\sym{***}&       0.310\sym{***}&       0.313\sym{***}\\
            &     (0.019)         &     (0.020)         &     (0.009)         &     (0.009)         &     (0.009)         &     (0.009)         &     (0.009)         &     (0.009)         &     (0.009)         &     (0.009)         \\
\midrule
Observations&\multicolumn{1}{c}{4470}         &\multicolumn{1}{c}{4470}         &\multicolumn{1}{c}{4465}         &\multicolumn{1}{c}{4465}         &\multicolumn{1}{c}{4301}         &\multicolumn{1}{c}{4301}         &\multicolumn{1}{c}{4445}         &\multicolumn{1}{c}{4445}         &\multicolumn{1}{c}{4281}         &\multicolumn{1}{c}{4281}         \\
R\textsuperscript{2}&\multicolumn{1}{c}{0.040}         &\multicolumn{1}{c}{0.093}         &\multicolumn{1}{c}{0.899}         &\multicolumn{1}{c}{0.900}         &\multicolumn{1}{c}{0.916}         &\multicolumn{1}{c}{0.916}         &\multicolumn{1}{c}{0.900}         &\multicolumn{1}{c}{0.901}         &\multicolumn{1}{c}{0.917}         &\multicolumn{1}{c}{0.917}         \\
Firm FE     &\multicolumn{1}{c}{No}         &\multicolumn{1}{c}{No}         &\multicolumn{1}{c}{Yes}         &\multicolumn{1}{c}{Yes}         &\multicolumn{1}{c}{Yes}         &\multicolumn{1}{c}{Yes}         &\multicolumn{1}{c}{Yes}         &\multicolumn{1}{c}{Yes}         &\multicolumn{1}{c}{Yes}         &\multicolumn{1}{c}{Yes}         \\
Year FE     &\multicolumn{1}{c}{No}         &\multicolumn{1}{c}{No}         &\multicolumn{1}{c}{Yes}         &\multicolumn{1}{c}{Yes}         &\multicolumn{1}{c}{No}         &\multicolumn{1}{c}{No}         &\multicolumn{1}{c}{Yes}         &\multicolumn{1}{c}{Yes}         &\multicolumn{1}{c}{No}         &\multicolumn{1}{c}{No}         \\
Sector $\times$ Year FE&\multicolumn{1}{c}{No}         &\multicolumn{1}{c}{No}         &\multicolumn{1}{c}{No}         &\multicolumn{1}{c}{No}         &\multicolumn{1}{c}{Yes}         &\multicolumn{1}{c}{Yes}         &\multicolumn{1}{c}{No}         &\multicolumn{1}{c}{No}         &\multicolumn{1}{c}{Yes}         &\multicolumn{1}{c}{Yes}         \\
ALGN \& PAYC excluded&\multicolumn{1}{c}{No}         &\multicolumn{1}{c}{No}         &\multicolumn{1}{c}{No}         &\multicolumn{1}{c}{No}         &\multicolumn{1}{c}{No}         &\multicolumn{1}{c}{No}         &\multicolumn{1}{c}{Yes}         &\multicolumn{1}{c}{Yes}         &\multicolumn{1}{c}{Yes}         &\multicolumn{1}{c}{Yes}         \\
\bottomrule
\multicolumn{11}{l}{\footnotesize Standard errors in parentheses}\\
\multicolumn{11}{l}{\footnotesize \sym{*} \(p<0.1\), \sym{**} \(p<0.05\), \sym{***} \(p<0.01\)}\\
\end{tabular}
}

}
\vspace{0.1in}
\caption*{\footnotesize \textit{Note:} 
The dependent variable is the logarithm of Tobin's Q. Columns (7) to (10) exclude Align Technology (ALGN) and Paycom (PAYC), whose market valuations dropped significantly since 2022 mainly due to weaker consumer demand and sharp growth slowdown respectively, which happened in parallel with their AI adoption. 
}
    \label{tab:reg_tobq_10yrs}
\end{table}

\paragraph{Headcounts. } As is shown in column (4) of \cref{tab:reg_hc_10yrs}, in non-technology sectors, the employment sizes of firms with AI adoption level 2 to 4 are on average around 2 to 6 percentage points smaller than those with minimal AI adoption, while firms with deep AI integration have around 20\% higher headcount. In technology sector, firms with AI adoption levels 2 to 5 are on average 7.3, 10.0, 24.7, 36.4 percentage points larger than those with minimal AI adoption. These patterns are robust to the inclusion of sector-year fixed effects (column (6)).

These findings reinforce our observations in \cref{subsec:obs_hc} that deeper AI adoption concentrates among larger firms. We lean towards the reverse causality story that larger firms tend to be early adopters of AI for a mix of economic, organizational, and strategic reasons: 
\begin{enumerate}
    \item Economies of scale: AI adoption often involves substantial upfront costs---data infrastructure, model development, integration, and talent. Large firms can spread these fixed costs over a bigger revenue base, making adoption more cost-effective.
    \item Data readiness: large firms typically have extensive and diverse datasets across business lines and better data infrastructure, and therefore enjoy a natural advantage, since AI performance improves with data scale and quality.
    \item Financial and organizational capacity: large firms can allocate greater budgets to experimentation and R\&D, are generally more tolerant of failed projects, and have better access to key AI resources, including computing hardware, partnerships with leading AI vendors, and the ability to attract and acquire top AI talent.
    \item Strategic incentives: large firms, especially those in technology sector, often face more competitive pressure to adopt AI to maintain or extend their market power and meet customer expectations. 
\end{enumerate}

\begin{table}[p]
    \centering
    \caption{\bf AI Adoption and Headcount from 2016 to 2025}
    % \hspace*{-2.5cm}
\scalebox{0.65}{
{
\def\sym#1{\ifmmode^{#1}\else\(^{#1}\)\fi}
\begin{tabular}{l*{6}{D{.}{.}{-1}}}
\toprule
\multicolumn{7}{c}{Dependent Variable: Log Headcount}\\
            &\multicolumn{1}{c}{(1)}         &\multicolumn{1}{c}{(2)}         &\multicolumn{1}{c}{(3)}         &\multicolumn{1}{c}{(4)}         &\multicolumn{1}{c}{(5)}         &\multicolumn{1}{c}{(6)}         \\
\midrule
AI(2): Exploration&       0.159\sym{*}  &       0.145\sym{*}  &      -0.044\sym{***}&      -0.059\sym{***}&      -0.040\sym{**} &      -0.060\sym{***}\\
            &     (0.083)         &     (0.087)         &     (0.016)         &     (0.017)         &     (0.016)         &     (0.017)         \\
\addlinespace[1.1em]
AI(3): In Pilot Phase&       0.241\sym{***}&       0.308\sym{***}&      -0.003         &      -0.024\sym{*}  &      -0.012         &      -0.031\sym{**} \\
            &     (0.053)         &     (0.059)         &     (0.013)         &     (0.014)         &     (0.013)         &     (0.014)         \\
\addlinespace[1.1em]
AI(4): Used in Production&       0.418\sym{***}&       0.391\sym{***}&       0.023         &      -0.057\sym{*}  &       0.001         &      -0.082\sym{***}\\
            &     (0.129)         &     (0.150)         &     (0.026)         &     (0.030)         &     (0.026)         &     (0.030)         \\
\addlinespace[1.1em]
AI(5): Deep Integration&       0.703\sym{***}&       0.106         &       0.254\sym{***}&       0.199\sym{***}&       0.176\sym{***}&       0.125\sym{***}\\
            &     (0.105)         &     (0.221)         &     (0.029)         &     (0.045)         &     (0.030)         &     (0.046)         \\
\addlinespace[1.1em]
AI(2) $\times$ Tech&                     &       0.161         &                     &       0.132\sym{***}&                     &       0.155\sym{***}\\
            &                     &     (0.280)         &                     &     (0.050)         &                     &     (0.051)         \\
\addlinespace[1.1em]
AI(3) $\times$ Tech&                     &      -0.112         &                     &       0.124\sym{***}&                     &       0.112\sym{***}\\
            &                     &     (0.142)         &                     &     (0.030)         &                     &     (0.034)         \\
\addlinespace[1.1em]
AI(4) $\times$ Tech&                     &       0.285         &                     &       0.304\sym{***}&                     &       0.340\sym{***}\\
            &                     &     (0.297)         &                     &     (0.054)         &                     &     (0.059)         \\
\addlinespace[1.1em]
AI(5) $\times$ Tech&                     &       1.006\sym{***}&                     &       0.165\sym{***}&                     &       0.202\sym{***}\\
            &                     &     (0.264)         &                     &     (0.057)         &                     &     (0.066)         \\
\addlinespace[1.1em]
Tech        &                     &      -0.279\sym{***}&                     &                     &                     &                     \\
            &                     &     (0.094)         &                     &                     &                     &                     \\
\addlinespace[1.1em]
constant    &       2.962\sym{***}&       2.987\sym{***}&       3.062\sym{***}&       3.057\sym{***}&       3.070\sym{***}&       3.063\sym{***}\\
            &     (0.027)         &     (0.028)         &     (0.006)         &     (0.006)         &     (0.006)         &     (0.006)         \\
\midrule
Observations&\multicolumn{1}{c}{4470}         &\multicolumn{1}{c}{4470}         &\multicolumn{1}{c}{4465}         &\multicolumn{1}{c}{4465}         &\multicolumn{1}{c}{4301}         &\multicolumn{1}{c}{4301}         \\
R\textsuperscript{2}&\multicolumn{1}{c}{0.014}         &\multicolumn{1}{c}{0.021}         &\multicolumn{1}{c}{0.977}         &\multicolumn{1}{c}{0.978}         &\multicolumn{1}{c}{0.981}         &\multicolumn{1}{c}{0.981}         \\
Firm FE     &\multicolumn{1}{c}{No}         &\multicolumn{1}{c}{No}         &\multicolumn{1}{c}{Yes}         &\multicolumn{1}{c}{Yes}         &\multicolumn{1}{c}{Yes}         &\multicolumn{1}{c}{Yes}         \\
Year FE     &\multicolumn{1}{c}{No}         &\multicolumn{1}{c}{No}         &\multicolumn{1}{c}{Yes}         &\multicolumn{1}{c}{Yes}         &\multicolumn{1}{c}{No}         &\multicolumn{1}{c}{No}         \\
Sector $\times$ Year FE&\multicolumn{1}{c}{No}         &\multicolumn{1}{c}{No}         &\multicolumn{1}{c}{No}         &\multicolumn{1}{c}{No}         &\multicolumn{1}{c}{Yes}         &\multicolumn{1}{c}{Yes}         \\
\bottomrule
\multicolumn{7}{l}{\footnotesize Standard errors in parentheses}\\
\multicolumn{7}{l}{\footnotesize \sym{*} \(p<0.1\), \sym{**} \(p<0.05\), \sym{***} \(p<0.01\)}\\
\end{tabular}
}

}
\vspace{0.1in}
% \caption*{\footnotesize \textit{Note:} 
% The dependent variable is the logarithm of headcount. 
% }
    \label{tab:reg_hc_10yrs}
\end{table}

In the meantime, we acknowledge that continued AI adoption may reshape skill demand in the labor market --- job destruction and creation may co-exist and may exhibit heterogeneous patterns across sectors, similar to robot adoption where firms lay off production workers and hire tech workers when they adopt industrial robots \citep{humlum2019robot}. 
These dynamics worth further investigation but are beyond the scope of this paper. 

\paragraph{Productivity. }As shown in columns (4) and (6) of \cref{tab:reg_prod_10yrs}, we don't find meaningful productivity gains due to AI adoption. A few possible explanations include:
\begin{enumerate}
    \item Organizational frictions: firms face coordination challenges when rolling out AI tools and workflows developed within IT units to business and operational teams, due to misaligned incentives, concerns over risk and accountability, limited AI literacy among end users, and the tendency for AI tools to be overly generic or insufficiently user-friendly.\footnote{\url{https://executive.mit.edu/blog/beyond-the-algorithm-bridging-the-last-mile-of-ai-adoption.html}} Hence, the practical usage might be low.
    \item Non-AI bottlenecks: AI often improves specific tasks (e.g., coding, customer service), but firm-level productivity may depend on bottlenecks elsewhere, and gains in one function may not translate into overall output increases \citep{chan2026paradox}.
    \item Lagged responses: AI may improve worker efficiency, but firms may not immediately adjust headcount or other inputs, leading to a temporary disconnect between AI adoption and productivity metrics. Over time, as firms optimize their workforce and processes around AI capabilities, stronger productivity effects may emerge \citep{brynjolfsson2021productivity}. 
\end{enumerate}

\begin{table}[p]
    \centering
    \caption{\bf AI Adoption and Productivity from 2016 to 2025}
    % \hspace*{-2.5cm}
\scalebox{0.65}{
{
\def\sym#1{\ifmmode^{#1}\else\(^{#1}\)\fi}
\begin{tabular}{l*{6}{D{.}{.}{-1}}}
\toprule
\multicolumn{7}{c}{Dependent Variable: Log Revenue per Employee}\\
            &\multicolumn{1}{c}{(1)}         &\multicolumn{1}{c}{(2)}         &\multicolumn{1}{c}{(3)}         &\multicolumn{1}{c}{(4)}         &\multicolumn{1}{c}{(5)}         &\multicolumn{1}{c}{(6)}         \\
\midrule
AI(2): Exploration&       0.047         &       0.036         &      -0.045\sym{***}&      -0.041\sym{***}&      -0.036\sym{***}&      -0.031\sym{**} \\
            &     (0.055)         &     (0.058)         &     (0.014)         &     (0.014)         &     (0.013)         &     (0.014)         \\
\addlinespace[1.1em]
AI(3): In Pilot Phase&      -0.149\sym{***}&      -0.122\sym{***}&      -0.051\sym{***}&      -0.043\sym{***}&      -0.046\sym{***}&      -0.038\sym{***}\\
            &     (0.035)         &     (0.039)         &     (0.011)         &     (0.012)         &     (0.011)         &     (0.012)         \\
\addlinespace[1.1em]
AI(4): Used in Production&      -0.121         &      -0.053         &      -0.027         &      -0.012         &      -0.036\sym{*}  &      -0.028         \\
            &     (0.085)         &     (0.099)         &     (0.022)         &     (0.025)         &     (0.021)         &     (0.025)         \\
\addlinespace[1.1em]
AI(5): Deep Integration&      -0.067         &      -0.284\sym{*}  &      -0.031         &      -0.001         &      -0.041         &      -0.013         \\
            &     (0.069)         &     (0.146)         &     (0.024)         &     (0.038)         &     (0.025)         &     (0.038)         \\
\addlinespace[1.1em]
AI(2) $\times$ Tech&                     &       0.143         &                     &      -0.033         &                     &      -0.036         \\
            &                     &     (0.185)         &                     &     (0.042)         &                     &     (0.043)         \\
\addlinespace[1.1em]
AI(3) $\times$ Tech&                     &       0.084         &                     &      -0.050\sym{**} &                     &      -0.046         \\
            &                     &     (0.094)         &                     &     (0.025)         &                     &     (0.028)         \\
\addlinespace[1.1em]
AI(4) $\times$ Tech&                     &      -0.032         &                     &      -0.065         &                     &      -0.045         \\
            &                     &     (0.196)         &                     &     (0.045)         &                     &     (0.049)         \\
\addlinespace[1.1em]
AI(5) $\times$ Tech&                     &       0.564\sym{***}&                     &      -0.074         &                     &      -0.081         \\
            &                     &     (0.174)         &                     &     (0.048)         &                     &     (0.055)         \\
\addlinespace[1.1em]
Tech        &                     &      -0.326\sym{***}&                     &                     &                     &                     \\
            &                     &     (0.062)         &                     &                     &                     &                     \\
\addlinespace[1.1em]
constant    &       6.343\sym{***}&       6.372\sym{***}&       6.323\sym{***}&       6.325\sym{***}&       6.318\sym{***}&       6.320\sym{***}\\
            &     (0.018)         &     (0.018)         &     (0.005)         &     (0.005)         &     (0.005)         &     (0.005)         \\
\midrule
Observations&\multicolumn{1}{c}{4470}         &\multicolumn{1}{c}{4470}         &\multicolumn{1}{c}{4465}         &\multicolumn{1}{c}{4465}         &\multicolumn{1}{c}{4301}         &\multicolumn{1}{c}{4301}         \\
R\textsuperscript{2}&\multicolumn{1}{c}{0.005}         &\multicolumn{1}{c}{0.015}         &\multicolumn{1}{c}{0.963}         &\multicolumn{1}{c}{0.963}         &\multicolumn{1}{c}{0.970}         &\multicolumn{1}{c}{0.970}         \\
Firm FE     &\multicolumn{1}{c}{No}         &\multicolumn{1}{c}{No}         &\multicolumn{1}{c}{Yes}         &\multicolumn{1}{c}{Yes}         &\multicolumn{1}{c}{Yes}         &\multicolumn{1}{c}{Yes}         \\
Year FE     &\multicolumn{1}{c}{No}         &\multicolumn{1}{c}{No}         &\multicolumn{1}{c}{Yes}         &\multicolumn{1}{c}{Yes}         &\multicolumn{1}{c}{No}         &\multicolumn{1}{c}{No}         \\
Sector $\times$ Year FE&\multicolumn{1}{c}{No}         &\multicolumn{1}{c}{No}         &\multicolumn{1}{c}{No}         &\multicolumn{1}{c}{No}         &\multicolumn{1}{c}{Yes}         &\multicolumn{1}{c}{Yes}         \\
\bottomrule
\multicolumn{7}{l}{\footnotesize Standard errors in parentheses}\\
\multicolumn{7}{l}{\footnotesize \sym{*} \(p<0.1\), \sym{**} \(p<0.05\), \sym{***} \(p<0.01\)}\\
\end{tabular}
}

}
\vspace{0.1in}
% \caption*{\footnotesize \textit{Note:} 
% The dependent variable is the logarithm of productivity. 
% }
    \label{tab:reg_prod_10yrs}
\end{table}

\section{Conclusion}\label{sec:conclusion}

In 2025, 21\% of S\&P 500 enterprises had deeply integrated AI into their business processes or were using AI in the production of goods and delivery of services, scoring a 4 or 5 in our rubric. With adoption accelerating among non-technology firms, deep AI adoption has grown from 1\% in 2022 to 5.7\% in 2025.  However, aggressive adoption in the technology sector accounts for two-thirds of 2025's advanced adoption. 

Our novel measure of deep AI adoption is designed to measure enterprise deployment in the performance of tasks in the execution of business processes. The use of generative AI models as a stand-alone work tool may increase worker productivity, but by itself might not be adequate in all cases to perform a wide range of tasks, for instance searching for new molecules in the pharmaceutical industry and managing supply chains in the retail industry. 

%For this reason, firms that score a 2 in the rubric are only exploring the possibly of adoption and those scoring a 3 are pilot-testing AI in products or processes but are not yet emphasizing financial results. Scores of 4 and 5 mean use of AI in production (4) and deeply embedded in business processes (5).

Regressing adoption on financial outcome shows differential effects across measures and sectors. Across all firms, we see profitability following a j-curve with AI adoption; those early in the integration process show 1\% to 3\% lower profitability than non-adopters, whereas those deeply embedding AI show around 6\% higher profitability. 
% Those firms using AI in production (score=4) or having deeply integrated AI into business processes (score=5) report higher net profits margins, whereas those at lower levels of adoption  with a J-curve effect. 
Conversely, there is no significant association between AI adoption and productivity, as measured by revenue per employee. 

There is also no significant correlation between AI adoption and capex. We show that the well-known substantial technology sector capex expenditures are limited to fewer than a half dozen firms among S\&P 500 firms over ten years. Those firms buying AI model services, even in the technology sector, engage in intermediate purchases, not capital spending. Finally, not surprisingly, there is a statistically significant relationship between the value of Tobin's q in the technology sector and AI adoption. Financial market participants seem to have been more willing to reward firms in the technology sector for the deep adoption of AI than firms in other sectors.

Throughout we are careful to observe that we are not making statements about causality, only descriptive relations between AI adoption and financial measures. Future work will take an endogenous growth perspective. The question to be explored is the marginal contribution of AI adoption, controlling for innovation, organizational transformation, and competitive pressures. \cite{aghion1992growth} marked a paradigm shift in the growth literature, featuring a rich interplay between competition and innovation in a process of creative destruction. Their endogenous growth theory suggests that the long-run growth rate determined by the frequency and size of innovations and the degree of market power enjoyed by innovators.
% makes a connection between the disruptive transformation process and technological change.

Similarly, \cite{bresnahan2002it}, using firm-level data, found that IT-enabled organizational change, rather than just technology adoption, constitutes key skill-biased technical change, boosting productivity. The returns to digital technologies are larger when firms pair information technology with complementary organizational practices and skills. 

If AI is to fulfill its promise as General Purpose Technology, transformation at a global scale will be required, business leaders will take new risks, and workers will be willing to find new skills.

\newpage
\bibliographystyle{plainnat}
\bibliography{reference}

\newpage

\appendix
\newpage
\section{The list of S\&P 500 firms}\label{apx:sp500scores}

{%
\setlength\LTleft{-1.5cm}
\setlength\LTright{0pt}
\setlength{\tabcolsep}{3pt}
\renewcommand{\arraystretch}{0.97}
\footnotesize
\begin{longtable}{p{5.5cm} p{4.2cm} c c c c c c c c c c}
\caption{\textbf{AI Scores by Company (2016--2025)}} \label{tab:sp500_ai_scores_2016_2025} \\
\toprule
\textbf{Company Name} & \textbf{Industry Group} & 2016 & 2017 & 2018 & 2019 & 2020 & 2021 & 2022 & 2023 & 2024 & 2025 \\
\midrule
\endfirsthead
\multicolumn{12}{c}{\tablename\ \thetable{} -- continued from previous page} \\
\toprule
\textbf{Company Name} & \textbf{Industry Group} & 2016 & 2017 & 2018 & 2019 & 2020 & 2021 & 2022 & 2023 & 2024 & 2025 \\
\midrule
\endhead
\midrule
\multicolumn{12}{r}{\textit{Continued on next page}} \\
\endfoot
\bottomrule
\endlastfoot
3M CO & Capital Goods & 1 & 1 & 1 & 3 & 1 & 1 & 2 & 1 & 2 & 3 \\
ABBOTT LABORATORIES & Health Care Equipment \& Services & 1 & 1 & 1 & 1 & 1 & 1 & 1 & 1 & 1 & 1 \\
ABBVIE INC & Pharmaceuticals, Biotechnology \& Life Sciences & 1 & 1 & 1 & 1 & 1 & 1 & 1 & 1 & 3 & 3 \\
ACCENTURE PLC & Software \& Services & 3 & 5 & 3 & 4 & 5 & 5 & 5 & 5 & 5 & 5 \\
ADOBE INC & Software \& Services & 3 & 5 & 5 & 5 & 5 & 5 & 5 & 5 & 5 & 5 \\
ADVANCED MICRO DEVICES & Semiconductors \& Semiconductor Equipment & 3 & 3 & 3 & 3 & 3 & 3 & 4 & 5 & 5 & 5 \\
AES CORP (THE) & Utilities & 1 & 1 & 1 & 1 & 1 & 1 & 3 & 3 & 3 & 4 \\
AFLAC INC & Insurance & 1 & 1 & 1 & 1 & 1 & 1 & 1 & 2 & 2 & 3 \\
AGILENT TECHNOLOGIES INC & Pharmaceuticals, Biotechnology \& Life Sciences & 1 & 1 & 1 & 2 & 3 & 2 & 2 & 3 & 3 & 3 \\
AIR PRODUCTS \& CHEMICALS INC & Materials & 1 & 1 & 1 & 1 & 1 & 1 & 1 & 1 & 1 & 1 \\
AIRBNB INC & Consumer Services & --- & --- & --- & --- & 3 & 3 & 3 & 3 & 3 & 3 \\
AKAMAI TECHNOLOGIES INC & Software \& Services & 1 & 1 & 1 & 1 & 1 & 1 & 1 & 1 & 3 & 4 \\
ALBEMARLE CORP & Materials & 1 & 1 & 1 & 1 & 1 & 1 & 1 & 1 & 1 & 3 \\
ALEXANDRIA R E EQUITIES INC & Real Estate & 1 & 1 & 1 & 1 & 1 & 1 & 1 & 1 & 1 & 3 \\
ALIGN TECHNOLOGY INC & Health Care Equipment \& Services & 1 & 1 & 1 & 1 & 1 & 1 & 3 & 3 & 5 & 5 \\
ALLEGION PLC & Capital Goods & 1 & 1 & 1 & 1 & 1 & 2 & 2 & 2 & 2 & 2 \\
ALLIANT ENERGY CORP & Utilities & 1 & 1 & 1 & 1 & 1 & 1 & 1 & 1 & 1 & 1 \\
ALLSTATE CORP & Insurance & 1 & 1 & 1 & 1 & 1 & 1 & 2 & 2 & 3 & 3 \\
ALPHABET INC & --- & 5 & 5 & 5 & 5 & 5 & 5 & 5 & 5 & 5 & 5 \\
ALTRIA GROUP INC & Household \& Personal Products & 1 & 1 & 1 & 1 & 1 & 1 & 1 & 1 & 1 & 1 \\
AMAZON.COM INC & Retailing & 1 & 1 & 1 & 3 & 5 & 5 & 5 & 5 & 5 & 5 \\
AMCOR PLC & Materials & --- & --- & --- & 1 & 1 & 1 & 1 & 1 & 1 & 1 \\
AMEREN CORP & Utilities & 1 & 1 & 1 & 1 & 1 & 1 & 1 & 1 & 1 & 1 \\
AMERICAN ELECTRIC POWER CO & Utilities & 1 & 1 & 1 & 1 & 1 & 1 & 1 & 1 & 2 & 3 \\
AMERICAN EXPRESS CO & Diversified Financials & 1 & 1 & 2 & 2 & 3 & 2 & 3 & 3 & 3 & 3 \\
AMERICAN INTERNATIONAL GROUP & Insurance & 1 & 1 & 1 & 1 & 1 & 1 & 1 & 2 & 2 & 3 \\
AMERICAN TOWER CORP & Real Estate & 1 & 1 & 1 & 1 & 1 & 1 & 1 & 2 & 2 & 4 \\
AMERICAN WATER WORKS CO INC & Utilities & 1 & 2 & 2 & 2 & 3 & 3 & 1 & 1 & 1 & 1 \\
AMERIPRISE FINANCIAL INC & Diversified Financials & 1 & 1 & 1 & 1 & 1 & 1 & 1 & 2 & 2 & 2 \\
AMETEK INC & Capital Goods & 3 & 1 & 2 & 2 & 1 & 1 & 1 & 1 & 1 & 3 \\
AMGEN INC & Pharmaceuticals, Biotechnology \& Life Sciences & 1 & 1 & 1 & 1 & 1 & 1 & 1 & 3 & 3 & 3 \\
AMPHENOL CORP & Technology Hardware \& Equipment & 1 & 1 & 1 & 1 & 1 & 1 & 1 & 1 & 4 & 5 \\
ANALOG DEVICES INC & Semiconductors \& Semiconductor Equipment & 1 & 1 & 1 & 1 & 1 & 1 & 1 & 1 & 3 & 3 \\
AON PLC & Insurance & 1 & 1 & 1 & 1 & 1 & 1 & 2 & 2 & 3 & 3 \\
APA CORP & Energy & --- & --- & --- & --- & --- & 1 & 1 & 1 & 1 & 1 \\
APOLLO GLOBAL MGMT INC & Diversified Financials & --- & --- & --- & --- & --- & --- & 2 & 1 & 2 & 3 \\
APPLE INC & Technology Hardware \& Equipment & 1 & 1 & 1 & 1 & 1 & 1 & 1 & 1 & 1 & 1 \\
APPLIED MATERIALS INC & Semiconductors \& Semiconductor Equipment & 1 & 1 & 1 & 1 & 1 & 1 & 2 & 2 & 3 & 3 \\
APPLOVIN CORP & Software \& Services & --- & --- & --- & --- & --- & 5 & 5 & 5 & 5 & 5 \\
APTIV PLC & Automobiles \& Components & 1 & --- & 3 & 5 & 5 & 3 & 3 & 1 & 3 & 3 \\
ARCH CAPITAL GROUP LTD & Insurance & 1 & 1 & 1 & 1 & 1 & 1 & 1 & 2 & 3 & 3 \\
ARCHER-DANIELS-MIDLAND CO & Household \& Personal Products & 1 & 1 & 1 & 1 & 1 & 1 & 1 & 2 & 3 & 3 \\
ARES MANAGEMENT CORP & Diversified Financials & 1 & 1 & 1 & 1 & 1 & 1 & 1 & 1 & 2 & 3 \\
ARISTA NETWORKS INC & Technology Hardware \& Equipment & 1 & 1 & 3 & 3 & 3 & 3 & 3 & 4 & 5 & 5 \\
ARTHUR J GALLAGHER \& CO & Insurance & 1 & 1 & 2 & 2 & 1 & 2 & 2 & 3 & 4 & 3 \\
ASSURANT INC & Insurance & 1 & 2 & 1 & 1 & 1 & 1 & 1 & 2 & 3 & 3 \\
ATMOS ENERGY CORP & Utilities & 1 & 1 & 1 & 1 & 1 & 1 & 1 & 1 & 1 & 1 \\
AT\&T INC & Telecommunication Services & 1 & 1 & 1 & 1 & 1 & 1 & 1 & 2 & 4 & 4 \\
AUTODESK INC & Software \& Services & 1 & 1 & 1 & 1 & 1 & 3 & 3 & 3 & 3 & 4 \\
AUTOMATIC DATA PROCESSING & Commercial \& Professional Services & 1 & 1 & 1 & 3 & 3 & 3 & 3 & 5 & 5 & 5 \\
AUTOZONE INC & Retailing & 1 & 1 & 1 & 1 & 1 & 1 & 1 & 1 & 1 & 2 \\
AVALONBAY COMMUNITIES INC & Real Estate & 1 & 1 & 1 & 1 & 1 & 1 & 1 & 3 & 3 & 3 \\
AVERY DENNISON CORP & Materials & 1 & 1 & 1 & 1 & --- & 1 & 1 & 1 & 2 & 2 \\
AXON ENTERPRISE INC & Capital Goods & 3 & 3 & 3 & 3 & 3 & 2 & 1 & 2 & 3 & 5 \\
BAKER HUGHES CO & Energy & --- & 3 & 1 & 3 & 3 & 3 & 3 & 3 & 3 & 4 \\
BALL CORP & Materials & 1 & 1 & 1 & 1 & 1 & 1 & 1 & 1 & 1 & 2 \\
BANK OF AMERICA CORP & Banks & 1 & 1 & 2 & 3 & 3 & 3 & 3 & 2 & 3 & 3 \\
BANK OF NEW YORK MELLON CORP & Diversified Financials & 1 & 1 & 1 & 1 & 1 & 3 & 2 & 3 & 3 & 5 \\
BAXTER INTERNATIONAL INC & Health Care Equipment \& Services & 1 & 1 & 1 & 1 & 1 & 1 & 1 & 2 & 3 & 3 \\
BECTON DICKINSON \& CO & Health Care Equipment \& Services & 1 & 1 & 1 & 1 & 1 & 1 & 1 & 3 & 3 & 3 \\
BERKLEY (W R) CORP & Insurance & 1 & 1 & 1 & 1 & 1 & 1 & 1 & 1 & 1 & 2 \\
BERKSHIRE HATHAWAY & Diversified Financials & 1 & 1 & 1 & 1 & 1 & 1 & 1 & 1 & 1 & 1 \\
BEST BUY CO INC & Retailing & 1 & 1 & 1 & 1 & 1 & 2 & 3 & 2 & 1 & 3 \\
BIO-TECHNE CORP & Pharmaceuticals, Biotechnology \& Life Sciences & 1 & 1 & 1 & 1 & 1 & 1 & 1 & 1 & 1 & 1 \\
BIOGEN INC & Pharmaceuticals, Biotechnology \& Life Sciences & 1 & 1 & 1 & 1 & 1 & 2 & 1 & 1 & 3 & 3 \\
BLACKROCK INC & Diversified Financials & --- & --- & --- & --- & --- & --- & --- & --- & 3 & 4 \\
BLACKSTONE INC & Diversified Financials & 1 & 1 & 1 & 1 & 1 & 2 & 2 & 2 & 3 & 3 \\
BLOCK INC & Diversified Financials & 4 & 3 & 4 & 3 & 3 & 3 & 3 & 3 & 3 & 4 \\
BOEING CO & Capital Goods & 1 & 1 & 2 & 3 & 1 & 1 & 1 & 1 & 1 & 1 \\
BOOKING HOLDINGS INC & Consumer Services & 1 & 1 & 1 & 3 & 3 & 3 & 3 & 3 & 3 & 4 \\
BOSTON SCIENTIFIC CORP & Health Care Equipment \& Services & 1 & 1 & 1 & 1 & 1 & 1 & 1 & 2 & 2 & 3 \\
BRISTOL-MYERS SQUIBB CO & Pharmaceuticals, Biotechnology \& Life Sciences & 1 & 1 & 1 & 1 & 1 & 1 & 1 & 1 & 3 & 3 \\
BROADCOM INC & Semiconductors \& Semiconductor Equipment & --- & --- & 1 & 1 & 3 & 3 & 3 & 3 & 5 & 5 \\
BROADRIDGE FINANCIAL SOLUTNS & Commercial \& Professional Services & 1 & 1 & 2 & 2 & 2 & 3 & 3 & 3 & 4 & 4 \\
BROWN FORMAN CORP & Household \& Personal Products & 1 & 1 & 1 & 1 & 1 & 1 & 1 & 1 & 1 & 1 \\
BROWN \& BROWN INC & Insurance & 1 & 1 & 1 & 1 & 1 & 1 & 1 & 1 & 3 & 3 \\
BUILDERS FIRSTSOURCE & Capital Goods & 1 & 1 & 1 & 1 & 1 & 1 & 1 & 1 & 1 & 2 \\
BUNGE GLOBAL SA & Household \& Personal Products & --- & --- & --- & --- & --- & --- & --- & 1 & 3 & 3 \\
BXP INC & Real Estate & 1 & 1 & 1 & 1 & 1 & 1 & 1 & 1 & 3 & 3 \\
C H ROBINSON WORLDWIDE INC & Transportation & 1 & 3 & 3 & 3 & 3 & 3 & 3 & 3 & 5 & 5 \\
CADENCE DESIGN SYSTEMS INC & Software \& Services & 1 & 1 & 3 & 3 & --- & 3 & 3 & 5 & 5 & 5 \\
CAESARS ENTERTAINMENT INC & Consumer Services & 1 & 1 & 1 & 1 & 1 & 1 & 1 & 1 & 1 & 1 \\
CAMDEN PROPERTY TRUST & Real Estate & 1 & 1 & 1 & 1 & 1 & 1 & 1 & 1 & 1 & 2 \\
CAMPBELL'S CO (THE) & Household \& Personal Products & 1 & 1 & 1 & 1 & 1 & 1 & 1 & 1 & 1 & 1 \\
CAPITAL ONE FINANCIAL CORP & Diversified Financials & 3 & 3 & 3 & 3 & 3 & 3 & 2 & 4 & 3 & 4 \\
CARDINAL HEALTH INC & Health Care Equipment \& Services & 1 & 1 & 1 & 1 & 1 & 1 & 1 & 1 & 3 & 2 \\
CARMAX INC & Retailing & 1 & 1 & 1 & 1 & 3 & 3 & 4 & 3 & 3 & 4 \\
CARNIVAL CORPORATION \& PLC & Consumer Services & 1 & 1 & 1 & 1 & 1 & 1 & 1 & 1 & 1 & 4 \\
CARRIER GLOBAL CORP & Capital Goods & --- & --- & --- & --- & 3 & 3 & 3 & 3 & 1 & 1 \\
CARVANA CO & Retailing & --- & 4 & 3 & 3 & 4 & 4 & 3 & 3 & 3 & 3 \\
CATERPILLAR INC & Capital Goods & 1 & 1 & 1 & 1 & 1 & 1 & 1 & 1 & 1 & 1 \\
CBOE GLOBAL MARKETS INC & Diversified Financials & 1 & 1 & 1 & 1 & 1 & 2 & 1 & 2 & 2 & 2 \\
CBRE GROUP INC & --- & 1 & 1 & 1 & 1 & 1 & 1 & 1 & 1 & 3 & 2 \\
CDW CORP & Technology Hardware \& Equipment & 1 & 1 & 1 & 1 & 1 & 1 & 1 & 3 & 3 & 3 \\
CENCORA INC & Health Care Equipment \& Services & 1 & 1 & 1 & 1 & 1 & 1 & 1 & 1 & 1 & 3 \\
CENTENE CORP & Health Care Equipment \& Services & 3 & 3 & 3 & 3 & 3 & 3 & 3 & 1 & 3 & 4 \\
CENTERPOINT ENERGY INC & Utilities & 1 & 1 & 1 & 1 & 1 & 1 & 1 & 3 & 2 & 2 \\
CF INDUSTRIES HOLDINGS INC & Materials & 1 & 1 & 1 & 1 & 1 & 1 & 1 & 1 & 1 & 1 \\
CHARLES RIVER LABS INTL INC & Pharmaceuticals, Biotechnology \& Life Sciences & 1 & 1 & 2 & 3 & 3 & 3 & 3 & 3 & 3 & 3 \\
CHARTER COMMUNICATIONS INC & --- & 1 & 1 & 1 & 1 & 1 & 1 & 1 & 3 & 3 & 3 \\
CHEVRON CORP & Energy & 1 & 1 & 1 & 1 & 1 & 1 & 1 & 3 & 3 & 3 \\
CHIPOTLE MEXICAN GRILL INC & Consumer Services & 3 & 3 & 4 & 2 & 1 & 1 & 1 & 1 & 1 & 2 \\
CHUBB LTD & Insurance & 1 & 1 & 1 & 1 & 1 & 1 & 1 & 2 & 3 & 3 \\
CHURCH \& DWIGHT INC & Consumer Staples Distribution \& Retail & 1 & 1 & 1 & 1 & 1 & 1 & 1 & 1 & 1 & 2 \\
CIGNA GROUP (THE) & Health Care Equipment \& Services & --- & --- & 4 & 1 & 4 & 4 & 3 & 4 & 4 & 4 \\
CINCINNATI FINANCIAL CORP & Insurance & 3 & 3 & 3 & 3 & 3 & 3 & 2 & 3 & 3 & 4 \\
CINTAS CORP & Commercial \& Professional Services & 1 & 1 & 1 & 1 & 1 & 1 & 1 & 1 & 1 & 1 \\
CISCO SYSTEMS INC & Technology Hardware \& Equipment & 1 & 3 & 3 & 3 & 3 & 3 & 3 & 5 & 3 & 5 \\
CITIGROUP INC & Banks & 1 & 1 & 1 & 1 & 2 & 1 & 1 & 2 & 3 & 3 \\
CITIZENS FINANCIAL GROUP INC & Banks & 1 & 1 & 1 & 1 & 3 & 3 & 3 & 2 & 3 & 3 \\
CLOROX CO/DE & Consumer Staples Distribution \& Retail & 1 & 1 & 1 & 1 & 1 & 1 & 1 & 1 & 2 & 1 \\
CME GROUP INC & Diversified Financials & 1 & 1 & 1 & 1 & 1 & 1 & 1 & 1 & 1 & 3 \\
CMS ENERGY CORP & Utilities & 1 & 1 & 1 & 1 & 1 & 1 & 1 & 1 & 1 & 1 \\
COCA-COLA CO & Household \& Personal Products & 1 & 1 & 2 & 3 & 3 & 2 & 3 & 4 & 3 & 3 \\
COGNIZANT TECH SOLUTIONS & Software \& Services & 1 & 5 & 3 & 3 & 3 & 3 & 3 & 4 & 5 & 5 \\
COINBASE GLOBAL INC & Diversified Financials & --- & --- & --- & --- & --- & 3 & 1 & 3 & 3 & 3 \\
COLGATE-PALMOLIVE CO & Consumer Staples Distribution \& Retail & 1 & 1 & 1 & 1 & 1 & 1 & 1 & 2 & 2 & 3 \\
COMCAST CORP & Telecommunication Services & 1 & 1 & 1 & 1 & 1 & 1 & 1 & 2 & 1 & 2 \\
COMFORT SYSTEMS USA INC & Capital Goods & 1 & 1 & 1 & 1 & 1 & 1 & 1 & 1 & 1 & 1 \\
CONAGRA BRANDS INC & Household \& Personal Products & 1 & 1 & 1 & 1 & 1 & 1 & 1 & 1 & 1 & 2 \\
CONOCOPHILLIPS & Energy & 1 & 1 & 1 & 3 & 3 & 1 & 1 & 1 & 1 & 1 \\
CONSOLIDATED EDISON INC & Utilities & 1 & 1 & 1 & 1 & 1 & 1 & 1 & 1 & 1 & 1 \\
CONSTELLATION BRANDS & Household \& Personal Products & 1 & 1 & 1 & 1 & 1 & 1 & 1 & 1 & 2 & 2 \\
CONSTELLATION ENE CORP & Utilities & --- & --- & --- & --- & --- & 3 & 1 & 1 & 1 & 3 \\
COOPER COS INC (THE) & Health Care Equipment \& Services & 1 & 1 & 1 & 1 & 1 & 1 & 1 & 1 & 1 & 1 \\
COPART INC & Commercial \& Professional Services & 1 & 1 & 1 & 3 & 3 & 3 & 3 & 3 & 3 & 3 \\
CORNING INC & Technology Hardware \& Equipment & 1 & 1 & 1 & 1 & 1 & 1 & 1 & 1 & 4 & 4 \\
CORPAY INC & Diversified Financials & 1 & 1 & 1 & 1 & 1 & 1 & 3 & 3 & 3 & 3 \\
CORTEVA INC & Materials & --- & --- & --- & 1 & 1 & 1 & 1 & 1 & 2 & 1 \\
COSTAR GROUP INC & --- & 1 & 1 & 1 & 1 & 1 & 1 & 1 & 1 & 1 & 5 \\
COSTCO WHOLESALE CORP & Food, Beverage \& Tobacco & 1 & 1 & 1 & 1 & 1 & 1 & 1 & 1 & 1 & 1 \\
COTERRA ENERGY INC & Energy & 1 & 1 & 1 & 1 & 1 & 1 & 1 & 1 & 1 & 1 \\
CRH PLC & Materials & --- & --- & --- & --- & --- & --- & --- & 2 & 2 & 3 \\
CROWDSTRIKE HOLDINGS INC & Software \& Services & --- & --- & --- & --- & 5 & 5 & 5 & 5 & 5 & 5 \\
CROWN CASTLE INC & Real Estate & 1 & 1 & 1 & 1 & 1 & 1 & 1 & 1 & 1 & 2 \\
CSX CORP & Transportation & 1 & 1 & 1 & 1 & 1 & 1 & 1 & 1 & 1 & 1 \\
CUMMINS INC & Capital Goods & 1 & 1 & 1 & 1 & 1 & 1 & 1 & 1 & 2 & 3 \\
CVS HEALTH CORP & Health Care Equipment \& Services & 1 & 1 & 1 & 2 & 3 & 3 & 3 & 5 & 3 & 3 \\
D R HORTON INC & Consumer Durables \& Apparel & 1 & 1 & 1 & 1 & 1 & 1 & 1 & 1 & 1 & 3 \\
DANAHER CORP & Pharmaceuticals, Biotechnology \& Life Sciences & 1 & 2 & 1 & 1 & 1 & 2 & 1 & 2 & 2 & 2 \\
DARDEN RESTAURANTS INC & Consumer Services & 1 & 1 & 1 & 1 & 1 & 1 & 1 & 1 & 1 & 1 \\
DATADOG INC & Software \& Services & --- & --- & --- & 3 & 3 & 3 & 3 & 3 & 4 & 5 \\
DAVITA INC & Health Care Equipment \& Services & 1 & 1 & 1 & 2 & 2 & 2 & 2 & 2 & 2 & 3 \\
DAYFORCE INC & Commercial \& Professional Services & --- & --- & 1 & 1 & 1 & 1 & 3 & 3 & 4 & --- \\
DECKERS OUTDOOR CORP & Consumer Durables \& Apparel & 1 & 1 & 1 & 1 & 1 & 1 & 1 & 1 & 3 & 3 \\
DEERE \& CO & Capital Goods & 1 & 3 & 3 & 3 & 1 & 1 & 3 & 3 & 3 & 3 \\
DELL TECHNOLOGIES INC & Technology Hardware \& Equipment & --- & 1 & 2 & 2 & 3 & 3 & 3 & 3 & 5 & 5 \\
DELTA AIR LINES INC & Transportation & 1 & 1 & 1 & 1 & 1 & 1 & 1 & 1 & 1 & 3 \\
DEVON ENERGY CORP & Energy & 1 & 4 & 1 & 2 & 2 & 2 & 2 & 2 & 2 & 3 \\
DEXCOM INC & Health Care Equipment \& Services & 2 & 2 & 2 & 2 & 2 & 3 & 2 & 2 & 3 & 3 \\
DIAMONDBACK ENERGY INC & Energy & 1 & 1 & 1 & 1 & 1 & 1 & 1 & 1 & 1 & 1 \\
DIGITAL REALTY TRUST INC & Real Estate & 1 & 1 & 1 & 1 & 1 & 1 & 1 & 3 & 3 & 2 \\
DISNEY (WALT) CO & --- & --- & --- & --- & 1 & 1 & 1 & 1 & 1 & 1 & 1 \\
DOLLAR GENERAL CORP & Food, Beverage \& Tobacco & 1 & 1 & 1 & 1 & 1 & 1 & 1 & 1 & 1 & 2 \\
DOLLAR TREE INC & Food, Beverage \& Tobacco & 1 & 1 & 1 & 1 & 1 & 1 & 1 & 1 & 1 & 1 \\
DOMINION ENERGY INC & Utilities & 1 & 1 & 1 & 1 & 1 & 1 & 1 & 1 & 1 & 1 \\
DOMINO'S PIZZA INC & Consumer Services & 1 & 1 & 1 & 1 & --- & 1 & 1 & 2 & 2 & 1 \\
DOORDASH INC & Consumer Services & --- & --- & --- & --- & 3 & 1 & 2 & 1 & 2 & 3 \\
DOVER CORP & Capital Goods & 1 & 1 & 2 & 1 & 3 & 3 & 3 & 4 & 3 & 3 \\
DOW INC & Materials & --- & --- & --- & 1 & 1 & 1 & 4 & 3 & 2 & 3 \\
DTE ENERGY CO & Utilities & 1 & 1 & 1 & 1 & 1 & 1 & 1 & 1 & 1 & 1 \\
DUKE ENERGY CORP & Utilities & 1 & 1 & 1 & 1 & 1 & 1 & 1 & 1 & 2 & 3 \\
DUPONT DE NEMOURS INC & Materials & --- & 1 & 1 & 1 & 1 & 1 & 1 & 1 & 4 & 2 \\
EASTMAN CHEMICAL CO & Materials & 1 & 1 & 1 & 1 & 1 & 3 & 3 & 3 & 1 & 3 \\
EATON CORP PLC & Capital Goods & 1 & 1 & 1 & 1 & 1 & 1 & 1 & 1 & 1 & 2 \\
EBAY INC & Retailing & 1 & 2 & 3 & 3 & 3 & 3 & 3 & 3 & 3 & 3 \\
ECOLAB INC & Materials & 1 & 1 & 1 & 1 & 2 & 2 & 3 & 1 & 3 & 3 \\
EDISON INTERNATIONAL & Utilities & 1 & 1 & 1 & 1 & 1 & 1 & 1 & 1 & 1 & 1 \\
EDWARDS LIFESCIENCES CORP & Health Care Equipment \& Services & 1 & 1 & 1 & 1 & 1 & 3 & 1 & 2 & 1 & 3 \\
ELECTRONIC ARTS INC & --- & 1 & 1 & 1 & 1 & 1 & 1 & 3 & 3 & 1 & 3 \\
ELEVANCE HEALTH INC & Health Care Equipment \& Services & 3 & 3 & 3 & 3 & 3 & 3 & 2 & 3 & 3 & 3 \\
EMCOR GROUP INC & Capital Goods & 1 & 1 & 1 & 1 & 1 & 1 & 1 & 2 & 2 & 2 \\
EMERSON ELECTRIC CO & Capital Goods & 1 & 1 & 1 & 1 & 1 & 1 & 4 & 1 & 1 & 3 \\
ENPHASE ENERGY INC & Semiconductors \& Semiconductor Equipment & 1 & 1 & 1 & 1 & 1 & 1 & 1 & 1 & 2 & 4 \\
ENTERGY CORP & Utilities & 1 & 1 & 1 & 1 & 1 & 1 & 1 & 1 & 2 & 1 \\
EOG RESOURCES INC & Energy & 1 & 1 & 1 & 1 & 1 & 1 & 1 & 1 & 1 & 1 \\
EPAM SYSTEMS INC & Software \& Services & 1 & 1 & 3 & 3 & 3 & 3 & 3 & 5 & 3 & 5 \\
EQT CORP & Energy & 1 & 1 & 1 & 1 & 1 & 1 & 1 & 1 & 1 & 1 \\
EQUIFAX INC & Commercial \& Professional Services & 1 & 1 & 3 & 3 & 3 & 3 & 3 & 3 & 3 & 3 \\
EQUINIX INC & Real Estate & 1 & 1 & 1 & 1 & 1 & 1 & 1 & 3 & 5 & 5 \\
EQUITY RESIDENTIAL & Real Estate & 1 & 1 & 1 & 1 & 1 & 1 & 1 & 3 & 3 & 3 \\
ERIE INDEMNITY CO & Insurance & 1 & 1 & 1 & 1 & 1 & 1 & 1 & 2 & 2 & 2 \\
ESSEX PROPERTY TRUST & Real Estate & 1 & 1 & 1 & 1 & 1 & 1 & 1 & 1 & 3 & 4 \\
ESTEE LAUDER COMPANIES INC & Consumer Staples Distribution \& Retail & 1 & 1 & 1 & 1 & 1 & 1 & 1 & 1 & 3 & 3 \\
EVEREST GROUP LTD & Insurance & 1 & 1 & 1 & 1 & 1 & 1 & 1 & 1 & 1 & 1 \\
EVERGY INC & Utilities & --- & --- & 1 & 1 & 1 & 1 & 1 & 1 & 3 & 3 \\
EVERSOURCE ENERGY & Utilities & 1 & 1 & 1 & 1 & 1 & 1 & 1 & 1 & 2 & 3 \\
EXELON CORP & Utilities & 1 & 1 & 1 & 1 & 1 & 1 & 1 & 1 & 1 & 1 \\
EXPAND ENERGY CORP & Energy & 1 & 1 & 1 & 1 & 1 & 1 & 1 & 1 & 1 & 1 \\
EXPEDIA GROUP INC & Consumer Services & 1 & 2 & 2 & 3 & 3 & 3 & 3 & 3 & 3 & 3 \\
EXPEDITORS INTL WASH INC & Transportation & 1 & 1 & 1 & 1 & 1 & 1 & 1 & 1 & 1 & 2 \\
EXTRA SPACE STORAGE INC & Real Estate & 1 & 1 & 1 & 1 & 1 & 1 & 1 & 1 & 1 & 3 \\
EXXON MOBIL CORP & Energy & 1 & 1 & 1 & 1 & 1 & 1 & 1 & 1 & 1 & 2 \\
F5 INC & Technology Hardware \& Equipment & 1 & 1 & 1 & 1 & 3 & 3 & 3 & 3 & 4 & 5 \\
FACTSET RESEARCH SYSTEMS INC & Diversified Financials & 1 & 1 & 1 & 1 & 1 & 3 & 3 & 3 & 3 & 4 \\
FAIR ISAAC CORP & Software \& Services & 5 & 5 & 5 & 4 & 5 & 5 & 5 & 3 & 3 & 5 \\
FASTENAL CO & Capital Goods & 1 & 1 & 1 & 1 & 1 & 1 & 1 & 1 & 1 & 3 \\
FEDERAL REALTY INVESTMENT TR & Real Estate & 1 & 1 & 1 & 1 & 1 & 1 & 1 & 1 & 1 & 2 \\
FEDEX CORP & Transportation & 1 & 1 & 1 & 3 & 3 & 3 & 3 & 4 & 4 & 4 \\
FIDELITY NATIONAL INFO SVCS & Diversified Financials & 3 & 3 & 3 & 3 & 3 & 3 & 3 & 3 & 3 & 4 \\
FIFTH THIRD BANCORP & Banks & 1 & 1 & 1 & 1 & 1 & 1 & 1 & 4 & 5 & 4 \\
FIRST SOLAR INC & Semiconductors \& Semiconductor Equipment & 1 & 1 & 1 & 1 & 1 & 1 & 1 & 1 & 3 & 3 \\
FIRSTENERGY CORP & Utilities & 1 & 1 & 1 & 1 & 1 & 1 & 1 & 1 & 1 & 2 \\
FISERV INC & Diversified Financials & 1 & 1 & 1 & 1 & 1 & 1 & 1 & 3 & 3 & 3 \\
FORD MOTOR CO & Automobiles \& Components & 1 & 1 & 1 & 1 & 3 & 3 & 3 & 2 & 2 & 1 \\
FORTINET INC & Software \& Services & 1 & 3 & 3 & 4 & 3 & 3 & 3 & 3 & 5 & 3 \\
FORTIVE CORP & Capital Goods & 1 & 1 & 1 & 1 & 1 & 1 & 1 & 2 & 2 & 3 \\
FOX CORP & --- & --- & --- & --- & 1 & 1 & 1 & 1 & 1 & 2 & 2 \\
FRANKLIN RESOURCES INC & Diversified Financials & 1 & 1 & 1 & 1 & 1 & 1 & 1 & 1 & 2 & 3 \\
FREEPORT-MCMORAN INC & Materials & 1 & 1 & 1 & 4 & 3 & 1 & 1 & 2 & 3 & 3 \\
GARMIN LTD & Consumer Durables \& Apparel & 1 & 1 & 1 & 1 & 1 & 1 & 1 & 1 & 1 & 1 \\
GARTNER INC & Software \& Services & 1 & 1 & 1 & 1 & 1 & 1 & 2 & 3 & 3 & 3 \\
GE AEROSPACE & Capital Goods & 3 & 3 & 1 & 3 & 3 & 3 & 3 & 2 & 2 & 1 \\
GE HEALTHCARE TECHNOLOGI INC & Health Care Equipment \& Services & --- & --- & --- & --- & --- & --- & 3 & 3 & 5 & 5 \\
GE VERNOVA INC & Capital Goods & --- & --- & --- & --- & --- & --- & --- & --- & 1 & 3 \\
GEN DIGITAL INC & Software \& Services & 1 & 3 & 3 & 3 & 1 & 1 & 1 & 3 & 3 & 3 \\
GENERAC HOLDINGS INC & Capital Goods & 1 & 1 & 1 & 1 & 1 & 1 & 1 & 1 & 2 & 1 \\
GENERAL DYNAMICS CORP & Capital Goods & 1 & 1 & 1 & 3 & 3 & 3 & 3 & 3 & 3 & 4 \\
GENERAL MILLS INC & Household \& Personal Products & 1 & 1 & 1 & 1 & 1 & 1 & 1 & 1 & 1 & 1 \\
GENERAL MOTORS CO & Automobiles \& Components & 1 & 1 & 1 & 1 & 1 & 1 & 1 & 1 & 3 & 3 \\
GENUINE PARTS CO & Retailing & 1 & 1 & 1 & 1 & 1 & 1 & 1 & 2 & 3 & 3 \\
GILEAD SCIENCES INC & Pharmaceuticals, Biotechnology \& Life Sciences & 1 & 1 & 1 & 1 & 1 & 1 & 1 & 1 & 1 & 2 \\
GLOBAL PAYMENTS INC & Diversified Financials & 1 & 1 & 1 & 1 & 1 & 1 & 1 & 2 & 2 & 3 \\
GLOBE LIFE INC & Insurance & 1 & 1 & 1 & 1 & 1 & 1 & 1 & 1 & 2 & 2 \\
GODADDY INC & Software \& Services & 1 & 1 & 1 & 2 & 2 & 2 & 2 & 3 & 4 & 5 \\
GOLDMAN SACHS GROUP INC & Diversified Financials & 1 & 1 & 1 & 1 & 1 & 1 & 1 & 2 & 3 & 3 \\
GRAINGER (W W) INC & Capital Goods & 1 & 1 & 1 & 1 & 1 & 1 & 1 & 3 & 3 & 3 \\
HALLIBURTON CO & Energy & 1 & 1 & 1 & 1 & 3 & 3 & 3 & 3 & 3 & 3 \\
HARTFORD INSURANCE GROUP INC & Insurance & 3 & 3 & 2 & 2 & 2 & 3 & 2 & 3 & 4 & 3 \\
HASBRO INC & Consumer Durables \& Apparel & 1 & 1 & 1 & 1 & 1 & 1 & 1 & 1 & 2 & 3 \\
HCA HEALTHCARE INC & Health Care Equipment \& Services & 1 & 1 & 1 & 1 & 1 & 1 & 3 & 3 & 3 & 3 \\
HEALTHPEAK PROPERTIES INC & Real Estate & 1 & 1 & 1 & 1 & 1 & 1 & 1 & 1 & 3 & 3 \\
HENRY (JACK) \& ASSOCIATES & Diversified Financials & 1 & 1 & 1 & 1 & 1 & 1 & 3 & 3 & 3 & 3 \\
HENRY SCHEIN INC & Health Care Equipment \& Services & 1 & --- & 1 & 1 & 1 & 1 & 1 & 1 & 3 & 3 \\
HERSHEY CO & Household \& Personal Products & 1 & 1 & 1 & 1 & 1 & 1 & 1 & 2 & 2 & 2 \\
HEWLETT PACKARD ENTERPRISE & Technology Hardware \& Equipment & 1 & 3 & 3 & 3 & 3 & 3 & 3 & 4 & 5 & 5 \\
HILTON WORLDWIDE HOLDINGS & Consumer Services & 1 & 1 & 1 & 1 & 1 & 1 & 1 & 3 & 3 & 3 \\
HOLOGIC INC & Health Care Equipment \& Services & 1 & 1 & 1 & 1 & 3 & 3 & 3 & 3 & 4 & 3 \\
HOME DEPOT INC & Retailing & 1 & 1 & 1 & 3 & 1 & 1 & 1 & 1 & 3 & 3 \\
HONEYWELL INTERNATIONAL INC & Capital Goods & 1 & 1 & 1 & 1 & 1 & 1 & 3 & 1 & 4 & 3 \\
HORMEL FOODS CORP & Household \& Personal Products & 1 & 1 & 1 & 1 & 1 & 1 & 1 & 1 & 1 & 1 \\
HOST HOTELS \& RESORTS INC & Real Estate & 1 & 1 & 1 & 1 & 1 & 1 & 1 & 1 & 2 & 2 \\
HOWMET AEROSPACE INC & Capital Goods & 1 & 1 & 1 & 1 & 1 & 1 & 1 & 1 & 1 & 2 \\
HP INC & Technology Hardware \& Equipment & 1 & 1 & 1 & 1 & 1 & 1 & 1 & 3 & 3 & 5 \\
HUBBELL INC & Capital Goods & 1 & 1 & 1 & 1 & 1 & 1 & 1 & 1 & 1 & 3 \\
HUMANA INC & Health Care Equipment \& Services & 3 & 3 & 3 & 3 & 3 & 1 & 1 & 1 & 2 & 2 \\
HUNT (JB) TRANSPRT SVCS INC & Transportation & 1 & 1 & 1 & 1 & 1 & 1 & 1 & 1 & 1 & 2 \\
HUNTINGTON BANCSHARES & Banks & 1 & 1 & 1 & 1 & 1 & 1 & 1 & 1 & 3 & 3 \\
HUNTINGTON INGALLS IND INC & Capital Goods & 1 & 1 & 1 & 3 & 3 & 3 & 3 & 3 & 3 & 3 \\
IDEX CORP & Capital Goods & 1 & 1 & 1 & 1 & 1 & 1 & 1 & 1 & 2 & 2 \\
IDEXX LABS INC & Health Care Equipment \& Services & 3 & 3 & 3 & 3 & 1 & 1 & 1 & 3 & 5 & 4 \\
ILLINOIS TOOL WORKS & Capital Goods & 1 & 1 & 1 & 1 & 1 & 1 & 1 & 1 & 1 & 2 \\
INCYTE CORP & Pharmaceuticals, Biotechnology \& Life Sciences & 1 & 1 & 1 & 1 & 1 & 1 & 1 & 1 & 1 & 3 \\
INGERSOLL RAND INC & Capital Goods & --- & 1 & 1 & 1 & 1 & 1 & 1 & 1 & 3 & 3 \\
INSULET CORP & Health Care Equipment \& Services & 1 & 1 & 1 & 1 & 1 & 1 & 1 & 1 & 3 & 3 \\
INTEL CORP & Semiconductors \& Semiconductor Equipment & 3 & 5 & 5 & 5 & 4 & 5 & 5 & 5 & 5 & 5 \\
INTERACTIVE BROKERS GROUP & Diversified Financials & 1 & 1 & 1 & 1 & 1 & 1 & 1 & 1 & 3 & 3 \\
INTERCONTINENTAL EXCHANGE & Diversified Financials & 1 & 1 & 1 & 1 & 3 & 3 & 3 & 3 & 3 & 3 \\
INTERPUBLIC GROUP OF COS & --- & 1 & 1 & 1 & 1 & 1 & 1 & 1 & 3 & 5 & --- \\
INTL BUSINESS MACHINES CORP & Software \& Services & 5 & 5 & 5 & 5 & 5 & 5 & 5 & 5 & 5 & 5 \\
INTL FLAVORS \& FRAGRANCES & Materials & 1 & 1 & 1 & 1 & 1 & 1 & 1 & 1 & 3 & 3 \\
INTL PAPER CO & Materials & 1 & 1 & 1 & 1 & 1 & 1 & 1 & 1 & 1 & 2 \\
INTUIT INC & Software \& Services & 1 & 3 & 3 & 3 & 5 & 5 & 3 & 5 & 5 & 5 \\
INTUITIVE SURGICAL INC & Health Care Equipment \& Services & 1 & 1 & 1 & 2 & 3 & 3 & 3 & 3 & 3 & 4 \\
INVESCO LTD & Diversified Financials & 1 & 1 & 1 & 1 & 1 & 1 & 1 & 1 & 3 & 3 \\
INVITATION HOMES INC & Real Estate & 2 & 2 & 1 & 1 & 2 & 2 & 1 & 2 & 3 & 3 \\
IQVIA HOLDINGS INC & Pharmaceuticals, Biotechnology \& Life Sciences & 1 & 1 & 3 & 3 & 5 & 5 & 5 & 3 & 5 & 5 \\
IRON MOUNTAIN INC & Real Estate & 1 & 1 & 1 & 3 & 3 & 3 & 2 & 3 & 3 & 3 \\
JABIL INC & Technology Hardware \& Equipment & 1 & 1 & 1 & 1 & 1 & 1 & 1 & 1 & 2 & 3 \\
JACOBS SOLUTIONS INC & Commercial \& Professional Services & 1 & 1 & 1 & 3 & 3 & 3 & 3 & 3 & 3 & 3 \\
JOHNSON CONTROLS INTL PLC & Capital Goods & 1 & 1 & 1 & 1 & 3 & 3 & 3 & 3 & 3 & 3 \\
JOHNSON \& JOHNSON & Pharmaceuticals, Biotechnology \& Life Sciences & 1 & 1 & 1 & 1 & --- & 1 & 1 & 3 & 3 & 3 \\
JPMORGAN CHASE \& CO & Banks & 1 & 1 & 1 & 1 & 1 & 1 & 1 & 3 & 4 & 3 \\
KELLANOVA & Household \& Personal Products & 1 & 1 & 1 & 1 & --- & 1 & 1 & 1 & 1 & --- \\
KENVUE INC & Consumer Staples Distribution \& Retail & --- & --- & --- & --- & --- & --- & --- & 2 & 2 & 2 \\
KEURIG DR PEPPER INC & Household \& Personal Products & 1 & 1 & 1 & 1 & 1 & 3 & 3 & 3 & 3 & 3 \\
KEYCORP & Banks & 1 & 1 & 1 & 1 & 1 & 1 & 1 & 1 & 3 & 3 \\
KEYSIGHT TECHNOLOGIES INC & Technology Hardware \& Equipment & 1 & 1 & 1 & 1 & 3 & 3 & 3 & 3 & 3 & 4 \\
KIMBERLY-CLARK CORP & Consumer Staples Distribution \& Retail & 1 & 1 & 1 & 1 & 1 & 1 & 1 & 1 & 1 & 1 \\
KIMCO REALTY CORP & Real Estate & 1 & 1 & 1 & 1 & 1 & 1 & 1 & 3 & 3 & 3 \\
KINDER MORGAN INC & Energy & 1 & 1 & 1 & 1 & 1 & 1 & 1 & 1 & 1 & 1 \\
KKR \& CO INC & Diversified Financials & 1 & 1 & 1 & 1 & 1 & 1 & 1 & 3 & 3 & 3 \\
KLA CORP & Semiconductors \& Semiconductor Equipment & 1 & 3 & 1 & 3 & --- & 2 & 1 & 1 & 3 & 4 \\
KRAFT HEINZ CO & Household \& Personal Products & 1 & 1 & 1 & 1 & 1 & 2 & 2 & 1 & 1 & 3 \\
KROGER CO & Food, Beverage \& Tobacco & 1 & 1 & 1 & 1 & 1 & 1 & 1 & 1 & 1 & 3 \\
L3HARRIS TECHNOLOGIES INC & Capital Goods & 1 & 2 & 3 & 3 & --- & 1 & 1 & 1 & --- & 3 \\
LABCORP HOLDINGS INC & Health Care Equipment \& Services & 1 & 1 & 3 & 3 & 3 & 3 & 3 & 3 & 3 & 3 \\
LAM RESEARCH CORP & Semiconductors \& Semiconductor Equipment & 1 & 1 & 1 & 1 & 2 & 1 & 1 & 1 & 2 & 1 \\
LAMB WESTON HOLDINGS INC & Household \& Personal Products & --- & 1 & 1 & 1 & 1 & 1 & 1 & 2 & 2 & 1 \\
LAS VEGAS SANDS CORP & Consumer Services & 1 & 1 & 1 & 1 & 1 & 1 & 1 & 1 & 1 & 1 \\
LEIDOS HOLDINGS INC & Commercial \& Professional Services & 3 & 1 & 1 & --- & 1 & 3 & 3 & 3 & --- & 3 \\
LENNAR CORP & Consumer Durables \& Apparel & 1 & 1 & 1 & 1 & 1 & 1 & 1 & 1 & 1 & 2 \\
LENNOX INTERNATIONAL INC & Capital Goods & 1 & 1 & 1 & 1 & 1 & 1 & 1 & 1 & 1 & 3 \\
LILLY (ELI) \& CO & Pharmaceuticals, Biotechnology \& Life Sciences & 1 & 1 & 1 & 1 & 1 & 1 & 1 & 3 & 3 & 3 \\
LINDE PLC & Materials & --- & 1 & 1 & 1 & 1 & 1 & 1 & 1 & 1 & 2 \\
LIVE NATION ENTERTAINMENT & --- & 1 & 1 & 1 & 1 & 1 & 1 & 1 & 1 & 1 & 1 \\
LKQ CORP & Retailing & 1 & 1 & 1 & 1 & 3 & 1 & 3 & 4 & 1 & 1 \\
LOCKHEED MARTIN CORP & Capital Goods & 1 & 1 & 1 & 1 & 2 & 2 & 2 & 2 & 3 & 3 \\
LOEWS CORP & Insurance & 1 & 1 & 1 & 1 & 1 & 1 & 1 & 1 & 3 & 2 \\
LOWE'S COS INC & Retailing & 1 & 1 & 1 & 1 & 1 & 1 & 1 & 1 & 2 & 3 \\
LULULEMON ATHLETICA INC & Consumer Durables \& Apparel & 1 & 1 & 1 & 1 & 1 & 1 & 1 & 1 & 1 & 1 \\
LYONDELLBASELL INDUSTRIES & Materials & 1 & 1 & 1 & 1 & 1 & 1 & 1 & 1 & 2 & 2 \\
MARATHON PETROLEUM CORP & Energy & 1 & 1 & 1 & 1 & 1 & 1 & 1 & 1 & 3 & 3 \\
MARKETAXESS HOLDINGS INC & Diversified Financials & 1 & 1 & 3 & 3 & 3 & 1 & 1 & 3 & 5 & 5 \\
MARRIOTT INTL INC & Consumer Services & 1 & 1 & 1 & 1 & 1 & 1 & 1 & 1 & 3 & 3 \\
MARSH & Insurance & 1 & 2 & 2 & 1 & 2 & 3 & 3 & 3 & 3 & 4 \\
MARTIN MARIETTA MATERIALS & Materials & 1 & 1 & 1 & 1 & 3 & 1 & 1 & 1 & 1 & 1 \\
MASCO CORP & Capital Goods & 1 & 1 & 1 & 1 & 1 & 1 & 1 & 1 & 1 & 1 \\
MASTERCARD INC & Diversified Financials & 4 & 5 & 5 & 3 & 4 & 4 & 3 & 4 & 5 & 5 \\
MATCH GROUP INC & --- & 1 & 1 & 1 & 1 & 1 & 1 & 1 & 3 & 3 & 3 \\
MCCORMICK \& CO INC & Household \& Personal Products & 1 & 1 & 1 & 1 & 1 & 1 & 1 & 1 & 1 & 2 \\
MCDONALD'S CORP & Consumer Services & 1 & 1 & 1 & 1 & 1 & 1 & 1 & 3 & 3 & 3 \\
MCKESSON CORP & Health Care Equipment \& Services & 1 & 1 & 1 & 1 & 1 & 1 & 1 & 1 & 2 & 2 \\
MEDTRONIC PLC & Health Care Equipment \& Services & 1 & 1 & 1 & 1 & 3 & 3 & 3 & 4 & 4 & 4 \\
MERCK \& CO INC & Pharmaceuticals, Biotechnology \& Life Sciences & 1 & 1 & 1 & 1 & 1 & 1 & 1 & 3 & 3 & 3 \\
META PLATFORMS INC & --- & 2 & 2 & 2 & 3 & 5 & 3 & 5 & 5 & 5 & 5 \\
METLIFE INC & Insurance & 1 & 1 & 1 & 1 & 2 & 2 & 1 & 2 & 2 & 3 \\
METTLER-TOLEDO INTL INC & Pharmaceuticals, Biotechnology \& Life Sciences & 1 & 1 & 1 & 1 & 1 & 1 & 1 & 3 & 3 & 3 \\
MGM RESORTS INTERNATIONAL & Consumer Services & 1 & 1 & 1 & 1 & 1 & 1 & 1 & 1 & 3 & 3 \\
MICROCHIP TECHNOLOGY INC & Semiconductors \& Semiconductor Equipment & 1 & 1 & 1 & 1 & 1 & 1 & 2 & 1 & 2 & 3 \\
MICRON TECHNOLOGY INC & Semiconductors \& Semiconductor Equipment & 1 & 2 & 2 & 3 & 3 & 3 & 3 & 4 & 5 & 5 \\
MICROSOFT CORP & Software \& Services & 3 & 5 & 5 & 5 & --- & 5 & 5 & 5 & 5 & 5 \\
MID-AMERICA APT CMNTYS INC & Real Estate & 1 & 1 & 1 & 1 & 1 & 1 & 1 & 1 & 1 & 2 \\
MODERNA INC & Pharmaceuticals, Biotechnology \& Life Sciences & --- & --- & 3 & 3 & 3 & 5 & 4 & 5 & 5 & 5 \\
MOHAWK INDUSTRIES INC & Consumer Durables \& Apparel & 1 & 1 & 1 & 1 & 1 & 1 & 1 & 1 & 1 & 3 \\
MOLINA HEALTHCARE INC & Health Care Equipment \& Services & 1 & 1 & 3 & 1 & 1 & 1 & 1 & 3 & 3 & 4 \\
MOLSON COORS BEVERAGE CO & Household \& Personal Products & 1 & 1 & 1 & 1 & 1 & 1 & 1 & 2 & 2 & 3 \\
MONDELEZ INTERNATIONAL INC & Household \& Personal Products & 1 & 1 & 1 & 1 & 1 & 1 & 1 & 1 & 2 & 2 \\
MONOLITHIC POWER SYSTEMS INC & Semiconductors \& Semiconductor Equipment & 1 & 1 & 1 & 1 & 1 & 1 & 1 & 4 & 4 & 2 \\
MONSTER BEVERAGE CORP & Household \& Personal Products & 1 & 1 & 1 & 1 & 1 & 1 & 1 & 1 & 1 & 3 \\
MOODY'S CORP & Diversified Financials & 1 & 1 & 2 & 1 & 1 & 1 & 1 & 3 & 4 & 4 \\
MORGAN STANLEY & Diversified Financials & 1 & 1 & 1 & 1 & 1 & 1 & 1 & 1 & 1 & 2 \\
MOSAIC CO & Materials & 1 & 1 & 1 & 1 & 1 & 1 & 1 & 1 & 1 & 1 \\
MOTOROLA SOLUTIONS INC & Technology Hardware \& Equipment & 1 & 1 & 3 & 3 & 3 & 3 & 3 & 3 & 3 & 5 \\
MSCI INC & Diversified Financials & 1 & 1 & 4 & 3 & 3 & 3 & 3 & 3 & 3 & 3 \\
M\&T BANK CORP & Banks & 1 & 1 & 1 & 1 & 1 & 1 & 1 & 1 & 1 & 1 \\
NASDAQ INC & Diversified Financials & 1 & 3 & 3 & 3 & 3 & 1 & 3 & 3 & 5 & 3 \\
NETAPP INC & Technology Hardware \& Equipment & 1 & 1 & 3 & 3 & 3 & 3 & 3 & 4 & 3 & 4 \\
NETFLIX INC & --- & 1 & 1 & 1 & 1 & 1 & 1 & 1 & 1 & 1 & 1 \\
NEWMONT CORP & Materials & 1 & 1 & 1 & 1 & 1 & 1 & 1 & 3 & 3 & 3 \\
NEWS CORP & --- & 1 & 2 & 3 & 1 & 1 & 1 & 2 & 1 & 3 & 3 \\
NEXTERA ENERGY INC & Utilities & 1 & 1 & 1 & 1 & 1 & 1 & 1 & 1 & 1 & 3 \\
NIKE INC  -CL B & Consumer Durables \& Apparel & 1 & 1 & 1 & 1 & 1 & 1 & 1 & 1 & 1 & 3 \\
NISOURCE INC & Utilities & 1 & 1 & 1 & 1 & 1 & 1 & 1 & 1 & 1 & 1 \\
NORDSON CORP & Capital Goods & 1 & 1 & 1 & 1 & 1 & 1 & 1 & 1 & 1 & 1 \\
NORFOLK SOUTHERN CORP & Transportation & 1 & 1 & 1 & 1 & 1 & 1 & 1 & 1 & 1 & 3 \\
NORTHERN TRUST CORP & Diversified Financials & 1 & 1 & 1 & 2 & 2 & 2 & 1 & 2 & 4 & 3 \\
NORTHROP GRUMMAN CORP & Capital Goods & 1 & 1 & 1 & 1 & 1 & 3 & 3 & 3 & 3 & 2 \\
NORWEGIAN CRUISE LINE HLDGS & Consumer Services & 1 & 1 & 1 & 1 & 1 & 1 & 1 & 1 & 3 & 3 \\
NRG ENERGY INC & Utilities & 1 & 1 & 1 & 1 & 1 & 1 & 1 & 3 & 3 & 3 \\
NUCOR CORP & Materials & 1 & 1 & 1 & 1 & 1 & 1 & 1 & 1 & 1 & 1 \\
NVIDIA CORP & Semiconductors \& Semiconductor Equipment & 5 & 5 & 5 & 5 & 5 & 5 & 5 & 5 & 5 & 5 \\
NVR INC & Consumer Durables \& Apparel & 1 & 1 & 1 & 1 & 1 & 1 & 1 & 1 & 1 & 1 \\
NXP SEMICONDUCTORS NV & Semiconductors \& Semiconductor Equipment & --- & --- & --- & 3 & 2 & 3 & 3 & 3 & 3 & 5 \\
O'REILLY AUTOMOTIVE INC & Retailing & 1 & 1 & 1 & 1 & 1 & 1 & 1 & 1 & 1 & 1 \\
OCCIDENTAL PETROLEUM CORP & Energy & 1 & 1 & 1 & 1 & 1 & 1 & 1 & 1 & 1 & 3 \\
OLD DOMINION FREIGHT & Transportation & 1 & 1 & 1 & 1 & 1 & 1 & 1 & 1 & 3 & 3 \\
OMNICOM GROUP INC & --- & 1 & 1 & 1 & 1 & 1 & 1 & 1 & 2 & 3 & 3 \\
ON SEMICONDUCTOR CORP & Semiconductors \& Semiconductor Equipment & 1 & 3 & 3 & 2 & 1 & 1 & 1 & 2 & 3 & 3 \\
ONEOK INC & Energy & 1 & 1 & 1 & 1 & 1 & 1 & 1 & 1 & 1 & 1 \\
ORACLE CORP & Software \& Services & 1 & 1 & 4 & 4 & 4 & 4 & 3 & 4 & 5 & 5 \\
OTIS WORLDWIDE CORP & Capital Goods & --- & --- & --- & --- & 1 & 1 & 1 & 1 & 1 & 2 \\
PACCAR INC & Capital Goods & 1 & 2 & 1 & 1 & 1 & 1 & 1 & 1 & 1 & 4 \\
PACKAGING CORP OF AMERICA & Materials & --- & 1 & 1 & 1 & 1 & 1 & 1 & 1 & 1 & 1 \\
PALANTIR TECHNOLOG INC & Software \& Services & --- & --- & --- & --- & 3 & 3 & 5 & 3 & 5 & 5 \\
PALO ALTO NETWORKS INC & Software \& Services & 1 & 1 & 3 & 3 & 5 & 5 & 3 & 5 & 5 & 5 \\
PARAMOUNT SKYDANCE CORP & --- & --- & --- & --- & --- & --- & --- & --- & --- & --- & 2 \\
PARKER-HANNIFIN CORP & Capital Goods & 1 & 1 & 1 & 1 & 1 & 1 & 1 & 1 & 1 & 1 \\
PAYCHEX INC & Commercial \& Professional Services & 1 & 1 & 1 & 3 & 3 & 3 & 3 & 4 & 4 & 4 \\
PAYCOM SOFTWARE INC & Commercial \& Professional Services & 1 & 1 & 3 & 3 & 3 & 3 & 3 & 3 & 4 & 5 \\
PAYPAL HOLDINGS INC & Diversified Financials & 1 & 1 & 1 & 3 & 3 & 3 & 3 & 1 & 3 & 3 \\
PENTAIR PLC & Capital Goods & 1 & 1 & 1 & 1 & 1 & 1 & 1 & 1 & 1 & 3 \\
PEPSICO INC & Household \& Personal Products & 1 & 1 & 1 & 2 & 1 & 3 & 2 & 3 & 4 & 4 \\
PFIZER INC & Pharmaceuticals, Biotechnology \& Life Sciences & 1 & 1 & 1 & 1 & 1 & 1 & 1 & 1 & 2 & 4 \\
PG\&E CORP & Utilities & 1 & 1 & 1 & 1 & 1 & 1 & 1 & 1 & 1 & 1 \\
PHILIP MORRIS INTERNATIONAL & Household \& Personal Products & 1 & 1 & 1 & 1 & 1 & 1 & 2 & 1 & 1 & 3 \\
PHILLIPS 66 & Energy & 1 & 1 & 1 & 1 & 1 & 1 & 1 & 1 & 1 & 1 \\
PINNACLE WEST CAPITAL CORP & Utilities & 1 & 1 & 1 & 1 & 1 & 1 & 1 & 2 & 2 & 3 \\
PNC FINANCIAL SVCS GROUP INC & Banks & 1 & 2 & 2 & 3 & 3 & 3 & 3 & 4 & 3 & 3 \\
POOL CORP & Retailing & 1 & 1 & 1 & 1 & 1 & 1 & 1 & 1 & 1 & 2 \\
PPG INDUSTRIES INC & Materials & 1 & 1 & 1 & 1 & 1 & 1 & 1 & 1 & 1 & 3 \\
PPL CORP & Utilities & 1 & 1 & 1 & 1 & 1 & 1 & 1 & 1 & 3 & 3 \\
PRICE (T. ROWE) GROUP & Diversified Financials & 1 & 1 & 1 & 1 & 1 & 1 & 1 & 2 & 3 & 3 \\
PRINCIPAL FINANCIAL GRP INC & Insurance & 1 & 1 & 1 & 1 & 1 & 1 & 1 & 1 & 1 & 2 \\
PROCTER \& GAMBLE CO & Consumer Staples Distribution \& Retail & 1 & 1 & 1 & 1 & 1 & 1 & 1 & 1 & 2 & 2 \\
PROGRESSIVE CORP-OHIO & Insurance & 1 & 3 & 4 & 3 & 1 & 3 & 3 & 3 & 3 & 3 \\
PROLOGIS INC & Real Estate & 1 & 1 & 1 & 1 & 1 & 1 & 1 & 1 & 1 & 1 \\
PRUDENTIAL FINANCIAL INC & Insurance & 1 & 1 & 1 & 1 & 1 & 1 & 2 & 2 & 2 & 2 \\
PTC INC & Software \& Services & 3 & 3 & 3 & 3 & 3 & 3 & 1 & 1 & 3 & 3 \\
PUBLIC SERVICE ENTRP GRP INC & Utilities & 1 & 1 & 1 & 1 & 1 & 1 & 1 & 2 & 2 & 3 \\
PUBLIC STORAGE & Real Estate & 1 & 1 & 1 & 1 & 1 & 1 & 1 & 1 & 3 & 3 \\
PULTEGROUP INC & Consumer Durables \& Apparel & 1 & 1 & 1 & 1 & 1 & 1 & 1 & 1 & 2 & 2 \\
QORVO INC & Semiconductors \& Semiconductor Equipment & 1 & 1 & 1 & 1 & 1 & 1 & 1 & 1 & 1 & 3 \\
QUALCOMM INC & Semiconductors \& Semiconductor Equipment & 2 & 3 & 3 & 3 & 3 & 3 & 3 & 5 & 5 & 5 \\
QUANTA SERVICES INC & Capital Goods & 1 & 1 & 1 & 1 & 1 & 1 & 1 & 1 & 1 & 1 \\
QUEST DIAGNOSTICS INC & Health Care Equipment \& Services & 1 & 1 & 1 & 3 & 3 & 3 & 3 & 4 & 4 & 4 \\
RALPH LAUREN CORP & Consumer Durables \& Apparel & 1 & 1 & 1 & 1 & 1 & 1 & 1 & 4 & 3 & 3 \\
RAYMOND JAMES FINANCIAL INC & Diversified Financials & 1 & 1 & 1 & 1 & 1 & 1 & 1 & 1 & 2 & 3 \\
REALTY INCOME CORP & Real Estate & 1 & 1 & 1 & 1 & 1 & 1 & 1 & 1 & 1 & 3 \\
REGENCY CENTERS CORP & Real Estate & 1 & 1 & 1 & 1 & 1 & 1 & 1 & 1 & 1 & 2 \\
REGENERON PHARMACEUTICALS & Pharmaceuticals, Biotechnology \& Life Sciences & 1 & 1 & 1 & 1 & 1 & 1 & 1 & 2 & 2 & 3 \\
REGIONS FINANCIAL CORP & Banks & 1 & 1 & 1 & 1 & 1 & 3 & 2 & 3 & 3 & 3 \\
REPUBLIC SERVICES INC & Commercial \& Professional Services & 1 & 1 & 1 & 1 & 1 & 1 & 1 & 1 & 1 & 3 \\
RESMED INC & Health Care Equipment \& Services & 1 & 1 & 1 & 1 & 1 & 1 & 3 & 3 & 3 & 4 \\
REVVITY INC & Pharmaceuticals, Biotechnology \& Life Sciences & 1 & 1 & 1 & 1 & --- & 1 & 1 & 3 & 3 & 3 \\
ROBINHOOD MARKETS INC & Diversified Financials & --- & --- & --- & --- & --- & 3 & 3 & 3 & 3 & 3 \\
ROCKWELL AUTOMATION & Capital Goods & 1 & 1 & 1 & 1 & 1 & 3 & 3 & 3 & 3 & 3 \\
ROLLINS INC & Commercial \& Professional Services & 1 & 1 & 1 & 1 & 1 & 1 & 1 & 1 & 1 & 3 \\
ROPER TECHNOLOGIES INC & Software \& Services & 1 & 1 & 1 & 1 & 1 & 1 & 1 & 3 & 3 & 3 \\
ROSS STORES INC & Retailing & 1 & 1 & 1 & 1 & 1 & 1 & 1 & 1 & 1 & 1 \\
ROYAL CARIBBEAN GROUP & Consumer Services & 1 & 1 & 1 & 3 & 3 & 3 & 3 & 3 & 3 & 4 \\
RTX CORP & Capital Goods & 1 & 1 & 1 & 1 & 1 & 1 & 3 & 3 & 3 & 3 \\
SALESFORCE INC & Software \& Services & 3 & 3 & 3 & 3 & 3 & 3 & 3 & 3 & 5 & 5 \\
SANDISK CORP & Technology Hardware \& Equipment & --- & --- & --- & --- & --- & --- & --- & --- & --- & 3 \\
SBA COMMUNICATIONS CORP & Real Estate & 1 & 1 & 1 & 1 & 1 & 1 & 1 & 1 & 1 & 2 \\
SCHWAB (CHARLES) CORP & Diversified Financials & 1 & 1 & 1 & 1 & 1 & 1 & 1 & 1 & 3 & 3 \\
SEAGATE TECHNOLOGY HOLDINGS & Technology Hardware \& Equipment & 1 & 1 & 3 & 3 & --- & 3 & 3 & 3 & 3 & 3 \\
SEMPRA ENERGY & Utilities & 1 & 1 & 1 & 1 & 1 & 1 & 1 & 1 & 2 & 3 \\
SERVICENOW INC & Software \& Services & 1 & --- & 1 & 3 & 3 & 3 & 3 & 5 & 5 & 5 \\
SHERWIN-WILLIAMS CO & Materials & 1 & 1 & 1 & 1 & 1 & 1 & 1 & 1 & 1 & 1 \\
SIMON PROPERTY GROUP INC & Real Estate & 1 & 1 & 1 & 1 & 1 & 1 & 1 & 1 & 3 & 3 \\
SKYWORKS SOLUTIONS INC & Semiconductors \& Semiconductor Equipment & 1 & 1 & 3 & 1 & 1 & 1 & 1 & 2 & 2 & 2 \\
SLB LTD & Energy & 1 & 1 & 1 & 1 & 1 & 3 & 4 & 4 & 4 & 5 \\
SMITH (A.O.) & Capital Goods & 1 & 1 & 1 & 1 & 1 & 1 & 1 & 1 & 1 & 1 \\
SMUCKER (JM) CO & Household \& Personal Products & 1 & 1 & 1 & 1 & 1 & 1 & 1 & 1 & 2 & 2 \\
SMURFIT WESTROCK PLC & Materials & --- & --- & --- & --- & --- & --- & --- & --- & 3 & 3 \\
SNAP-ON INC & Capital Goods & 1 & 1 & 1 & 1 & --- & 1 & 1 & 1 & 1 & --- \\
SOLVENTUM CORP & Health Care Equipment \& Services & --- & --- & --- & --- & --- & --- & --- & --- & 2 & 3 \\
SOUTHERN CO & Utilities & 1 & 1 & 1 & 1 & 1 & 1 & 1 & 1 & 1 & 1 \\
SOUTHWEST AIRLINES & Transportation & 1 & 1 & 1 & 1 & 1 & 1 & 1 & 2 & 4 & 4 \\
STANLEY BLACK \& DECKER INC & Capital Goods & 1 & 2 & 2 & 3 & --- & 3 & 3 & 1 & 2 & --- \\
STARBUCKS CORP & Consumer Services & 1 & 1 & 1 & 1 & 1 & 1 & 1 & 1 & 3 & 2 \\
STATE STREET CORP & Diversified Financials & 1 & 1 & 1 & 1 & 2 & 3 & 1 & 3 & 3 & 4 \\
STEEL DYNAMICS INC & Materials & 1 & 1 & 1 & 1 & 1 & 1 & 1 & 1 & 1 & 1 \\
STERIS PLC & Health Care Equipment \& Services & --- & --- & --- & 1 & 1 & 1 & 1 & 1 & 3 & 3 \\
STRYKER CORP & Health Care Equipment \& Services & 1 & 1 & 1 & 1 & 1 & 1 & 1 & 3 & 3 & 3 \\
SUPER MICRO COMPUTER INC & Technology Hardware \& Equipment & 1 & 3 & --- & 3 & 3 & 3 & 3 & 3 & 4 & 5 \\
SYNCHRONY FINANCIAL & Diversified Financials & 1 & 3 & 3 & 3 & 4 & 3 & 3 & 3 & 3 & 3 \\
SYNOPSYS INC & Software \& Services & 1 & 1 & 1 & 3 & 3 & 3 & 3 & 4 & 5 & 5 \\
SYSCO CORP & Food, Beverage \& Tobacco & 1 & 1 & 1 & 1 & 1 & 1 & 1 & 1 & 3 & 3 \\
S\&P GLOBAL INC & Diversified Financials & 1 & 3 & 3 & 3 & 3 & 3 & 3 & 5 & 5 & 4 \\
T-MOBILE US INC & Telecommunication Services & 1 & 1 & 1 & 1 & 1 & 1 & 1 & 2 & 3 & 4 \\
TAKE-TWO INTERACTIVE SFTWR & --- & 1 & 1 & 1 & 1 & 1 & 1 & 1 & 1 & 1 & 1 \\
TAPESTRY INC & Consumer Durables \& Apparel & 1 & 1 & 1 & 1 & 1 & 1 & 1 & 3 & 3 & 3 \\
TARGA RESOURCES CORP & Energy & 1 & 1 & 1 & 1 & 1 & 1 & 1 & 1 & 1 & 1 \\
TARGET CORP & Food, Beverage \& Tobacco & 1 & 1 & 1 & 1 & 1 & 1 & 1 & 1 & 2 & 3 \\
TE CONNECTIVITY PLC & Technology Hardware \& Equipment & 1 & 1 & 1 & 1 & 1 & 1 & 2 & 3 & 4 & 4 \\
TELEDYNE TECHNOLOGIES INC & Technology Hardware \& Equipment & 1 & 3 & 1 & 1 & --- & 1 & 3 & 3 & 3 & 3 \\
TERADYNE INC & Semiconductors \& Semiconductor Equipment & 1 & 1 & 1 & 1 & 1 & 1 & 1 & 3 & 2 & 5 \\
TESLA INC & Automobiles \& Components & 3 & 3 & 3 & 5 & 3 & 5 & 4 & 4 & 5 & 5 \\
TEXAS INSTRUMENTS INC & Semiconductors \& Semiconductor Equipment & 1 & 1 & 1 & 1 & 1 & 1 & 1 & 1 & 1 & 1 \\
TEXAS PACIFIC LAND CORP & Energy & --- & --- & --- & --- & 1 & 1 & 1 & 1 & 1 & 1 \\
TEXTRON INC & Capital Goods & 1 & 1 & 1 & --- & 1 & 1 & 1 & 1 & 1 & --- \\
THERMO FISHER SCIENTIFIC INC & Pharmaceuticals, Biotechnology \& Life Sciences & 1 & 1 & 1 & 1 & 1 & 1 & 2 & 1 & 1 & 3 \\
TJX COS INC (THE) & Retailing & 1 & 1 & 1 & 1 & 1 & 1 & 1 & 1 & 1 & 1 \\
TKO GROUP HOLDINGS INC & --- & --- & --- & --- & --- & --- & --- & --- & 1 & 1 & 2 \\
TRACTOR SUPPLY CO & Retailing & 1 & 2 & 2 & 2 & 2 & 2 & 2 & 3 & 3 & 3 \\
TRADE DESK INC & --- & 3 & 3 & 3 & 3 & 3 & 3 & 4 & 3 & 3 & 5 \\
TRANE TECHNOLOGIES PLC & Capital Goods & 1 & 1 & 1 & 1 & 1 & 1 & 1 & 1 & 3 & 3 \\
TRANSDIGM GROUP INC & Capital Goods & 1 & 1 & 1 & 1 & 1 & 1 & 1 & 1 & 1 & 1 \\
TRAVELERS COS INC & Insurance & 1 & 1 & 3 & 2 & 2 & 3 & 3 & 3 & 3 & 3 \\
TRIMBLE INC & Software \& Services & 3 & 3 & 3 & --- & 3 & 3 & 3 & 3 & --- & 5 \\
TRUIST FINANCIAL CORP & Banks & 3 & 2 & 3 & 3 & 1 & 3 & 1 & 3 & 3 & 3 \\
TYLER TECHNOLOGIES INC & Software \& Services & 1 & 1 & 1 & 1 & 1 & 1 & 1 & 3 & 3 & 3 \\
TYSON FOODS INC  -CL A & Household \& Personal Products & 2 & 1 & 2 & 2 & 2 & 2 & 3 & 4 & 1 & 1 \\
UBER TECHNOLOGIES INC & Transportation & --- & --- & --- & 1 & 3 & 3 & 3 & 3 & 2 & 3 \\
UDR INC & Real Estate & 1 & 1 & 1 & 1 & 1 & 1 & 2 & 1 & 1 & 1 \\
ULTA BEAUTY INC & Retailing & 1 & 1 & 1 & 3 & 3 & 3 & 3 & 3 & 3 & 3 \\
UNION PACIFIC CORP & Transportation & 1 & 1 & 1 & 1 & 1 & 1 & 1 & 1 & 1 & 1 \\
UNITED AIRLINES HOLDINGS INC & Transportation & 1 & 1 & 1 & 1 & 1 & 1 & 1 & 3 & 3 & 3 \\
UNITED PARCEL SERVICE INC & Transportation & 1 & 1 & 1 & 1 & 1 & 1 & 1 & 1 & 1 & 3 \\
UNITED RENTALS INC & Capital Goods & 1 & 1 & 1 & 1 & 1 & 1 & 1 & 1 & 1 & 3 \\
UNITEDHEALTH GROUP INC & Health Care Equipment \& Services & 3 & 3 & 3 & 3 & 3 & 3 & 2 & 3 & 3 & 3 \\
UNIVERSAL HEALTH SVCS INC & Health Care Equipment \& Services & 1 & 1 & 1 & 1 & 1 & 1 & 1 & 1 & 1 & 2 \\
US BANCORP & Banks & 1 & 1 & 1 & 1 & 1 & 1 & 1 & 2 & 3 & 3 \\
VALERO ENERGY CORP & Energy & 1 & 1 & 1 & 1 & 1 & 1 & 1 & 1 & 2 & 2 \\
VENTAS INC & Real Estate & 1 & 1 & 1 & 1 & 1 & 1 & 1 & 1 & 1 & 3 \\
VERALTO CORP & Commercial \& Professional Services & --- & --- & --- & --- & --- & --- & --- & 1 & 1 & 2 \\
VERISIGN INC & Software \& Services & 1 & 1 & 1 & 1 & 1 & 1 & 1 & 1 & 1 & 3 \\
VERISK ANALYTICS INC & Commercial \& Professional Services & 3 & 5 & 4 & 3 & 3 & 5 & 4 & 4 & 4 & 4 \\
VERIZON COMMUNICATIONS INC & Telecommunication Services & 1 & 1 & 1 & 1 & 1 & 1 & 1 & 1 & 4 & 3 \\
VERTEX PHARMACEUTICALS INC & Pharmaceuticals, Biotechnology \& Life Sciences & 1 & 1 & 1 & 1 & 1 & 1 & 1 & 1 & 1 & 2 \\
VIATRIS INC & Pharmaceuticals, Biotechnology \& Life Sciences & --- & --- & --- & 1 & 1 & 1 & 1 & 2 & 3 & 2 \\
VICI PROPERTIES INC & Real Estate & --- & 1 & 1 & 1 & 1 & 1 & 1 & 1 & 1 & 1 \\
VISA INC & Diversified Financials & 1 & 4 & 3 & 3 & 4 & 1 & 1 & 3 & 5 & 4 \\
VISTRA CORP & Utilities & --- & 1 & 1 & 1 & 1 & 1 & 1 & 2 & 2 & 2 \\
VULCAN MATERIALS CO & Materials & 1 & 1 & 1 & 1 & 1 & 1 & 1 & 1 & 1 & 2 \\
WABTEC CORP & Capital Goods & 1 & 1 & 1 & 1 & 3 & 3 & 4 & 2 & 2 & 3 \\
WALGREENS BOOTS ALLIANCE INC & Food, Beverage \& Tobacco & 1 & 1 & 1 & 1 & 1 & 1 & 1 & 1 & 2 & --- \\
WALMART INC & Food, Beverage \& Tobacco & 1 & 1 & 1 & 1 & 1 & 1 & 1 & 1 & 2 & 3 \\
WARNER BROS DISCOVERY INC & --- & 1 & 1 & 1 & 1 & 1 & 1 & 1 & 1 & 1 & 1 \\
WASTE MANAGEMENT INC & Commercial \& Professional Services & 1 & 1 & 1 & 1 & 1 & 1 & 2 & 2 & 2 & 2 \\
WATERS CORP & Pharmaceuticals, Biotechnology \& Life Sciences & 1 & 1 & 1 & 1 & 1 & 1 & 1 & 1 & 2 & 3 \\
WEC ENERGY GROUP INC & Utilities & 1 & 1 & 1 & 1 & 1 & 1 & 1 & 1 & 2 & 2 \\
WELLS FARGO \& CO & Banks & 1 & 1 & 3 & 3 & 3 & 3 & 1 & 3 & 3 & 3 \\
WELLTOWER INC & Real Estate & 1 & 1 & 1 & 1 & 1 & 2 & 2 & 1 & 3 & 3 \\
WEST PHARMACEUTICAL SVSC INC & Pharmaceuticals, Biotechnology \& Life Sciences & 1 & 1 & 1 & 1 & 1 & 1 & 1 & 1 & 1 & 1 \\
WESTERN DIGITAL CORP & Technology Hardware \& Equipment & 1 & 1 & 1 & 1 & 1 & 1 & 1 & 1 & 3 & 2 \\
WEYERHAEUSER CO & Real Estate & 1 & 1 & 1 & 1 & 1 & 3 & 3 & 3 & 3 & 3 \\
WILLIAMS COS INC & Energy & 1 & 1 & 1 & 1 & 1 & 1 & 1 & 1 & 2 & 2 \\
WILLIAMS-SONOMA INC & Retailing & 1 & 1 & 2 & 1 & 1 & 1 & 1 & 2 & 3 & 3 \\
WILLIS TOWERS WATSON PLC & Insurance & 1 & 1 & 1 & 1 & 1 & 1 & 1 & 2 & 2 & 3 \\
WORKDAY INC & Software \& Services & 3 & 3 & 3 & 3 & 3 & 3 & 3 & 3 & 5 & 5 \\
WYNN RESORTS LTD & Consumer Services & 1 & 1 & 1 & 1 & 1 & 1 & 1 & 1 & 1 & 1 \\
XCEL ENERGY INC & Utilities & 1 & 1 & 1 & 1 & 1 & 1 & 2 & 3 & 3 & 1 \\
XYLEM INC & Capital Goods & 1 & 1 & 1 & 1 & 1 & 1 & 1 & 1 & 2 & 3 \\
YUM BRANDS INC & Consumer Services & 1 & 1 & 1 & 1 & 1 & 3 & 3 & 1 & 4 & 3 \\
ZEBRA TECHNOLOGIES CP  -CL A & Technology Hardware \& Equipment & 1 & 1 & 1 & 3 & 3 & 3 & 3 & 3 & 3 & 4 \\
ZIMMER BIOMET HOLDINGS INC & Health Care Equipment \& Services & 1 & 1 & 1 & 1 & 2 & 2 & 3 & 3 & 3 & 4 \\
ZOETIS INC & Pharmaceuticals, Biotechnology \& Life Sciences & 1 & 1 & 1 & 1 & 3 & 3 & 3 & 3 & 3 & 3 \\
\end{longtable}

}

\newpage
\section{The list of AI-relevant keywords}\label{apx:aikeywords}

artificial intelligence,
machine learning,
AI,
ML,
computer vision,
machine vision,
deep learning,
virtual agents,
image recognition,
natural language processing,
speech recognition,
pattern recognition,
object recognition,
neural networks,
AI chatbot,
supervised learning,
text mining,
unsupervised learning,
image processing,
Mahout,
recommender systems,
support vector machines,
random forests,
latent semantic analysis,
sentiment analysis,
opinion mining,
latent Dirichlet allocation,
predictive models,
kernel methods,
Keras,
gradient boosting,
OpenCV,
XGBoost,
Libsvm,
Word2vec,
machine translation,
sentiment classification,
transformer,
adversarial network,
decision tree,
reinforcement learning,
genetic algorithm,
word embedding,
feature extraction,
image segmentation,
principal component analysis,
recognition system,
object detection,
learning model,
sentiment analysis

\newpage
\section{Detailed Summary Statistics}\label{apx:detailstats}

\begin{table}[htbp]
\caption{\bf Detailed Summary Statistics}
    \centering    \begin{tabular}{llllllllll}
\toprule
 &  & count & mean & std & min & 25\% & 50\% & 75\% & max \\
Variable & AI Score &  &  &  &  &  &  &  &  \\
\midrule
\multirow[t]{5}{*}{Headcount} & 1 & 2844 & 53.54 & 138.77 & 0.02 & 8.36 & 18.27 & 50.00 & 2300.00 \\
 & 2 & 326 & 51.19 & 128.37 & 0.16 & 10.20 & 22.61 & 60.00 & 2100.00 \\
 & 3 & 975 & 54.77 & 104.21 & 0.39 & 10.85 & 23.80 & 58.00 & 2100.00 \\
 & 4 & 128 & 64.70 & 93.44 & 1.69 & 10.67 & 27.25 & 79.32 & 510.00 \\
 & 5 & 197 & 136.50 & 283.11 & 0.87 & 11.86 & 39.80 & 121.10 & 1608.00 \\
\cline{1-10}
\multirow[t]{5}{*}{Net Profit Margin} & 1 & 2844 & 0.12 & 0.21 & -6.95 & 0.06 & 0.11 & 0.18 & 1.11 \\
 & 2 & 326 & 0.12 & 0.10 & -0.56 & 0.06 & 0.11 & 0.17 & 0.47 \\
 & 3 & 975 & 0.10 & 0.37 & -8.54 & 0.05 & 0.12 & 0.19 & 1.56 \\
 & 4 & 128 & 0.15 & 0.13 & -0.38 & 0.08 & 0.13 & 0.22 & 0.81 \\
 & 5 & 197 & 0.15 & 0.21 & -1.45 & 0.08 & 0.15 & 0.26 & 0.66 \\
\cline{1-10}
\multirow[t]{5}{*}{Capex-to-Revenue} & 1 & 2844 & 0.09 & 0.14 & 0.00 & 0.02 & 0.04 & 0.09 & 2.74 \\
 & 2 & 326 & 0.07 & 0.10 & 0.00 & 0.02 & 0.03 & 0.06 & 0.62 \\
 & 3 & 975 & 0.06 & 0.10 & 0.00 & 0.01 & 0.03 & 0.06 & 1.46 \\
 & 4 & 128 & 0.06 & 0.07 & 0.00 & 0.02 & 0.04 & 0.06 & 0.49 \\
 & 5 & 197 & 0.07 & 0.10 & 0.00 & 0.02 & 0.03 & 0.09 & 0.58 \\
\cline{1-10}
\multirow[t]{5}{*}{Tobin's Q} & 1 & 2844 & 1.98 & 2.11 & 0.00 & 0.70 & 1.32 & 2.48 & 22.85 \\
 & 2 & 326 & 1.89 & 1.86 & 0.05 & 0.66 & 1.33 & 2.43 & 12.54 \\
 & 3 & 975 & 2.53 & 2.79 & 0.06 & 0.80 & 1.56 & 3.22 & 27.60 \\
 & 4 & 128 & 2.65 & 2.72 & 0.10 & 0.76 & 1.58 & 3.92 & 16.09 \\
 & 5 & 197 & 5.31 & 5.88 & 0.13 & 1.88 & 3.57 & 6.49 & 47.75 \\
\cline{1-10}
\bottomrule
\end{tabular}

    \captionsetup{font=footnotesize}
    \vspace{0.1in}
    \caption*{\textit{Note: }\emph{Headcount} is in thousand, \emph{Net profit margin} is net income-to-revenue, \emph{Tobin's Q} is market value-to-book value.}
    \label{tab:detailed_summarystat}
\end{table}
\section{AI Score Distribution by Broad Sector}\label{sec:aiscore_value_bysector}

\begin{table}[H]
    \caption{\bf AI Adoption Score Distribution by Broad Sector}
    \centering
    \makebox[\textwidth][l]{\hspace*{2cm}\begin{tabular}{lrrrrr}
\toprule
 & \textbf{2021} & \textbf{2022} & \textbf{2023} & \textbf{2024} & \textbf{2025} \\
\midrule
\textbf{All Enterprises} \\
5 Deep Integration & 3.7\% & 3.1\% & 5.4\% & 9.0\% & 11.4\% \\
4 Used in Production & 0.8\% & 1.8\% & 4.8\% & 6.4\% & 10.0\% \\
3 In Pilot Phase & 22.1\% & 21.9\% & 26.2\% & 37.5\% & 45.2\% \\
2 Exploration & 4.9\% & 8.2\% & 13.5\% & 15.8\% & 15.3\% \\
1 No Mention & 68.5\% & 65.0\% & 50.1\% & 31.3\% & 18.1\% \\
% Total & 100.0\% & 100.0\% & 100.0\% & 100.0\% & 100.0\% \\
\midrule
\textbf{Technology}  \\
% \midrule
5 Deep Integration & 14.9\% & 13.5\% & 25.3\% & 36.0\% & 50.0\% \\
4 Used in Production & 1.4\% & 1.4\% & 10.7\% & 14.7\% & 11.8\% \\
3 In Pilot Phase & 39.2\% & 39.2\% & 33.3\% & 34.7\% & 27.6\% \\
2 Exploration & 2.7\% & 6.8\% & 8.0\% & 6.7\% & 5.3\% \\
1 No Mention & 41.9\% & 39.2\% & 22.7\% & 8.0\% & 5.3\% \\
% Total & 100.0\% & 100.0\% & 100.0\% & 100.0\% & 100.0\% \\
\midrule
\textbf{Financial}  \\
% \midrule
5 Deep Integration & 0.0\% & 0.0\% & 1.9\% & 9.4\% & 3.8\% \\
4 Used in Production & 2.0\% & 0.0\% & 7.7\% & 5.7\% & 18.9\% \\
3 In Pilot Phase & 37.3\% & 32.7\% & 42.3\% & 64.2\% & 67.9\% \\
2 Exploration & 7.8\% & 9.6\% & 21.2\% & 13.2\% & 5.7\% \\
1 No Mention & 52.9\% & 57.7\% & 26.9\% & 7.5\% & 3.8\% \\
Total & 100.0\% & 100.0\% & 100.0\% & 100.0\% & 100.0\% \\
\midrule
\textbf{All Other} \\
% \midrule
5 Deep Integration & 1.5\% & 1.0\% & 1.9\% & 4.5\% & 5.7\% \\
4 Used in Production & 0.5\% & 2.0\% & 4.3\% & 6.4\% & 10.9\% \\
3 In Pilot Phase & 20.7\% & 20.4\% & 25.7\% & 37.9\% & 47.5\% \\
2 Exploration & 4.9\% & 7.6\% & 14.7\% & 17.1\% & 17.1\% \\
1 No Mention & 72.3\% & 69.0\% & 53.4\% & 34.0\% & 18.8\% \\
% Total & 100.0\% & 100.0\% & 100.0\% & 100.0\% & 100.0\% \\
\bottomrule
\end{tabular}
}
    \captionsetup{font=footnotesize}
    \caption*{\textit{Note: }Values are the share (in \%) of firms in each AI adoption score category by year and sector grouping.}
    \label{tab:score_distribution_sector_2021_2025}
\end{table}

\end{document}